\documentclass[reqno]{amsart}

\usepackage[unicode,psdextra]{hyperref}
\hypersetup{
    colorlinks,
    linkcolor={red!50!black},
    citecolor={blue!50!black},
    urlcolor={blue!90!black}
}

\usepackage{graphicx}
\usepackage{xcolor}
\usepackage{xfrac}

\usepackage[english]{babel}
\usepackage[T1]{fontenc}
\usepackage{libertine}

\newcommand{\pd}{\partial}

\newcommand{\dd}{\mathrm{d}}
\newcommand{\ii}{\mathrm{i}}
\newcommand{\pdv}[2]{\frac{\partial #1}{\partial #2}}
\newcommand{\dv}[2]{\frac{\mathrm{d} #1}{\mathrm{d} #2}}
\newcommand{\Res}{\;\mathrm{Res}}
\newcommand{\erf}{\;\mathrm{erf}}

\makeatletter
\renewcommand\subsubsection{\@startsection{subsubsection}{3}%
  \z@{.5\linespacing\@plus.7\linespacing}{-.5em}%
  {\normalfont\bfseries}}
\makeatother

\makeatletter
\renewcommand\paragraph{\@startsection{paragraph}{4}%
  \z@{.5\linespacing\@plus.7\linespacing}{-.5em}%
  {\normalfont\bfseries}*}
\makeatother

\usepackage[
    backend=biber,
    style=phys,
    citestyle=numeric-comp,
    sorting=none,
    eprint=true,
    doi=false,
    url=false,
    biblabel=brackets
]{biblatex}

\usepackage[foot]{amsaddr}

\begin{document}

\title{First-passage properties of the jump process with a drift. Two exactly solvable cases.}
\author{Ivan N. Burenev}
\author{Satya N. Majumdar}
\email{inburenev@gmail.com}
\address{LPTMS, CNRS, Universit\'e Paris-Saclay, 91405 Orsay, France}

\begin{abstract}
We investigate the first-passage properties of a jump process with a constant drift, focusing on two key observables: the first-passage time $\tau$ and the number of jumps $n$ before the first-passage event.
By mapping the problem onto an effective discrete-time random walk, we derive an exact expression for the Laplace transform of the joint distribution of $\tau$ and $n$ using the generalized Pollaczek-Spitzer formula. 
This result is then used to analyze the first-passage properties for two exactly solvable cases: (i) both the inter-jump intervals and jump amplitudes are exponentially distributed, and (ii) the inter-jump intervals are exponentially distributed while all jumps have the same fixed amplitude.
We show the existence of two distinct regimes governed by the strength of the drift: (i) a \textit{survival regime}, where the process remains positive indefinitely with finite probability; (ii) an \textit{absorption regime}, where the first-passage eventually occurs; and (iii) a \textit{critical point} at the boundary between these two phases. 
We characterize the asymptotic behavior of survival probabilities in each regime: they decay exponentially to a constant in the survival regime, vanish exponentially fast in the absorption regime, and exhibit power-law decay at the critical point. Furthermore, in the absorption regime, we derive large deviation forms for the marginal distributions of $\tau$ and $n$. The analytical predictions are validated through extensive numerical simulations.
\end{abstract}

\maketitle

\vspace{-.5cm}

\tableofcontents

\section{Introduction}\label{sec:intro}

\par First-passage problems concern the time at which a stochastic process first reaches a boundary or a target. They have a long-standing history, with significant contributions dating back to the 1950s and since then they have been widely studied across various contexts in both mathematics and physics (see e{.}g{.}, \cite{BMS-13} for a physical perspective and \cite{AS-15} for a more mathematical point of view). Despite their apparent simplicity, obtaining exact analytical results remains a challenging task, even for simple one-dimensional systems~\cite{Redner2001,MajumdarSchehr2024}. In this work, we analyze a minimal yet nontrivial model in which a persistent drift competes with discrete stochastic jumps, leading to rich first-passage dynamics.

\par
\begin{figure}[h]
\includegraphics[width=\linewidth]{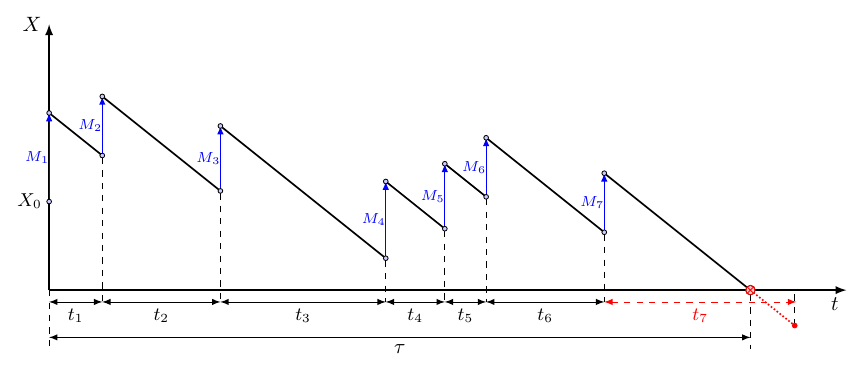}
\caption{
An example of the trajectory. Starting at $X_0$ the process instantaneously undergoes a jump $M_1$, then moves toward the origin with constant velocity $\alpha$ for the time $t_1$, when the next jump $M_2$ occurs. This pattern continues until the process crosses the origin at time $\tau$ after $n$ jumps (here $n=7$). The inter-jump intervals $\{t_1,\ldots,t_7\}$ are \textit{i.i.d.} random variables drawn from $p(t)$, and the jump amplitudes $\{M_1,\ldots,M_7\}$ are also \textit{i.i.d.} random variables following $q(M)$.}
\label{fig:example_trajectory}
\end{figure}

 We consider a jump process with a constant drift, taking place on the positive semi-axis, where the system evolves through a combination of ballistic motion with constant velocity and random jumps. We assume the following:
\begin{enumerate}
    \item The process, starting at $X_0\ge0$, undergoes two alternative moves: jump by a random amount and then a linear decrease with the speed $-\alpha$ during a random time. The first jump occurs at $t=0$ (see Fig.~\ref{fig:example_trajectory}).
    \item The time intervals $t_j$ between consecutive jumps are independent identically distributed (\textit{i.i.d.}) random variables drawn from a distribution $p(t)$.
    \item The amplitudes of the jumps $M_j\ge0$ are also \textit{i.i.d.} following a distribution $q(M)$.
    \item The process stops after first crossing of the origin to the negative side. 
\end{enumerate}

\par The primary goal of this paper is to analyze the first-passage properties of such a process by focusing on two observables: the time $\tau$ at which the process first crosses the origin, and the number of jumps $n$ that occur prior to that. Both $\tau$ and $n$ are random variables that are, in general, correlated. Their statistics are governed by four parameters: the initial position $X_0$, the drift velocity $\alpha$ and, most importantly,  two probability distributions, $p(t)$ and $q(M)$.

\par Although seemingly simple at first glance, the jump process with a constant drift appears in various fields under different names.
In mathematics, it is known as the Markovian growth-collapse process \cites{EK-04,BPSZ-06,LS-11}; in risk theory, it is referred to as the (dual) risk model \cites{GS-05,AGS-07,Asmussen2010,YS-14}; in queueing theory it corresponds to the G/G/1 queue \cites{Bhat2015}, to name a few. 
In the physics literature it arises, for instance, in the studies of avalanches \cites{LLRM-14,SMR-21,JBMM-25}. Alternatively, it can be interpreted as a system with partial resetting, an extension of the well-known stochastic resetting introduced in \cite{EM-11} (see \cite{EMS-20} for a review), which has recently attracted growing attention \cites{DCSM-21,TRS-22,HKPS-23,BCHPM-23,H-23,OL-24,BFHMS-24,OG-24,BCGSTP-25,MOPR-25}.

\par A standard approach to studying first-passage properties relies on renewal equations \cites{Redner2001,MajumdarSchehr2024}. However, this leads to Wiener-Hopf integral equations, which are notoriously difficult to solve. Exact results are generally limited to special cases, typically when the inter-jump intervals follow an exponential or a Gamma distribution.

\par We propose a different approach, wherein we map the original process onto an effective discrete-time random walk. Using the generalized Pollaczek-Spitzer formula \cite{P-52,S-56,S-57}, a key result in the theory of discrete-time random walks, we derive a closed-form expression for the (triple) Laplace transform of the joint probability distribution of $\tau$ and $n$ (with respect to $\tau$, $n$, and $X_0$). 
In principle, this expression can be used to extract asymptotic results via techniques developed in \cites{CM-05,MCZ-05,ZMC-07,MMS-13,MMS-14,MMS-17,BMS-21b,MSV-12,MMS-18,BMS-21} (see \cite{M-09} for a pedagogical review). 
However, it is important to emphasize that the effective random walk in our case is neither symmetric, as in \cites{CM-05,MCZ-05,ZMC-07,MMS-13,MMS-14,MMS-17,BMS-21b}, nor does it have a constant drift, as in \cites{MSV-12,MMS-18,BMS-21}. 
Consequently, the asymptotic analysis is significantly more complex and falls beyond the scope of the present paper. 
Instead, here we focus on two exactly solvable cases where both approaches can be carried out in a well-controlled manner, yielding explicit analytical results. Specifically, we consider the probability distributions
\begin{align}
    \label{eq:case1 p,q,=}
    \text{Case I:}\qquad 
        &p(t) = \beta e^{-\beta t},\quad q(M) = \gamma e^{-\gamma M},\\
    \label{eq:case2 p,q,=}
    \text{Case II:}\qquad 
        &p(t) = \beta e^{-\beta t},\quad q(M) = \delta(M-M_0).
\end{align}
In both cases the distribution of the time intervals is exponential, hence the model falls within the realm of compound Poisson processes, for which there is quite a well-formed body of theory in the literature \cites{P-59,Z-91,S-93,DP-97,PSZ-99} (see \cite{Kyprianou2014} for a comprehensive review). Importantly, however, the effective random walk approach we propose does not rely on $p(t)$ being exponential and remains applicable to more general waiting-time distributions. 
The examples \eqref{eq:case1 p,q,=} and \eqref{eq:case2 p,q,=} are chosen deliberately for two reasons: first, they allow for an exact treatment using both the renewal equation techniques and the random walk mapping, enabling a direct comparison between the two; second, they serve as a clear illustration of the machinery of the proposed approach.

\par We show that, depending on the strength of the drift, there are two distinct regimes: the \textit{absorption regime}, where the first-passage eventually happens; the \textit{survival regime}, where there is a finite probability that the process stays positive indefinitely; and the \textit{critical point}, which separates these two regimes. 
We obtain the large $\tau$ (large $n$) behavior of the survival probability, i.e., the probability that the process stays positive up to time $\tau$ (up to the $n$th jump).
In the \textit{survival regime}, these probabilities decay to a constant, while in the \textit{absorption regime}, they decay to zero, with both decays being exponentially fast. 
At the \textit{critical point}, the probabilities also vanish at long times, but more slowly, following a power-law decay. 
Furthermore, we compute more refined statistics: in the \textit{survival regime} we calculate the mean and the variance of $\tau$ and $n$ for the trajectories, in which the first-passage occurs, while in the \textit{absorption regime}, we compute the mean and the variance of $\tau$ and $n$, and also, using the large deviation theory (see \cites{T-09,T-18,BCKT-25} for a pedagogical review), find the $X_0\to\infty$ behavior of the corresponding marginal probability distributions.

\par The paper is organized as follows: 
Section~\ref{sec:model} provides a formal definition of the model, explains the origin of the different regimes, and presents our main results.
In Section~\ref{sec:framework}, we briefly recall the renewal equation approach and results known in the literature for the discrete-time random walks, and then, construct the mapping of the original process onto the effective discrete-time random walk. 
Sections~\ref{sec:exact1} and \ref{sec:exact2} focus on detailed computations for two exactly solvable cases, given by \eqref{eq:case1 p,q,=} and \eqref{eq:case2 p,q,=}, respectively, leading to the results outlined in Section~\ref{sec:model}. Finally, we conclude  in Section~\ref{sec:conclusion}. Additional details on the numerical simulations presented throughout the paper are provided in Appendix~\ref{sec:numerics}.

\section{The model and the main results}\label{sec:model}
\par First, we define the process shown in Fig.~\ref{fig:example_trajectory} in a more formal way. Mathematically, it can be expressed as
\begin{equation}\label{eq:X(t)=definition}
    X(t) = X_0 - \alpha \,  t + \sum_{j=1}^{n(t)}M_j,
\end{equation}
where $X_0$ is the initial position, $\alpha>0$ is the drift velocity, and $M_j$ are positive \textit{i.i.d.} random variables distributed with $q(M)$. Here, $n(t)$ is the number of jumps that occurred up to the time $t$. We assume that the first jump happens at $t=0$, hence $n(0)=1$, and that the inter-jump intervals $t_j$ are also \textit{i.i.d.} random variables drawn from $p(t)$.

\par Before going into the details, it is convenient to gain some intuition and speculate about the expected results. Consider the two extreme cases:
\begin{itemize}
    \item  If the drift is very strong ($\alpha\to\infty$), then the process rapidly moves toward the origin, and first-passage happens almost immediately. In this limit, we have  $\tau=0$ and $n=1$. 
    \item Conversely, if there is no drift ($\alpha=0$), then the process consists solely of positive jumps, so the first-passage never occurs.
\end{itemize}
This simple observation suggests the existence of two distinct regimes. For sufficiently small $\alpha$, the process remains positive indefinitely with finite probability. On the other hand, for sufficiently large $\alpha$, the process inevitably crosses the origin. We will refer to these two scenarios as the \textit{survival regime} and the \textit{absorption regime} respectively.
It is also natural to anticipate the existence of a critical value $\alpha_c$ that separates these two behaviors, leading to what we will call the \textit{critical point}.

\par As a matter of fact, the critical value $\alpha_c$ and the large $X_0$ behavior of the means of $\tau$ and $n$ in the \textit{absorption regime} can be determined by a simple heuristic argument. Suppose that $\alpha$ is large enough so that the process eventually crosses the origin. At this moment, we must have
\begin{equation}\label{eq:0=X0-first-passage condition}
    X(\tau) = X_0 - \alpha \tau + \sum_{j=1}^{n} M_j = 0.
\end{equation}
From \eqref{eq:0=X0-first-passage condition} it is clear that $n$ and $\tau$ are inherently correlated. Moreover, \eqref{eq:0=X0-first-passage condition} implies the relation between their means:
\begin{equation}\label{eq:E[tau]=X+E[n]}
    \alpha \, \mathbb{E}[\tau\,\vert\,X_0] 
    = X_0 + \langle M\rangle \, \mathbb{E}[n\,\vert\,X_0],
\end{equation}
where $\mathbb{E}[ \cdots \vert\,X_0]$ denotes averaging over all possible trajectories starting from $X_0$, and brackets $\langle \cdots \rangle$ represent  averaging over the distributions $p(t)$ and $q(M)$, 
\begin{equation}\label{eq:<...>=def}
    \langle f(M,t) \rangle
    = \int_{0}^{\infty} \dd M 
      \int_{0}^{\infty} \dd t\,
        f(M,t)\, q(M)\, p(t) .
\end{equation}
Another simple observation is that if the first-passage happens after $n$ jumps, then 
\begin{equation}\label{eq:sum t < tau < sum t}
     \sum_{j=1}^{n-1} t_j < \tau < \sum_{j=1}^{n} t_j.
\end{equation}
The further from the origin the process starts, the more jumps occur before the first-passage event. Therefore, as $X_0\to\infty$, we have $n\to\infty$, and both sums in \eqref{eq:sum t < tau < sum t} can be approximated by $n \langle t\rangle$. Hence,
\begin{equation}\label{eq:E[tau]~<t>E[n]}
    \mathbb{E}[\tau\,\vert\,X_0] \underset{X_0\to\infty}{\approx} 
\langle t\rangle\, \mathbb{E}[n\,\vert\,X_0].    
\end{equation}
By combining \eqref{eq:E[tau]~<t>E[n]} with \eqref{eq:E[tau]=X+E[n]}, we can easily find the asymptotic means of $\tau$ and $n$:
\begin{equation}\label{eq:E[t], E[n],X->inf, expected}
    \mathbb{E}[n\,\vert\,X_0] \underset{X_0\to\infty}{\approx} 
    -
    \frac{X_0}{\left\langle M - \alpha t\right\rangle},
    \qquad
    \mathbb{E}[\tau\,\vert\,X_0] 
    \underset{X_0\to\infty}{\approx} -  
    \frac{\langle t \rangle X_0}{\left\langle M - \alpha t \right\rangle}.
\end{equation}
Note that $\left\langle M - \alpha t \right\rangle$ is nothing but the average displacement per jump, and \eqref{eq:E[t], E[n],X->inf, expected} suggests that for the system to be in the absorption regime, this displacement must be negative, meaning that, on average,  the process moves toward the origin. This, in turn, suggests that the critical value of the drift velocity $\alpha_c$ is given by
\begin{equation}\label{eq:alpha_c=def}
     \alpha_c = \frac{\left\langle M\right\rangle}{\left\langle t\right\rangle }.
\end{equation} 
Having presented these heuristic arguments, we now proceed to the main results of the paper, which cannot be derived from the such a simple intuitive reasoning. 

\par To characterize the first-passage properties more precisely we need to introduce several key quantities that will be used in the analysis. Throughout the paper, and in line with common practice in the physics literature, we adopt a slight abuse of notation by using the same symbols to denote both random variables and their realizations. While this may introduce some ambiguity from a strictly mathematical perspective, we believe it significantly improves readability by keeping the notation more compact.

\par The first two quantities are the survival probabilities, defined as:
\begin{equation}\label{eq:S(n),S(tau)=def}
    S_N (n \, \vert\, X_0)
     \equiv 
     \sum_{k = n+1}^{\infty} \int_{0}^{\infty} \dd \tau\, \mathbb{P}[\tau,k\,\vert\,X_0]
    ,
    \quad
    S_T (\tau \,\vert\, X_0)
    \equiv
    \int_{\tau}^{\infty} \dd \tilde{\tau} \, 
    \sum_{n=0}^{\infty} \mathbb{P}[\tilde{\tau}, n \,\vert\,X_0],
\end{equation}
where $\mathbb{P}[\tau,n\,\vert\,X_0]$ is the joint probability distribution of the first-passage time $\tau$ and the number of jumps $n$ before the first-passage, given that the process starts at $X_0$. 
Specifically, $S_N(n\,\vert\,X_0)$ is the probability that the process has remained positive for the first $n$ jumps, while $S_T(\tau\,\vert\,X_0)$ is the probability that the process has stayed positive up to time $\tau$. Additionally, we define the probability that the process stays positive indefinitely as
\begin{equation}\label{eq:S_infty(X0)}
    S_{\infty}(X_0) \equiv 1- \sum_{n=0}^{\infty} 
    \int_{0}^{\infty}
    {\dd \tau}\, 
        \mathbb{P}[\tau,n\,\vert X_0].
\end{equation}
Finally, we introduce two marginal first-passage probability densities, 
\begin{equation}\label{eq:P[T], P[n]=def}
    \mathbb{P}_T [\tau \,\vert\,X_0] 
    \equiv 
    \sum_{n=0}^{\infty} 
        \mathbb{P}[\tau,n\,\vert\,X_0],
    \qquad
    \mathbb{P}_N [n\,\vert\,X_0] 
    \equiv
    \int_{0}^{\infty} 
    \dd \tau\,
           \mathbb{P}[\tau,n\,\vert\,X_0].
\end{equation}
These are directly related to the survival probabilities \eqref{eq:S(n),S(tau)=def} through:
\begin{equation}\label{eq:P[T]=pdS[T], P[n]=pdS[n]}
    \mathbb{P}_T[\tau\,\vert\,X_0] = - \partial_\tau S_T(\tau\,\vert\,X_0) ,
    \qquad
    \mathbb{P}_N[n\,\vert\,X_0] = S_N(n-1\,\vert\,X_0) - S_N(n\,\vert\,X_0).
\end{equation}

\par
\begin{figure}[h]
    \includegraphics{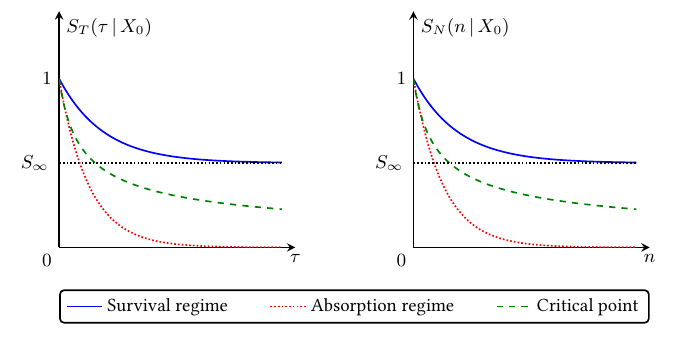}
    \caption{
    Schematic plots of the survival probabilities $S_T(\tau\,\vert\,X_0)$ (left) and $S_N(n\,\vert\,X_0)$ (right). 
    The solid blue lines correspond to the survival regime, where both survival probabilities exponentially decay to a constant value $S_\infty = S_\infty(X_0)$.
    The dotted red lines correspond to the absorption regime, where both survival probabilities exponentially decay to zero.
    The dashed purple lines correspond to the critical point, where the survival probabilities decay as a power law.
    }
    \label{fig:StSn=scheme}
\end{figure}

 In what follows, we show that there are indeed three scenarios with qualitatively different first-passage properties, which can be summarized as follows
\begin{enumerate}
    \item \textit{Absorption regime.} The drift is strong $\alpha>\alpha_c$, and the process eventually crosses the origin. 
    \begin{itemize}
        \item Both $S_T(\tau\,\vert\,X_0)$ and $S_N(n\,\vert\,X_0)$ decay exponentially, and $S_\infty(X_0) = 0$;
        \item We compute the decay rates for the survival probabilities and find the first two moments of $n$ and $\tau$, as well as the marginal probability distributions \eqref{eq:P[T], P[n]=def} for $X_0=0$ and in the limit $X_0\to\infty$.
    \end{itemize}
    \item \textit{Survival regime.} The drift is weak $\alpha<\alpha_c$, and the process may stay positive indefinitely.
    \begin{itemize}
        \item $S_T(\tau\,\vert\,X_0)$ and $S_N(n\,\vert\,X_0)$ approach a constant value $S_\infty(X_0) > 0$;
        \item We compute the decay rates of the survival probabilities and the first two moments of $\tau$  and  $n$  for the trajectories where the first-passage occurs.
    \end{itemize} 
    \item \textit{Critical point.} The drift is critical $\alpha=\alpha_c$, marking the transition between the two regimes above. 
    \begin{itemize}
        \item $S_T(\tau\,\vert\,X_0)$ and $S_N(n\,\vert\,X_0)$ decay as a power-law, and $S_\infty(X_0) = 0$.
        \item We determine the prefactor of the power-law decay for both survival probabilities and analyze the continuous limit behavior as $X_0\to\infty$ and $\tau\to\infty$ (or $n\to\infty$).
    \end{itemize} 
\end{enumerate}
More formally, we show that in the \textit{absorption regime}, the survival probabilities decay exponentially
\begin{equation}\label{eq:res_absorption S=isinglike}
    \alpha > \alpha_c : \quad
    \left\{
    \begin{aligned}
     & S_T(\tau\,\vert\,X_0) 
        \underset{\tau\to\infty}{\asymp} e^{ - \frac{\tau}{\xi_\tau(\alpha)}},\\
     & S_N(n\,\vert\,X_0) 
        \underset{n\to\infty}{\asymp} e^{ - \frac{n}{\xi_n(\alpha)}},   
    \end{aligned}\right.
\end{equation}
in the \textit{survival regime}, the survival probabilities asymptotically approach a constant with an exponential correction
\begin{equation}\label{eq:res_survival S=isinglike}
    \alpha < \alpha_c : \quad
    \left\{
    \begin{aligned}
     & S_T(\tau\,\vert\,X_0) - S_\infty(X_0)
        \underset{\tau\to\infty}{\asymp} e^{ - \frac{\tau}{\xi_\tau(\alpha)}},\\
     & S_N(n\,\vert\,X_0) - S_\infty(X_0)
        \underset{n\to\infty}{\asymp} e^{ - \frac{n}{\xi_n(\alpha)}}.  
    \end{aligned}\right.
\end{equation}
As the drift velocity approaches the critical value $\alpha\to\alpha_c$ (from either side), both $\xi_\tau(\alpha)$ and $\xi_n(\alpha)$ diverge as
\begin{equation}\label{eq:res_xi(alpha)_divergence}
    \xi_\tau(\alpha) \underset{\alpha\to\alpha_c}{\sim}
    \frac{1}{(\alpha-\alpha_c)^2},\qquad
    \xi_n(\alpha) \underset{\alpha\to\alpha_c}{\sim}
    \frac{1}{(\alpha-\alpha_c)^2},
\end{equation}
which leads to the power-law decay of the survival probabilities at the \textit{critical point}
\begin{equation}\label{eq:res_critical S=isinglike}
    \alpha = \alpha_c : \quad
    \left\{
    \begin{aligned}
     & S_T(\tau\,\vert\,X_0)
        \underset{\tau\to\infty}{\sim} \frac{c_\tau}{\sqrt{\tau}},\\
     & S_N(n\,\vert\,X_0)
        \underset{n\to\infty}{\sim} \frac{c_n}{\sqrt{n}}.   
    \end{aligned}\right.
\end{equation}
These different behaviors are illustrated schematically in Fig.~\ref{fig:StSn=scheme}. 

\par The exact expressions for $\xi_\tau(\alpha)$, $\xi_n(\alpha)$, $c_\tau$, $c_n$, and $S_\infty(X_0)$ depend on the distributions $p(t)$ and $q(M)$ and will be provided shortly, along with a more detailed description of the first-passage properties in Sec.~\ref{sec:Res_case1} and Sec.~\ref{sec:Res_case2}. Since these results are rather technical, it is convenient to make some general comments before presenting them.

\par First, although in this paper we obtain \eqref{eq:res_absorption S=isinglike}, \eqref{eq:res_survival S=isinglike},  \eqref{eq:res_xi(alpha)_divergence} and \eqref{eq:res_critical S=isinglike} in the two exactly solvable cases, it is reasonable to expect that this behavior is universal for arbitrary light-tailed distributions $p(t)$ and $q(M)$. Such analysis however is a non-trivial task on its own and falls beyond the scope of the present paper.

\par Second, to further develop intuition for these results, we take a brief detour and discuss the critical phenomena with the two dimensional Ising model serving as a representative example. 
This model is a classical model in statistical mechanics that consists of spins $\sigma_i\in\{-1,1\}$ located on the sites of the square lattice, with nearest-neighbor interactions between the spins. The probability of a given configuration is:
\begin{equation}\label{eq:Ising=def}
    \mathrm{Prob}[\{ \sigma\}] = \frac{1}{Z} e^{\frac{J}{T} \sum_{\langle i,j\rangle}\sigma_i \sigma_j},
\end{equation}
where $Z$ is the partition function, $J$ is the interaction strength, $T$ is the temperature, and the sum is over all neighboring sites of the lattice. This model is known to have two phases depending on the temperature: an ordered phase at low temperatures and a disordered phase at high temperatures, with a phase transition occurring at $T=T_c$. 

\par Above the critical temperature ($T>T_c$), thermal fluctuations dominate and $\langle \sigma\rangle=0$. The two-point correlation function in the scaling limit decays exponentially as 
\begin{equation}\label{eq:isingHighT<ss>}
    T>T_c : \qquad
    \left\langle \sigma_0 \sigma_\mathbf{r}\right\rangle 
    \underset{\left|\mathbf{r}\right|\to\infty}{\asymp} 
    e^{-\frac{\left|\mathbf{r}\right| }{\xi(T)}},
\end{equation}
where $\xi(T)$ is the temperature-dependent correlation length. This regime corresponds to complete disorder.

\par Below the critical temperature ($T<T_c$), the system exhibits spontaneous magnetization ($\langle \sigma\rangle\ne0$), and long-range ferromagnetic order. Specifically, in the scaling limit, the two-point correlation function decays to a constant as
\begin{equation}\label{eq:isingLowT<ss>}
    T<T_c : \qquad
    \left\langle \sigma_0 \sigma_\mathbf{r}\right\rangle 
        - \left\langle \sigma\right\rangle^2 
    \underset{\left|\mathbf{r}\right|\to\infty}{\asymp} 
    e^{-\frac{\left|\mathbf{r}\right| }{\xi(T)}}.
\end{equation}
This behavior reflects short-range order. While local regions maintain spin alignment, global coherence diminishes exponentially.

\par As the critical temperature is approached (from either side), the correlation length diverges as 
\begin{equation}\label{eq:Ising corr length}
    \xi(T) \underset{T\to T_c}{\sim} \frac{1}{\left| T-T_c \right|},
\end{equation}
leading to the power-law decay in the correlation function
\begin{equation}\label{eq:isingCritical<ss>}
    T=T_c : \qquad
    \left\langle \sigma_0 \sigma_\mathbf{r}\right\rangle 
    \underset{\left|\mathbf{r}\right|\to\infty}{\sim}
    \frac{1}{\left| \mathbf{r}\right|^{\frac{1}{4}}}.
\end{equation}
\par Comparing \eqref{eq:res_absorption S=isinglike}, \eqref{eq:res_survival S=isinglike} and \eqref{eq:res_critical S=isinglike} with \eqref{eq:isingHighT<ss>}, \eqref{eq:isingLowT<ss>} and \eqref{eq:isingCritical<ss>} we observe a clear analogy: 
the drift velocity $\alpha$ plays the role of the temperature; \textit{absorption} and \textit{survival} regimes resemble high- and low-temperature behaviors of the Ising model; $\xi_\tau(\alpha)$ and $\xi_n(\alpha)$ act as ``correlation lengths''. 
It is important to emphasize that this is merely an illustrative analogy, which can help build intuition, but there is no connection between the stochastic process \eqref{eq:X(t)=definition} and the Ising model. 
With this remark, we complete our detour and proceed to the results for the two exactly solvable cases.

\subsection{Results in the case I}\label{sec:Res_case1}
The first exactly solvable case we consider is the scenario in which both the inter-event times and jump sizes follow exponential distributions
\begin{equation}\label{eq:res_2exp p,q=def}
    p(t) = \beta e^{-\beta t},{}
        \qquad
    q(M) = \gamma e^{-\gamma M}.
\end{equation}
We begin by determining the explicit form of the double Laplace transform of the joint probability distribution $\mathbb{P}[\tau,n\,\vert\,X_0]$ (given by  \eqref{eq:LT P = Q = 2exp}). By analyzing this expression, we obtain the following results.

\par The critical drift, as anticipated in \eqref{eq:alpha_c=def}, is given by
\begin{equation}\label{eq:res_alpha_c=2exp}
    \alpha_c = \frac{\langle M\rangle}{\langle t \rangle} = \frac{\beta}{\gamma}.
\end{equation}
\paragraph{Survival regime.} For $\alpha<\alpha_c$, the survival probabilities tend to a constant with exponentially small correction, as described by \eqref{eq:res_survival S=isinglike}. The probability of the process remaining positive indefinitely is
\begin{equation}\label{eq:res_Sinf=2exp}
\alpha<\alpha_c: \qquad
    S_\infty(X_0) = 1 - \frac{\alpha}{\alpha_c} 
        \exp\left[ -\left(\frac{\alpha_c}{\alpha}-1\right)\gamma X_0 \right].
\end{equation}
The corresponding ``correlation lengths'' are
\begin{equation}\label{eq:res_corrLengths=2exp}
    \xi_\tau(\alpha) = \frac{1}{\gamma\, (\sqrt{\alpha} - \sqrt{\alpha_c})^2},
    \qquad
    \xi_n(\alpha) = \frac{1}{
        \log \frac{(\alpha+\alpha_c)^2}{4\,\alpha\,\alpha_c} }.
\end{equation}
For the trajectories where the first-passage does occur, we compute the conditional means
\begin{align}
\label{eq:res_2exp_meanT_survival}
    \alpha< \alpha_c : \quad
        & \mathbb{E}[\tau\,\vert\,X_0, \tau<\infty]
        = \frac{\alpha + \beta X_0}{\alpha(\beta - \alpha\gamma)},    \\
\label{eq:res_2exp_meanN_survival}
    \alpha< \alpha_c : \quad
        & \mathbb{E}[n\,\vert\,X_0, n<\infty]
        = \frac{\beta + \beta\gamma X_0}{\beta-\alpha\gamma},
\end{align}
and the conditional variances
\begin{align}
\label{eq:res_2exp_varT_survival}
\alpha< \alpha_c : \quad
        & \mathrm{Var}[\tau\,\vert\,X_0, \tau<\infty] 
            = \frac{\beta + \alpha\gamma + \beta \gamma X_0}
                   {(\beta-\alpha\gamma)^3}, \\
\label{eq:res_2exp_varN_survival}
\alpha< \alpha_c : \quad
        & \mathrm{Var}[n\,\vert\,X_0, n<\infty] 
            = \beta\gamma \,
            \frac{\alpha(\beta+\alpha\gamma) + (\alpha^2\gamma^2 + \beta^2)X_0}
                 {(\beta-\alpha\gamma)^3}.
\end{align}

\paragraph{Absorption regime.} For $\alpha>\alpha_c$, the survival probabilities decay exponentially, as in~\eqref{eq:res_absorption S=isinglike}, and the decay rates are again given by \eqref{eq:res_corrLengths=2exp}.
In this regime, we find the mean values of $\tau$ and $n$
\begin{equation}\label{eq:res_means_2exp}
    \alpha>\alpha_c : \quad
    \mathbb{E}[\tau\,\vert\,X_0 ] 
    = \frac{1 + \gamma X_0}{ \alpha \gamma - \beta},
    \qquad
    \mathbb{E}[n\,\vert\,X_0 ]  = 
    \gamma \, \frac{\alpha + \beta X_0}{\alpha \gamma - \beta},
\end{equation}
and the variances 
\begin{align}\label{eq:res_vars_2exp}
    \alpha>\alpha_c : \quad
    & \mathrm{Var}[\tau\,\vert\,X_0]  
    = \frac{\beta + \alpha\gamma + 2\beta\gamma X_0}{(\alpha\gamma - \beta)^3},\\
    \alpha>\alpha_c: \quad
    &\mathrm{Var}[n\,\vert\,X_0]  
    = \beta \gamma \, \frac{  \alpha(\alpha\gamma+\beta) + (\alpha^2\gamma^2 + \beta^2)X_0}{(\alpha\gamma - \beta)^3}.
\end{align}
Moreover, we compute the probability distribution in the two limiting cases. Specifically, for $X_0=0$ we obtain the exact form of the joint probability distribution
\begin{equation}\label{eq:res_P[X=0]_2exp}
    \alpha>\alpha_c: \quad 
    \mathbb{P}[\tau,n\,\vert\, X_0 = 0] = \frac{1}{\beta}
    e^{ -(\beta+\alpha\gamma)\tau }
    \; \frac{n}{(n!)^2}(\alpha\beta\gamma)^{n} \; \tau^{2n-2}.
\end{equation}
Conversely, in the limit $X_0\to\infty$, the marginal first-passage probability distributions \eqref{eq:P[T], P[n]=def} admit large deviation forms
\begin{align}\label{eq:res_2exp_P[T]=LDF}
    & \alpha>\alpha_c: \quad 
    \mathbb{P}_T[\tau\,\vert\,X_0]
    \underset{X_0\to\infty}{\asymp}
    e^{- X_0 \Phi(z)}, \qquad z = \frac{\alpha \tau}{X_0} - 1 ,
    \\
    \label{eq:res_2exp_P[n]=LDF}
    & \alpha>\alpha_c: \quad 
    \mathbb{P}_N[n\,\vert\,X_0]
    \underset{X_0\to\infty}{\asymp}
    e^{- X_0 \Psi(\nu)}, \qquad \nu = \frac{\alpha n}{\beta X_0},
\end{align}
where the explicit expressions for the rate functions $\Phi(z)$ and $\Psi(\nu)$ are given by \eqref{eq:Phi(z)=_2exp} and \eqref{eq:Psi(c)=_2exp} respectively.

\paragraph{Critical point.} When $\alpha\to\alpha_c$, i.e., when the drift approaches its critical value, the ``correlation lengths''  diverge as
\begin{equation}\label{eq:res_corrLengths divergence=2exp}
    \xi_\tau(\alpha) \underset{\alpha\to\alpha_c}{\sim}
    \frac{1}{(\alpha-\alpha_c)^2},\qquad
    \xi_n(\alpha) \underset{\alpha\to\alpha_c}{\sim}
    \frac{1}{(\alpha-\alpha_c)^2}.
\end{equation}
This divergence leads to a power-law behavior given in \eqref{eq:res_critical S=isinglike}, with coefficients
\begin{equation}\label{eq:res_constants_critical=2exp}
    c_\tau = \frac{1+\gamma X_0}{\sqrt{\pi \beta}},
    \qquad
    c_n = \frac{1+\gamma X_0}{\sqrt{\pi}}.
\end{equation}
At the critical point the process consists of identical jumps with finite moments and zero mean. Thus it naturally converges to Brownian motion in the scaling limit. For the survival probabilities, we have the following scaling behavior
\begin{align}\label{eq:res_2exp_S(T) critical scaling}
    \alpha=\alpha_c : \quad 
    & S_{T}(\tau\,\vert\,X_0) \sim
    \erf\left(
        \frac{1}{2}  \frac{\gamma X_0}{\sqrt{\beta \tau}}
    \right),
    \quad \tau \to \infty, 
    \quad \frac{X_0}{\sqrt{\tau}} \text{~--- fixed},\\
\label{eq:res_2exp_S(n) critical scaling}
    \alpha=\alpha_c : \quad     
    & S_N(n\,\vert\,X_0)
        \sim \erf\left( \frac{1}{2}\frac{\gamma X_0}{\sqrt{n}}  \right), 
    \quad n\to \infty, 
    \quad \frac{X_0}{\sqrt{n}}\text{~--- fixed}.
\end{align}
This concludes the results for the case of exponential distributions.

\subsection{Results in the case II}\label{sec:Res_case2}
In the second exactly solvable case, we consider an exponential distribution for the time intervals, while keeping the jumps fixed
\begin{equation}\label{eq:res_fixM p,q=def}
    p(t) = \beta e^{-\beta t},
    \qquad
    q(M) = \delta(M-M_0).
\end{equation}
With fixed jump sizes, the first-passage condition  \eqref{eq:0=X0-first-passage condition} simplifies to
\begin{equation}\label{eq:res_first-passage_cond_fixM}
    X_0 - \alpha \tau + n M_0 = 0, 
\end{equation}
implying a one-to-one correspondence between $n$ and $\tau$. This means that one quantity fully determines the other. However, for completeness, we present results for both $n$ and~$\tau$.

\par Again, by computing the explicit form \eqref{eq:LTP[tau,n|x]=fixM} of the double Laplace transform of the joint probability distribution $\mathbb{P}[\tau,n\,\vert\,X_0]$ and analyzing it, we obtain the results listed below.

\par The critical drift is given by
\begin{equation}
    \alpha_c = \frac{\langle M\rangle}{\langle t\rangle}=\beta M_0.
\end{equation}

\paragraph{Survival regime.} For $\alpha<\alpha_c$, the survival probabilities approach a constant with exponentially small corrections, as seen in \eqref{eq:res_survival S=isinglike}. The probability that the process never crosses the origin is
\begin{equation}\label{eq:res_fixM_S(X0)}
  \alpha < \alpha_c: \quad S_\infty(X_0) = 
  1 - e^{ - \Lambda(\alpha) \frac{X_0+M_0}{M_0} }, \qquad
    \Lambda(\alpha) = \frac{\alpha_c}{\alpha}
      + W_0\left[
        -\frac{\alpha_c}{\alpha}
            e^{-\frac{\alpha_c}{\alpha}}
      \right],
\end{equation}
where $W_0(z)$ is the principal branch of the Lambert $W$-function (see Sec.~\ref{sec:exact2} for details). The ``correlation lengths'' in \eqref{eq:res_survival S=isinglike}  are
\begin{equation}\label{eq:res_corrLengths=fixM}
    \xi_\tau(\alpha) = 
    \frac{M_0}{\alpha_c - \alpha 
        - \alpha \log \frac{\alpha_c}{\alpha}},
    \qquad
    \xi_n(\alpha) = 
    \frac{\alpha}{\alpha_c - \alpha - \alpha \log \frac{\alpha_c}{\alpha}}.
\end{equation}
We again compute the conditional means
\begin{align}
\label{eq:res_fixM_meanT_survival}
    \alpha< \alpha_c : \quad
        & \mathbb{E}[\tau\,\vert\,X_0, \tau<\infty]
        = \frac{X_0 + M_0} 
               {\alpha\left(1+ \Lambda(\alpha) - \frac{\beta M_0}{\alpha}\right)}
        ,
        \\
\label{eq:res_fixM_meanN_survival}
    \alpha< \alpha_c : \quad
        & \mathbb{E}[n\,\vert\,X_0, n<\infty]
        = \frac{X_0+M_0}{M_0\left(1+ \Lambda(\alpha) - \frac{\beta M_0}{\alpha}\right)} - \frac{X_0}{M_0}
        ,
\end{align}
and the variances
\begin{align}
\label{eq:res_fixM_varT_survival}
\alpha< \alpha_c : \quad
        & \mathrm{Var}[\tau\,\vert\,X_0, \tau<\infty] 
        =  \frac{M_0(X_0+M_0) 
                    \left(\frac{\beta M_0}{\alpha} - \Lambda(\alpha)\right)}
                {\alpha^2\left(1+\Lambda(\alpha)-\frac{\beta M_0}{\alpha}\right)^3}
        , \\
\label{eq:res_fixM_varN_survival}
\alpha< \alpha_c : \quad
        & \mathrm{Var}[n\,\vert\,X_0, n<\infty] 
         =  \frac{(X_0+M_0) 
                \left(\frac{\beta M_0}{\alpha} - \Lambda(\alpha)\right)}
            {M_0\left(1+\Lambda(\alpha)-\frac{\beta M_0}{\alpha}\right)^3}
            .
\end{align}

\paragraph{Absorption regime.} If $\alpha>\alpha_c$, then the survival probabilities vanish exponentially fast as in \eqref{eq:res_absorption S=isinglike} with the ``correlation lengths'' given by \eqref{eq:res_corrLengths=fixM}. The means of $\tau$ and $n$ are given by:
\begin{equation}\label{eq:res_mean_fixM}
    \alpha>\alpha_c : \quad
    \mathbb{E}[\tau\,\vert\,X_0 ] 
    = \frac{M_0+X_0}{\alpha - \beta M_0},
    \quad
    \mathbb{E}[n\,\vert\,X_0 ]  = \frac{\alpha + \beta X_0}{\alpha - \beta M_0}
    ,
\end{equation}
and for the variances we have
\begin{equation}\label{eq:res_var_fixM}
    \alpha>\alpha_c : \quad
    \mathrm{Var}[\tau\,\vert\,X_0]  
    = \frac{ X_0+M_0}{(\alpha-\beta M_0)^3}\, \beta M_0^2,
    \quad
    \mathrm{Var}[n\,\vert\,X_0]  
    = \frac{X_0 + M_0}{(\alpha-\beta M_0)^3} \, \beta \alpha^2
    .
\end{equation}
The joint probability distribution for $X_0=0$ is given by
\begin{equation}
    \mathbb{P}[\tau,n\,\vert\,X_0=0]
    = \frac{1}{n!} e^{-\beta \tau}
    \left( n\,\frac{\beta M_0}{\alpha}\right)^{n-1}
    \delta\left(
        \tau - n \frac{M_0}{\alpha}
    \right),
\end{equation}
where $\delta$-function enforces the relation \eqref{eq:res_first-passage_cond_fixM} between $n$ and $\tau$. 
In the limit $X_0\to\infty$ the marginal probability distributions admit large deviation forms
\begin{align}\label{eq:res_fixM_P[T]=LDF}
    & \alpha>\alpha_c: \quad 
    \mathbb{P}_T[\tau\,\vert\,X_0]
    \underset{X_0\to\infty}{\asymp}
    e^{- X_0 \Phi(z)}, \qquad z = \frac{\alpha \tau}{X_0} - 1 ,
    \\
    \label{eq:res_fixM_P[T]=LDF}
    & \alpha>\alpha_c: \quad 
    \mathbb{P}_N[n\,\vert\,X_0]
    \underset{X_0\to\infty}{\asymp}
    e^{- X_0 \Psi(\nu)}, \qquad \nu = \frac{\alpha n}{\beta X_0},
\end{align}
with the rate functions $\Phi(z)$ and $\Psi(\nu)$ given by \eqref{eq:Phi(z)=fixM} and \eqref{eq:Psi(nu)=ans_fixM} respectively. Note that the large deviation form is the same as in \eqref{eq:res_2exp_P[T]=LDF} and \eqref{eq:res_2exp_P[n]=LDF}, but the rate functions are different.

\paragraph{Critical point.} As $\alpha\to\alpha_c$, the ``correlation lengths'' diverge as
\begin{equation}\label{eq:res_corrLengths divergence=fixM}
    \xi_\tau(\alpha) \underset{\alpha\to\alpha_c}{\sim}
    \frac{1}{(\alpha-\alpha_c)^2},\qquad
    \xi_n(\alpha) \underset{\alpha\to\alpha_c}{\sim}
    \frac{1}{(\alpha-\alpha_c)^2}.
\end{equation}
This leads to a power-law behavior \eqref{eq:res_critical S=isinglike}, with coefficients
\begin{equation}\label{eq:res_constants_critical=fixM}
    c_\tau = \sqrt{\frac{2}{\pi\beta}} \left(1+\frac{X_0}{M_0}\right) ,
    \qquad
    c_n = \sqrt{\frac{2}{\pi}} \left(1+\frac{X_0}{M_0}\right).
\end{equation}
In the scaling limit, for the survival probabilities we have
\begin{align}\label{eq:res_fixM_S(T) critical scaling}
    \alpha=\alpha_c : \quad 
    & S_{T}(\tau\,\vert\,X_0) \sim
    \erf\left( \frac{X_0}{M_0} \frac{1}{\sqrt{2\beta\tau}} \right),
    \quad \tau \to \infty, 
    \quad \frac{X_0}{\sqrt{\tau}} \text{~--- fixed},\\
\label{eq:res_fixM_S(n) critical scaling}
    \alpha=\alpha_c : \quad     
    & S_N(n\,\vert\,X_0)
        \sim  \erf\left( \frac{X_0}{M_0} \frac{1}{\sqrt{2 n} }  \right), 
    \quad n\to \infty, 
    \quad \frac{X_0}{\sqrt{n}}\text{~--- fixed}.
\end{align}
This concludes the results for the second exactly solvable case.

\section{An effective discrete-time random walk}\label{sec:framework}
\par In this section we construct a mapping of the original process onto an effective discrete-time random walk performing the trick similar to that used in \cites{MDMS-20,MDMS-20b}  for the run-and-tumble particle in $d$-dimensions and in \cite{MMV-24} for the cost of excursions (see also \cite{SBEM-25}). 
This mapping enables us to circumvent solving the integral equations that arise in the renewal equation approach. 
Below, we first briefly recall the renewal equation method, summarize key results from the literature on the discrete-time random walks, and finally, construct the aforementioned mapping.

\subsection{Renewal equation}
The core idea behind the renewal equation formalism is a recurrence relation for the probability distribution $\mathbb{P}[\tau,n\,\vert\,X_0]$. Denote by $M_1$ the amplitude of the first jump (at $t=0$). The process then follows one of the two mutually exclusive scenarios:
\begin{enumerate}
    \item No further jumps occur ($n=1$). After the first jump the process follows purely ballistic motion with the velocity $\alpha$, leading to the first-passage time of $\tau=(X_0+M_1) / \alpha$.
    \item At least one additional jump occurs ($n>1$). Denoting by $t_1$ the time of the second jump, we end up with the same process starting from $X_0'=X_0+M_1-\alpha t_1$ and hence the joint probability distribution of $\tau$ and $n$ reduces to $\mathbb{P}[\tau-t_1, n-1\,\vert\,X_0']$.
\end{enumerate}
This simple argument results in the recurrence relation:
\begin{multline}\label{eq:renewal_P}
  \mathbb{P}[\tau,n\,\vert\,X_0] 
  =
    \int_{0}^{\infty} q(M_1) \dd M_1\; 
    \int_{\frac{X_0 + M_1}{\alpha}}^{\infty} p(t_1) \dd t_1\;
    \delta\left(\tau - \frac{1}{\alpha}(X_0 + M_1) \right) 
    \delta_{n,1}
    \\
    + 
    \int_{0}^{\infty} q(M_1) \dd M_1\;
    \int_{0}^{\frac{X_0 + M_1}{\alpha}} p(t_1) \dd t_1\;
    \mathbb{P}[\tau - t_1, n-1 \,\vert\, X_0 + M_1-\alpha t_1],
\end{multline}
where we have also accounted for the fact that $M_1$ and $t_1$ are random variables distributed with $p(t)$ and $q(M)$. 

\par In terms of the generating function 
\begin{equation}\label{eq:Q(rho,s|X0)=def}
    Q(\rho,s\,\vert\,X_0) \equiv
    \int_{0}^{\infty} \dd \tau \, e^{-\rho\tau}
    \sum_{n=0}^{\infty} s^n 
    \mathbb{P}[\tau,n\,\vert\, X_0]
\end{equation}
the renewal equation \eqref{eq:renewal_P} reads
\begin{multline}\label{eq:renewal_Q}
  Q(\rho,s\,\vert\, X_0) = 
  s 
  \int_{0}^{\infty} q(M_1) \dd M_1\; 
    e^{ - \rho \frac{X_0 + M_1}{\alpha}}
    \int_{\frac{X_0 + M_1}{\alpha}}^{\infty} p(t_1) \dd t_1
  \\
  + s
  \int_{0}^{\infty} q(M_1) \dd M_1\;
    \int_{0}^{\frac{X_0 + M_1}{\alpha}} p(t_1) \dd t_1
    \,
    e^{ - \rho t_1} 
    \,
    Q(\rho, s \,\vert\, X_0 + M_1-\alpha t_1).
\end{multline}
Equations \eqref{eq:renewal_P} and \eqref{eq:renewal_Q} are integral equations of the Wiener–Hopf type that commonly arise in the study of first-passage properties of continuous-time random walks (see e.g.~\cites{MV-13,DG-22}). Such equations are notoriously difficult to solve unless $p(t)$ and $q(M)$ take specific forms. In the both cases we consider, $p(t)$ is an exponential distribution, enabling us to transform \eqref{eq:renewal_Q} into a differential equation and subsequently solve it obtaining an explicit expression for $Q(\rho,s\,\vert\,X_0)$.

\subsection{Generalized Pollaczek–Spitzer formula} 
Consider now a discrete-time random walk, where the jumps $\eta$'s are distributed with some probability density $f_\text{d}(\eta)$
\begin{equation}\label{eq:RW=def}
    X_{j+1} = X_j + \eta_j, \qquad \eta_j \leftarrow f_\text{d}(\eta).
\end{equation}
We use the subscript ``d'' to avoid the confusion between the entities for the discrete-time random walk and their analogs for the original process \eqref{eq:X(t)=definition}. 

\par Denote by $\mathrm{P}^+_\text{d}[X_1,X_2\,\vert\,n]$ the constrained propagator, i.e., the probability that the random walk starting at $X_1$ arrives at $X_2$ after $n$ steps while always staying above the origin. Similarly, denote by $\mathrm{P}^-_\text{d}[X_1,X_2\,\vert\,n]$ the time reversed propagator, which corresponds to the change $\eta\mapsto-\eta$. These two propagators are related via
\begin{equation}\label{eq:P_c+=P_c-}
    \mathrm{P}_\text{d}^+[X_1,X_2\,\vert\,n] = \mathrm{P}^{-}_\text{d}[X_2,X_1\,\vert\,n].    
\end{equation} 
There is a closed expression for the Laplace transforms of these propagators which we will refer to as the generalized Pollaczek-Spitzer formula (see \cites{M-09,BMS-13} and references therein for a review). Specifically, 
\begin{equation}\label{eq:P=phiphi}
    \int_{0}^{\infty} \dd X_1\, 
        e^{-\lambda_1 X_1}
    \int_{0}^{\infty} \dd X_2\, 
        e^{- \lambda_2 X_2}
    \sum_{n=0}^{\infty} s^{n} \mathrm{P}^{\pm}_\text{d}[X_1,X_2\,\vert\,n] 
    = 
    \frac{\phi^{\mp}_\text{d}(\lambda_1;s) \;
          \phi^{\pm}_\text{d}(\lambda_2;s)}
         {\lambda_1 + \lambda_2},
\end{equation}
where
\begin{equation}\label{eq:phi^pm=RW}
    \phi^{\pm}_\text{d}(\lambda;s) = 
    \exp\left[-\frac{1}{2\pi}
        \int_{-\infty}^{\infty}
        \dd k \frac{1}{\lambda\pm\ii k}
        \log\left[
            1 - s F_\text{d}(k)
        \right]
    \right],
    \quad
    F_\text{d}(k) = \int_{-\infty}^{\infty} \dd\eta\, e^{\ii k \eta} f_\text{d}(\eta).
\end{equation} 
For the symmetric jump distribution the result \eqref{eq:P=phiphi} was first obtained in \cite{I-94}. A relatively simple proof of its generalization to the asymmetric jump distributions  can be found in Sec. III.A of \cite{MMS-18}. Strictly speaking, the derivation in \cite{MMS-18} was carried out for the case where $\eta$ is symmetric with a constant drift, but the very same derivation is valid for arbitrary asymmetric random walks. It is worth mentioning that although only the propagator $\mathrm{P}_\text{d}^+[X_1,X_2\,\vert\,n]$, will be used in the subsequent derivations, we provide the analogous expression for the reversed propagator $\mathrm{P}_\text{d}^-[X_1,X_2\,\vert\,n]$ for completeness. This also facilitates comparison with the proof presented in \cite{MMS-18}, where both propagators play an important role.

\par To extract the distribution of the number of jumps before the first-passage from the constrained propagator, note that the survival probability $S_\text{d}[n\,\vert\,X_0]$ can be obtained from the constrained propagator $\mathrm{P}^{+}_\text{d}[X_0,X_2\,\vert\,n]$ by integrating over the final position $X_2$ at step $n$, i.e.,
\begin{equation}\label{eq:S[n]=int propagator}
    S_\text{d}[n\,\vert\,X_0] = \int_{0}^{\infty} \dd X_2 \, \mathrm{P}^{+}_\text{d}[X_0, X_2\,\vert\,n].
\end{equation}
Thus we can find the Laplace transform of the survival probability by taking the $\lambda_2\to0$ limit in \eqref{eq:P=phiphi},
\begin{equation}\label{eq:S=phiphi}
    \int_{0}^{\infty} \dd X_0\, 
        e^{-\lambda X_0} 
    \sum_{n=0}^{\infty} s^{n} S_\text{d}[n\,\vert\,X_0] 
    = 
    \frac{1}{\lambda}
    \phi^{-}_\text{d}(\lambda;s) \;
          \phi^{+}_\text{d}(0;s).
\end{equation}
At the same time the probability distribution of the number of jumps $n$ before the first-passage $\mathbb{P}_\text{d}[n\,\vert\,X_0]$ can be expressed in terms of the survival probability as
\begin{equation}\label{eq:P[n|X]=S-S drw}
    \mathbb{P}_\text{d}[n\,\vert\,X_0] = S_\text{d}[n-1\,\vert\,X_0] - S_\text{d}[n\,\vert\,X_0].
\end{equation}
Substituting \eqref{eq:P[n|X]=S-S drw} into \eqref{eq:S=phiphi}, after a simple calculation, yields
\begin{equation}\label{eq:PS_formula_fpp=}
    \int_{0}^{\infty} \dd X_0\, e^{-\lambda X_0}
    \sum_{n=0}^{\infty} s^{n} \mathbb{P}_\text{d}[n\,\vert\,X_0] 
    =
    \frac{1}{\lambda} - \frac{1-s}{\lambda}\phi^{-}_\text{d}(\lambda;s) \phi^{+}_\text{d}(0;s).
\end{equation}
Now we construct a mapping of the original process onto a discrete random walk and use \eqref{eq:PS_formula_fpp=} to obtain an explicit expression for the Laplace transform of $\mathbb{P}[\tau,n\,\vert\,X_0]$.

\subsection{Mapping onto an effective discrete-time random walk}\label{sec:framework_RW}
\begin{figure}[h]
    \includegraphics[width=\linewidth]{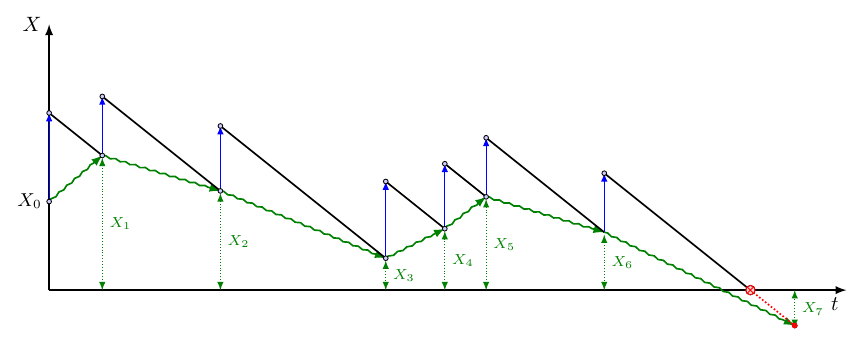}
    \caption{
    Schematic representation of the trajectory of the original process (black) from Fig.~\ref{fig:example_trajectory}, along with its effective random walk counterpart (green) given by \eqref{eq:X_j=effective walk}.
    The coordinate $X_j$ of the effective random walk corresponds to the position of the original process right before the $(j+1)$th jump. 
    Here, the first-passage occurs after $n=7$ steps, meaning that the first six positions $X_1,\ldots,X_6$ of the effective walk are positive, while $X_7$ is negative. 
    Note that the discrete-time walk does not include $t$ and the time axis is shown only to clarify the construction.
    }
    \label{fig:example_trajectory_RW}
\end{figure}

The idea behind the mapping of the original process onto an effective random walk is simple. Let $X_j$ be the position of the original process (described by \eqref{eq:X(t)=definition}) just before the $(j+1)$th jump (see Fig.~\ref{fig:example_trajectory_RW}). The position $X_j$ then evolves according to
\begin{equation}\label{eq:X_j=effective walk}
    X_{j+1} = X_j +\eta_j,\qquad
    \eta_j = M_j - \alpha t_j.
\end{equation}
This closely resembles a discrete-time random walk as in \eqref{eq:RW=def}, except that we have yet to define the probability distribution of the increments $\eta_j$.

\par A naive approach is to treat $\eta_j$ as the difference between two independent random variables: the jump amplitude $M_j$ drawn from $q(M)$ and the waiting time $t_j$ (multiplied by $\alpha$) drawn from $p(t)$. 
This simple mapping allows us to apply Pollazcek-Spitzer formula~\eqref{eq:PS_formula_fpp=} and extract the distribution of the number of jumps $n$ before the first-passage. However, this approach discards all information about the first-passage time $\tau$. To account for both $\tau$ and $n$, a more refined mapping is required. 

\par Both the jumps $M_j$ and the time intervals $t_j$ are positive, which allows us to express the joint distribution $\mathbb{P}[\tau, n\,\vert\,X_0]$ as
\begin{multline}\label{eq:P[tau,n]=int int int Formal}
    \mathbb{P}[\tau,n \,\vert\,X_0]
    = 
    \int_{0}^{\infty} \dd t_1\ldots \dd t_n
    \int_{0}^{\infty} \dd M_1\ldots \dd M_n \;
    \prod_{j=1}^{n} p(t_j) q(M_j)
    \\
    \delta\left(
        \tau - \frac{1}{\alpha} \left(X_0 + \sum_{j=1}^{n} M_j\right)
    \right)
    \theta(-X_n) \prod_{j=1}^{n-1}\theta(X_j),
\end{multline}
where $X_j$ are given by \eqref{eq:X_j=effective walk}. The $\delta$-function in \eqref{eq:P[tau,n]=int int int Formal} enforces the first-passage condition \eqref{eq:0=X0-first-passage condition}. 
The product of Heaviside step functions ensures that the first  $n-1$  coordinates $X_j$ are positive, while  $X_n$  is negative, i.e., that the first-passage occurs exactly after $n$ jumps.

\par Taking the Laplace transform with respect to $\tau$ and changing the integration variable in \eqref{eq:P[tau,n]=int int int Formal} from $t_j$ to $\eta_j$ via $\eta_j = M_j - \alpha t_j$ (so that $X_j$ is now function of $\{\eta_k\}_{k=1,\ldots,j-1}$), we obtain
\begin{multline}\label{eq:LT P[t,n]=int RW}
    \int_{0}^{\infty} \dd \tau\, e^{-\rho \tau}\, 
    \mathbb{P}[\tau, n \,\vert\, X_0] 
     = 
     e^{-\rho \frac{X_0}{\alpha}} \left[c(\rho)\right]^{n}
     \\ \times
     \int_{-\infty}^{\infty} \dd \eta_1\ldots \dd \eta_n
        \; \theta(-X_n) \prod_{j=1}^{n-1}\theta(X_j)\;
        \prod_{j=1}^{n} f(\eta_j;\rho)
        ,
\end{multline}
where 
\begin{gather}
\label{eq:f(eta)=def}
    f(\eta; \rho) = \frac{1}{\alpha\; c(\rho)}
        \int_0^{\infty} 
        \dd M \, e^{-\rho\frac{M}{\alpha}} q(M)\;
        p\left(\frac{1}{\alpha}\left(M-\eta\right)\right),
        \\
\label{eq:c(rho)=def}
    c(\rho) = \int_0^{\infty} \dd M \, 
    e^{-\rho\frac{M}{\alpha}} q(M).
\end{gather}
The functions $c(\rho)$ and $f(\eta;\rho)$ are chosen in such a way, that $f(\eta;\rho)$ is normalized to one, which can be easily verified. Integrating over $\eta$ we obtain
\begin{equation}\label{eq:int f(eta)=1}
    \int_{-\infty}^{\infty} f(\eta;\rho) \dd\eta =
    \frac{1}{\alpha\, c(\rho)} \int_{-\infty}^{\infty} \dd\eta  
    \int_0^{\infty} 
        \dd M \, e^{-\rho\frac{M}{\alpha}} q(M)\;
        p\left(\frac{1}{\alpha}\left(M-\eta\right)\right).
\end{equation}
Changing the variable of integration via $\eta\mapsto \tilde{\eta} = \frac{1}{\alpha}(M-\eta)$ yields
\begin{equation}\label{eq:int f(eta)=2}
    \int_{-\infty}^{\infty} f(\eta;\rho) \dd\eta =
    \frac{1}{c(\rho)}
    \int_0^{\infty} 
        \dd M \, e^{-\rho\frac{M}{\alpha}} q(M)\;
    \int_{-\infty}^{\infty} \dd\tilde{\eta}\,
        p\left(\tilde{\eta}\right)
    .
\end{equation}
Recall that $p(t)$ is a probability density and hence the second integral in \eqref{eq:int f(eta)=2} is one and comparing the remaining integral with \eqref{eq:c(rho)=def}, we immediately see that
\begin{equation}\label{eq:int f(eta)=3}
    \int_{-\infty}^{\infty} f(\eta;\rho) \dd\eta =
    \frac{1}{c(\rho)}
    \int_0^{\infty} 
        \dd M \, e^{-\rho\frac{M}{\alpha}} q(M)\;
    =1 
    .
\end{equation}

Since $f(\eta;\rho)$ is positive and normalized to one, it can be interpreted as a probability distribution of the increments $\eta_j$.
The integral \eqref{eq:LT P[t,n]=int RW} is then nothing but the probability that a discrete-time random walk  $X_{j+1} = X_{j} + \eta_j$ crosses the origin exactly between steps $n-1$ and $n$. This probability can be computed using the Pollaczek-Spitzer formula \eqref{eq:PS_formula_fpp=}, with the prefactors in \eqref{eq:LT P[t,n]=int RW} inducing the shift of $s$ and $\lambda$ in \eqref{eq:PS_formula_fpp=}, namely  $s\mapsto s c(\rho)$ and $\lambda\mapsto\lambda+\frac{\rho}{\alpha}$. 
Introducing the notation
\begin{equation}\label{eq:Q(rho,s|lambda)=def}
    \hat{Q}(\rho,s\,\vert\,\lambda) \equiv \int_{0}^{\infty}\dd X_0\,
    e^{-\lambda X_0}
    \int_{0}^{\infty} \dd \tau \, e^{-\rho \tau}
    \sum_{n=0}^{\infty} s^n\, 
    \mathbb{P}[\tau,n\,\vert\,X_0],
\end{equation}
and comparing \eqref{eq:LT P[t,n]=int RW} with \eqref{eq:PS_formula_fpp=},
we arrive at the closed expression for the triple Laplace transform of $\mathbb{P}[\tau,n\,\vert\,X_0]$:
\begin{equation}\label{eq:LTQ=generalIvanov}
    \hat{Q}(\rho,s\,\vert\,\lambda)
     = 
    \frac{1}{\lambda+\frac{\rho}{\alpha}}
    - 
    \frac{1 - s \, c(\rho)}{\lambda+\frac{\rho}{\alpha}}
    \;
    \phi^{-}\!\left( \lambda + \frac{\rho}{\alpha}; \rho, s \right)
    \phi^{+}\!\left( 0; \rho,s \right),
\end{equation}
where 
\begin{equation}\label{eq:phi^pm=def}
    \phi^{\pm}(\lambda;\rho,s) = 
    \exp\left[
        -\frac{1}{2\pi} \int_{-\infty}^{\infty}
        \dd k\; \frac{1}{\lambda\pm \ii k}
        \log\left[1 - s\, c(\rho) F(k;\rho)\right] 
    \right]
\end{equation}
and
\begin{equation}\label{eq:F(k)=def}
    F(k;\rho) = \int_{-\infty}^{\infty} 
    e^{\ii k \eta} f(\eta;\rho) \, \dd \eta.
\end{equation}
Rewriting $F(k;\rho)$ in terms of the original process rather than the effective random walk, we obtain
\begin{equation}\label{eq:F(k)=int p q}
    F(k;\rho) = \frac{1}{c(\rho)} 
    \int_{0}^{\infty} \dd M\, e^{ - \rho \frac{M}{\alpha} + \ii k M} q(M)
    \int_{0}^{\infty} \dd t\, e^{-\ii \alpha k t} p(t).
\end{equation}

\par Suppose, for a moment, that we are interested only in the number of jumps $n$ before the first-passage. This corresponds to setting $\rho=0$. In this case $c(\rho)=1$ and
\begin{multline}
    f(\eta;0) = \frac{1}{\alpha} \int_{0}^{\infty}\dd M\, q(M) \, p\left(\frac{1}{\alpha}(M-\eta)\right)
    \\ =
    \int_{0}^{\infty} \dd M\, q(M) \int_{0}^{\infty} \dd t\, p(t)\,
    \delta\left(
        \eta - (M-\alpha t)
    \right).
\end{multline}
Thus, for  $\rho = 0$, an increment  $\eta_j$  is simply the difference between the jump amplitude $M_j$ and the waiting time $t_j$, as we would expect from the ``naive'' mapping. 

\par In the rest of the paper we derive the results presented in Sec.~\ref{sec:model} from \eqref{eq:LTQ=generalIvanov}. Before diving in the computation let us make one final remark. The mapping to an effective random walk described in this section is essentially an intuitive way to motivate the representation \eqref{eq:P[tau,n]=int int int Formal}. In principle, one can substitute this representation into the renewal equation \eqref{eq:renewal_P} and then follow the derivation of the Pollaczek-Spitzer formula, as outlined, for example, in \cite{I-94}.

\section{Exactly solvable case I. Double exponential distribution}\label{sec:exact1}

\par 
In this section, we focus on a special case where both jumps and waiting times are exponentially distributed,
\begin{equation}\label{eq:p,q=2exp}
    p(t) = \beta e^{-\beta t},
    \qquad
    q(M) = \gamma e^{-\gamma M}. 
\end{equation}
In this scenario, the generating function $Q(\rho, s\,\vert\, X_0)$ defined in \eqref{eq:Q(rho,s|X0)=def} can be determined explicitly  through two approaches: either by solving the renewal equation \eqref{eq:renewal_Q} directly, or by inverting the Laplace transform \eqref{eq:LTQ=generalIvanov}. In the following, we perform both computations and demonstrate their equivalence. The resulting expressions are then used to analyze the first-passage properties.

\subsection{Renewal equation approach}\label{sec:exact1_renew}
With the probability distributions \eqref{eq:p,q=2exp}, renewal equation \eqref{eq:renewal_Q} simplifies to
\begin{multline}\label{eq:renewal_Q_2exp_1}
  Q(\rho,s\,\vert\, X_0) = s 
  \int_{0}^{\infty}\dd M\,
    \gamma e^{-\gamma M}  
    e^{ - \rho \frac{X_0 + M}{\alpha}}
    \int_{\frac{X_0 + M}{\alpha}}^{\infty} 
    \dd t\,
    \beta e^{-\beta t}
  \\
  + s
  \int_{0}^{\infty}\dd M\, \gamma e^{-\gamma M}
    \int_{0}^{\frac{X_0 + M}{\alpha}} 
    \dd t\,
    \beta e^{-(\beta+\rho) t} \,
    Q(\rho, s \,\vert\, X_0 + M-\alpha t).
\end{multline}
This integral equation gives a recurrence relation for the Laplace transform of the joint probability distribution $\mathbb{P}[\tau,n\,\vert\,X_0]$. This recurrence relation may be solved explicitly by reducing the integral equation to an ordinary differential equation. This is done in four steps. First we compute the integrals in the first term, change the variable of integration in the second via $t\mapsto y=X_0+M-\alpha t$ and multiply both sides by $e^{(\rho+\beta)X_0/\alpha}$. This results in:
\begin{multline}\label{eq:renewal_Q_2exp_3}
  Q(p,s\,\vert\, X_0)\,e^{\frac{\rho+\beta}{\alpha}X_0} =  
  \frac{s}
       {1 + \frac{\rho+\beta}{\alpha \gamma}}
  \\ + s \frac{\gamma \beta}{\alpha}
  \int_{0}^{\infty} e^{-\gamma M} \dd M\;
    \int_{0}^{X_0 + M} 
    e^{-\frac{\rho+\beta}{\alpha} (M-y)}
    Q(\rho, s \,\vert\, y)
    \dd y.
\end{multline}
Second step is to take the derivative with respect to $X_0$ essentially getting rid of the constant term and removing one integration
\begin{equation}\label{eq:renewal_Q_2exp_d1}
  \pdv{}{X_0} 
  \left[ Q(\rho,s\,\vert\, X_0)\,e^{\frac{\rho+\beta}{\alpha}X_0} \right]
  =
  s \frac{\gamma \beta}{\alpha}
  \int_{0}^{\infty} \dd M\, e^{-\gamma M}
    e^{\frac{\beta+\rho}{\alpha} X_0}
    Q(\rho, s \,\vert\, X_0 + M).
\end{equation}
At the thirds step, we once again extract the exponential factor from the integrand by multiplying both sides by $e^{-\frac{\beta+\rho}{\alpha}X_0}$ and change the variable of integration $M\mapsto Y = X_0 + M$. This simple calculation yields
\begin{equation}\label{eq:renewal_Q_2exp_d2}
  e^{-\frac{\rho+\beta}{\alpha} X_0}
  \pdv{}{X_0} 
  \left[ Q(\rho,s\,\vert\, X_0)\, e^{\frac{\rho+\beta}{\alpha}X_0} \right]
  =
  s \frac{\gamma \beta}{\alpha}
  \int_{X_0}^{\infty} \dd Y\, e^{-\gamma (Y- X_0)}
    Q(\rho, s \,\vert\, Y).
\end{equation}
Finally, we take yet another derivative with respect to $X_0$, and arrive at 
\begin{equation}\label{eq:renewal_Q_2exp_d3}
  e^{\gamma X_0}
  \pdv{}{X_0}\left[ 
  e^{-\frac{\rho+\beta}{\alpha} X_0 - \gamma X_0}
  \pdv{}{X_0} 
  \left[ Q(\rho,s\,\vert\, X_0)\, e^{\frac{\rho+\beta}{\alpha}X_0} \right]
  \right]
  =
  - s \frac{\gamma \beta}{\alpha}
    Q(\rho, s \,\vert\, X_0).
\end{equation}
This equation, though correct, is not yet in its most convenient form. By expanding and rearranging terms, we obtain a canonical form of a standard second-order linear differential equation:
\begin{equation}\label{eq:renewal_Q_2exp_d4}
  \left[
    \pdv{^2}{X_0^2} 
    + 
    \frac{\rho+\beta - \alpha \gamma}{\alpha} \pdv{}{X_0}
    -
    \gamma \, \frac{\rho+\beta(1-s)}{\alpha} 
  \right]Q(\rho,s\,\vert\,X_0) = 0.
\end{equation}
\par Now that we have derived the differential equation \eqref{eq:renewal_Q_2exp_d4}, we solve it explicitly. The general solution is a sum of two exponentials:
\begin{equation}\label{eq:Q(X0)=Ap+Am}
  Q(\rho,s\,\vert\, X_0) = 
  A_+ e^{ -\omega_+ X_0} + A_- e^{-\omega_- X_0},
\end{equation}
where
\begin{equation}\label{eq:w+-=}
  \omega_\pm =\frac{1}{2\alpha} \left( 
    \rho+\beta - \alpha\gamma
    \pm 
    \sqrt{(\rho+\beta-\alpha\gamma)^2 + 4\alpha\gamma\, (\rho+\beta(1-s))}
    \right).
\end{equation}
The only thing left is to fix the constants $A_\pm$. 
Note that $Q(\rho,s\,\vert\,X_0)$ is the Laplace transform of the probability distribution and hence it should not diverge as $X_0\to\infty$. Since $\omega_{-}<0$, this boundary condition implies that physical solutions are those with $A_{-}=0$. The constant $A_{+}$ is then found by substituting \eqref{eq:Q(X0)=Ap+Am} into \eqref{eq:renewal_Q_2exp_1}. This computation gives the final result:
\begin{equation}\label{eq:Q(X0)=2exp}
  Q(\rho,s\,\vert\,X_0) 
    = \frac{s}{1+\frac{\omega_+}{\gamma}} e^{ -\omega_+ X_0 }.
\end{equation}
Note, that the above derivation heavily relies on the exact form of the probability distributions and the renewal equation reduces to such simple differential equation only in the case where both $p(t)$ and $q(M)$ are exponential distributions.

\subsection{Effective random walk approach}\label{sec:exact1_RW}
Now we present an alternative derivation of \eqref{eq:Q(X0)=2exp} that relies on the generalized Pollaczek-Spitzer formula for the effective random walk \eqref{eq:LTQ=generalIvanov}. 
First, we construct an effective random walk given by \eqref{eq:f(eta)=def} and \eqref{eq:F(k)=int p q}. The direct computation shows that, for the probability distributions \eqref{eq:p,q=2exp}, the effective random walk is given by:
\begin{equation}
\label{eq:F(k)=2exp}
     F(k;\rho) = 
    \frac{\beta \gamma}{\alpha}
     \frac{1 + \frac{\rho}{\alpha\gamma}}
          { \left( k - \ii \frac{\beta}{\alpha} \right)
            \left( k + \ii \left(\gamma + \frac{\rho}{\alpha}\right)\right)}
    ,
    \qquad
    c(\rho) = 
     \frac{1}{1+\frac{\rho}{\alpha\gamma}}.
\end{equation}
The key ingredient in the effective random walk approach is the functions $\phi^{\pm}(\lambda;\rho,s)$. If both $p(t)$ and $q(M)$ are exponential distributions, then these functions can be found explicitly. 
\par First we integrate \eqref{eq:phi^pm=def} by parts to obtain a more convenient representation for the functions $\phi^{\pm}(\lambda;\rho,s)$. Specifically, we have:
\begin{equation}\label{eq:phi^pm=byparts}
    \phi^{\pm}(\lambda;\rho,s) = \exp\left[  
    \frac{1}{2\pi \ii} \int_{-\infty}^{\infty}
    \dd k\;  \mathcal{I}^{\pm}(k)
    \right],
\end{equation} 
where
\begin{equation}
\label{eq:I^pm=byparts}
    \mathcal{I}^\pm(k) = \mp \log[ k \mp \ii \lambda] 
    \frac{\partial_k F(k;\rho)}{\frac{1}{s\, c(\rho)}-  F(k;\rho)}.
\end{equation}
We then compute the integrals by extending them into the complex plane of $k$. Let us now describe the analytic structure of the integrands in \eqref{eq:phi^pm=byparts}.
There is a branch cut originating from the logarithmic term $\log[k\mp\ii\lambda]$. Using the exact form of $F(k;\rho)$ as in \eqref{eq:F(k)=2exp}, we see that the numerator $F'(k;\rho)$ has two poles at $k=\ii \beta/\alpha$ and $k=-\ii (\gamma+\rho/\alpha)$. The denominator gives rise to two more purely imaginary poles at
\begin{equation}\label{eq:k*_pm}
    k_\pm^*(\rho,s) =  \frac{\ii}{2\alpha}\left(
            \beta - \alpha\gamma  - \rho
            \pm \sqrt{(\rho+\beta+\alpha\gamma)^2 - 4s\, 
            \alpha\beta\gamma}
        \right).
\end{equation}
\begin{figure}[h]
\includegraphics{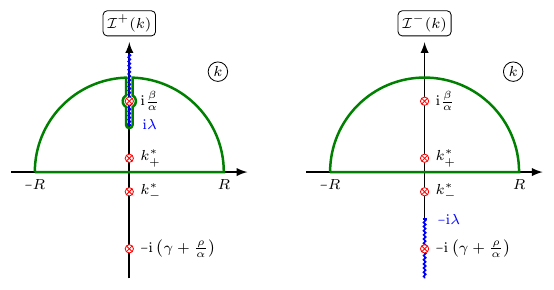}
\caption{Analytic structure of $\mathcal{I}^+(k)$ (left) and $\mathcal{I}^-(k)$ (right) in \eqref{eq:I^pm=byparts} for the probability distributions \eqref{eq:p,q=2exp}. The integrals over semi-circles vanish when $R\to\infty$.}
\label{fig:I^pm_2exp_anal_stracture}
\end{figure}
This analytic structure is schematically shown in Fig.~\ref{fig:I^pm_2exp_anal_stracture}. To compute the integrals we close the contour in the upper-half plane, and, since the integral over semi-circle vanishes at infinity, we find that $\mathcal{I}^{-}(k)$ is simply a sum of two residues
\begin{equation}\label{eq:int=Res+Res+bc_I-}
    \int_{-\infty}^{\infty} \dd k\; \mathcal{I}^-(k)
    = 
    2\pi \ii 
    \Res_{\ii\frac{\beta}{\alpha}}\left[\mathcal{I}^{-}(k)\right]
    +
    2\pi \ii 
    \Res_{k_+^*}\left[\mathcal{I}^{-}(k)\right].
\end{equation}
For $\mathcal{I}^{+}(k)$ we should also account for the branch cut, and hence
\begin{equation}\label{eq:int=Res+Res+bc_I+}
    \int_{-\infty}^{\infty} \dd k\; \mathcal{I}^+(k)
    = 
    2\pi \ii 
    \Res_{\ii\frac{\beta}{\alpha}}\left[\mathcal{I}^{+}(k)\right]
    +
    2\pi \ii 
    \Res_{k_+^*}\left[\mathcal{I}^{+}(k)\right] 
    -\int_{\text{b.c.}} \dd k\; \mathcal{I}^+(k),
\end{equation}
where ``$\text{b.c.}$'' stands for the integral over the branch cut, that is given by
\begin{equation}\label{eq:int_bc=2exp_def}
    -\int_{\text{b.c.}} \dd k\; \mathcal{I}^+(k)=
    -2\pi \ii 
        \int_{\ii\lambda}^{\ii \infty} 
        \frac{\partial_k F(k;\rho)}{\frac{1}{s\, c(\rho)} - F(k;\rho)} \dd k.
\end{equation}
All terms entering \eqref{eq:int=Res+Res+bc_I-} and \eqref{eq:int=Res+Res+bc_I+} can be found explicitly. The residues are straightforward to compute
\begin{equation}\label{eq:res=2exp}
    \Res_{\ii\frac{\beta}{\alpha}}\left[\mathcal{I}^{\pm}(k)\right]
    = \mp \log\left[ \ii \left(\frac{\beta}{\alpha} \mp \lambda\right) \right]
    ,\quad 
    \Res_{k_+^*}\left[\mathcal{I}^{\pm}(k)\right]
    = \pm \log\left[ \strut k_+^*(\rho,s) \mp \ii \lambda \right].
\end{equation}
The integrand in \eqref{eq:int_bc=2exp} is a full derivative, hence
\begin{equation}\label{eq:int_bc=2exp}
\int_{\text{b.c.}} \dd k\; \mathcal{I}^+(k)
    = - 2\pi \ii \log\left[1 - s\, c(\rho) F(k;\rho)\right] \Big|_{\ii \lambda}^{\ii \infty}
    = 2\pi \ii \log\left[1 - s\, c(\rho) F(\ii\lambda;\rho)\right]
    ,
\end{equation}
where we used the fact that $F(\ii \infty;\rho) = 0$. Finally we substitute \eqref{eq:int_bc=2exp} and \eqref{eq:int_bc=2exp_def} into \eqref{eq:int=Res+Res+bc_I-} and \eqref{eq:int=Res+Res+bc_I+}. The obtained results are then used in \eqref{eq:phi^pm=byparts} to find the closed forms of the functions $\phi^{\pm}(\lambda;\rho,s)$. The final expressions are:
\begin{equation}\label{eq:phi^pm=answer_2exp}
    \phi^{+}(\lambda;\rho,s) = 
    \frac{\lambda + \ii k_+^*(\rho,s)}{\lambda-\beta/\alpha}
    \frac{1}{1 - s\, c(\rho) F(\ii\lambda; \rho)},
    \quad
    \phi^{-}(\lambda;\rho;s) = 
    \frac{\lambda + \beta/\alpha}
         {\lambda - \ii k_+^*(\rho,s)}.
\end{equation}
Now we have all ingredients for the generalized Pollaczek-Spitzer formula. Since $F(k;\rho)$ is a Fourier transform of the probability distribution, we must have $F(0;\rho)=1$, thus substituting  \eqref{eq:phi^pm=answer_2exp} into \eqref{eq:LTQ=generalIvanov} yields
\begin{equation}\label{eq:dEs^nP[n|E]=2exp}
    \hat{Q}(\rho,s\,\vert\,\lambda)
    =
    \frac{1}{\lambda+\frac{\rho}{\alpha}} 
    + 
    \frac{1}{\lambda + \frac{\rho}{\alpha}}\;
    \frac{\ii k_+^*(\rho,s)}{\beta/\alpha}\;
    \frac{\lambda + \frac{\rho+\beta}{\alpha}}
         {\lambda +\frac{\rho}{\alpha}- \ii k_+^*(\rho,s)}.
\end{equation}
Recall that $\hat{Q}(\rho,s\,\vert\,\lambda)$ is the triple Laplace transform of the joint probability distribution $\mathbb{P}[\tau,n\,\vert\,X_0]$, as given by \eqref{eq:Q(rho,s|lambda)=def}. At the same time, using the renewal equation approach, we have found the double transform $Q(\rho,s\,\vert\,X_0)$ directly. Fortunately, \eqref{eq:dEs^nP[n|E]=2exp} can be easily inverted with respect to $X_0$ leading to 
\begin{equation}\label{eq:Q(X_0)=2exp_CTRW}
    Q(\rho,s\,\vert\,X_0) = 
    \left( 1+ \frac{\alpha}{\beta}\, \ii k_{+}^{*}(\rho,s) \right)
    \exp\left[ - \left(\frac{\rho}{\alpha} - \ii k_{+}^{*}(\rho,s) \right) X_0\right].
\end{equation}
At first glance, this result appears different from those obtained within the renewal equation approach \eqref{eq:Q(X0)=2exp} and it is essential to verify their equivalence. A simple computation using the exact forms of $k^*_+$ and $\omega_+$ as in \eqref{eq:k*_pm} and \eqref{eq:w+-=}, shows that
\begin{equation}
    \frac{\rho}{\alpha} - \ii k_{+}^{*}(\rho,s)
    = 
    \omega_{+} ,
    \qquad
    \left( 1+ \frac{\alpha}{\beta}\,\ii k_{+}^{*}(\rho,s)  \right) 
    =
    \frac{s}{1+\frac{\omega_{+}}{\gamma}}.
\end{equation}
Thus \eqref{eq:Q(X_0)=2exp_CTRW} coincides with \eqref{eq:Q(X0)=2exp}, confirming that the results obtained from the renewal equation approach and the effective random walk approach are indeed equivalent.

\subsection{First-passage properties}\label{sec:exact1_fpp}
Once $Q(\rho,s\,\vert\,X_0)$ is found by either solving the renewal equation \eqref{eq:Q(X0)=2exp} or inverting the Laplace transform in the generalized Pollaczek-Spitzer formula for the effective random walk \eqref{eq:Q(X_0)=2exp_CTRW}, the joint  distribution $\mathbb{P}[\tau,n\,\vert\,X_0]$ can, in principle, be obtained by inverting the double transform 
\begin{equation}\label{eq:LT P = Q = 2exp}
    \int_{0}^{\infty} \dd \tau e^{-\rho \tau}
    \sum_{n=0}^{\infty} s^{n} \, \mathbb{P}[\tau,n\,\vert\,X_0]
    = 
    Q(\rho,s\,\vert\,X_0)
    = 
    \frac{s}{1 + \frac{\omega_{+}}{\gamma}} e^{-\omega_{+} X_0},
\end{equation}
where  $\omega_{+}$ is given by \eqref{eq:w+-=}.  However, there is no closed-form expression for the inverse transform of~\eqref{eq:LT P = Q = 2exp}, except when $X_0=0$. In this case, first inverting $s\mapsto n$, and then  $\rho\mapsto\tau$, yields
\begin{equation}\label{eq:P[tau,n|0]=2exp}
    \mathbb{P}[\tau,n\,\vert\, X_0 = 0] = \frac{1}{\beta}
    e^{ -(\beta+\alpha\gamma)\tau }
    \; \frac{n}{(n!)^2}(\alpha\beta\gamma)^{n} \; \tau^{2n-2}.
\end{equation}
The exact form of the joint probability distribution for $X_0$ can be used to verify most of the results presented below for arbitrary $X_0$ (unless they concern the limit $X_0\to\infty$). However, since the computations are technical and the results can be obtained by simply setting $X_0=0$ in the general case, we leave these computations to the reader.

\par
\begin{figure}[h]
\includegraphics[width=.5\linewidth]{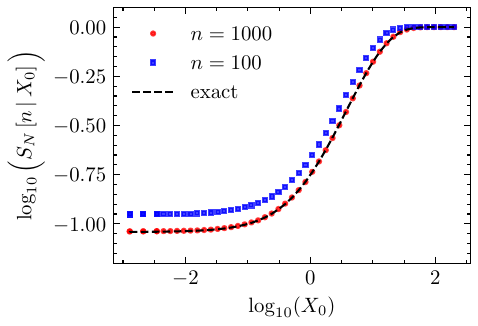}
\caption{
Survival probability $S_N(n\,\vert\,X_0)$ as a function of $X_0$ for $n=1000$ (red circles) and $n=100$ (blue squares) computed numerically and exact expression for $S_\infty(X_0)$ as in \eqref{eq:S_infty=2exp} (black dashed line).
The model parameters are $\alpha=1$,  $\beta=1{.}1$, $\gamma=1$. For the details of the simulation see Appendix~\ref{sec:numerics}.}\label{fig:Sinf_2exp}
\end{figure}

 The analysis for $X_0\ne0$ commences with computing $S_\infty(X_0)$, the probability of surviving indefinitely. As argued in Sec.~\ref{sec:model}, this is a robust quantity that can be used to distinguish the \textit{absorption regime} from the \textit{survival regime}. Recall that according to \eqref{eq:S_infty(X0)} this probability is given by:
\begin{equation}\label{eq:S_infty=2exp_exact}
    S_\infty(X_0) = 1 - \int_{0}^{\infty} \dd \tau
    \sum_{n=0}^{\infty}\, \mathbb{P}[\tau,n\,\vert\,X_0] 
    =  1 - Q(\rho,s\,\vert\,X_0)\Big|_{\rho=0,s=1},
\end{equation}
which is easily computed by fixing  $\rho=0$ and $s=1$ in \eqref{eq:LT P = Q = 2exp} to be
\begin{equation}\label{eq:S_infty=2exp}
    S_{\infty}(X_0) = 1 - \frac{2\alpha\gamma}{\beta + \alpha\gamma + \sqrt{(\beta-\alpha\gamma)^2}} e^{ -\frac{1}{2\alpha}
    \left(\beta - \alpha\gamma + \sqrt{(\beta-\alpha \gamma)^2}\right)X_0 }.
\end{equation}
From \eqref{eq:S_infty=2exp} it is clear that the behavior of $S_\infty(X_0)$ depends on whether $(\beta - \alpha\gamma)$ is positive or negative. This is a clear manifestation of the transition happening at $\alpha_c=\sfrac{\beta}{\gamma}$. Such transition is very natural as discussed in Sec.~\ref{sec:model}. 
The critical value of the drift coincides with that found by heuristic argument in \eqref{eq:alpha_c=def}, since for the considered probability distributions, $\left\langle M\right\rangle=\sfrac{1}{\gamma}$ and $\langle t\rangle=\sfrac{1}{\beta}$. Rewriting \eqref{eq:S_infty=2exp_exact} in terms of $\alpha_c$ yields
\begin{equation}\label{eq:S_infty=2exp_cases}
  S_\infty(X_0) 
  =
  \left\{
  \begin{aligned}
    &1 - \frac{\alpha}{\alpha_c} 
        \exp\left[-\left(\frac{\alpha_c}{\alpha}-1\right)\gamma X_0\right], 
        && \alpha < \alpha_c,\\
    &0,  && \alpha \ge \alpha_c.\\
  \end{aligned}\right.,
  \qquad
  \alpha_c = \frac{\beta}{\gamma}.
\end{equation}
In other words, there are three distinct cases 
\begin{enumerate}
    \item $\alpha<\alpha_c$, 
        \textit{survival regime},
    \item $\alpha=\alpha_c$,
        \textit{critical point}, 
    \item $\alpha>\alpha_c$,
        \textit{absorption regime}.
\end{enumerate}
See Fig.~\ref{fig:Sinf_2exp} for the numerical verification of \eqref{eq:S_infty=2exp_cases}. 

To get a more precise description, we look at the survival probabilities $S_N(n\,\vert\,X_0)$ and $S_T(\tau\,\vert\,X_0)$. Using \eqref{eq:P[T]=pdS[T], P[n]=pdS[n]} we express the Laplace transform of the survival probability $S_T(\tau\,\vert\,X_0)$ in terms of $Q(\rho,s\,\vert\,X_0)$ as
\begin{equation}
    \label{eq:S(tau)=1/rho(1-Q)}
    \int_{0}^{\infty}\dd\tau\, e^{-\rho\tau} S_T(\tau\,\vert\,X_0)
    =
    \left. \frac{1}{\rho}\left(1 - Q(\rho,s\,\vert\,X_0)\right)\right|_{s=1},
\end{equation}
similarly, for $S_N(n\,\vert\,X_0)$ we have
\begin{equation}
    \label{eq:S(n)=1/(1-s) (1-Q)}
    \sum_{n=0}^{\infty} s^n\, S_N(n\,\vert\,X_0)
    = \left. \frac{1}{1-s}\left(1 - Q(\rho,s\,\vert\,X_0)\right) \right|_{\rho=0}.
\end{equation}
Formally inverting \eqref{eq:S(tau)=1/rho(1-Q)} and \eqref{eq:S(n)=1/(1-s) (1-Q)} we obtain
\begin{align}
\label{eq:S_T=inverseLT}
    &S_T(\tau\,\vert\,X_0) = \frac{1}{2\pi \ii} \int_{\mathcal{C}_1} \dd \rho \; e^{\rho \tau} 
        \frac{1}{\rho}\left(1 - Q(\rho,s\,\vert\,X_0)\right)
    \Big|_{s=1},
    \\
\label{eq:S_N=inverseT}
    &S_N(n\,\vert\,X_0) =
    \frac{1}{2\pi\ii} 
    \oint_{\mathcal{C}_0} \dd s\, 
    \frac{1}{s^{n+1}} \frac{1}{1-s} (1 - Q(\rho,s\,\vert\,X_0))
    \Big|_{\rho=0},
\end{align}
where $\mathcal{C}_1$ is a vertical contour in the $\rho$ plane, chosen so that all singularities of the integrand lie to its left, and $\mathcal{C}_0$ is a counterclockwise circle around the origin in the $s$ plane. 

\par The behavior of $S_N(n\,\vert\,X_0)$ and $S_T(\tau\,\vert\, X_0)$ for $\tau\to\infty$ and $n\to\infty$ can now be inferred from the analytic structure of the integrands \eqref{eq:S_T=inverseLT} and \eqref{eq:S_N=inverseT} in the $\rho$ and $s$ complex planes. Below, we analyze three cases separately.

\subsubsection{Survival regime}\label{sec:caseI_survival}
If $\alpha<\alpha_c$ ($\beta>\alpha\gamma$), then the drift is weak and there is a finite probability of the process never crossing the origin $S_\infty(X_0)>0$. In this regime, we expect that both $S_T(\tau\,\vert\,X_0)$ and $S_N(n\,\vert\,X_0)$ tend to a constant $S_\infty(X_0)$ with exponential corrections as in \eqref{eq:res_survival S=isinglike}. Below we analyze the inverse transforms \eqref{eq:S_T=inverseLT} and  \eqref{eq:S_N=inverseT} and show that the behavior stated in \eqref{eq:res_survival S=isinglike} is indeed correct. Additionally we compute the ``correlation lengths'' presented in \eqref{eq:res_corrLengths=2exp} and the conditional means and variances \eqref{eq:res_2exp_meanT_survival}, \eqref{eq:res_2exp_meanN_survival}, \eqref{eq:res_2exp_varT_survival}, and \eqref{eq:res_2exp_varN_survival}.

\par
\begin{figure}[h]
    \includegraphics{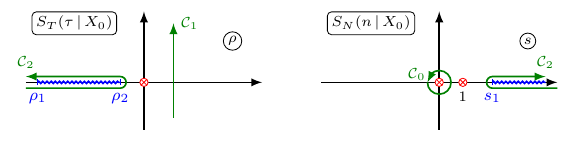}
    \caption{Analytic structure of the Laplace transform of $S_T(\tau\,\vert\,X_0)$ (left) and of the generating function of $S_N(n\,\vert\,X_0)$ (right) in the survival regime for the distributions \eqref{eq:p,q=2exp}. }\label{fig:S_survival_2exp}
\end{figure}

 We start with $S_T(\tau\,\vert\,X_0)$. The explicit form of $Q(\rho,s\,\vert\,X_0)$ given by \eqref{eq:LT P = Q = 2exp} implies that the integrand in the inverse transform \eqref{eq:S_T=inverseLT} has a pole at $\rho=0$ and a branch cut
\begin{equation}\label{eq:rho=branchcut 2exp}
\rho \in [\rho_1,\rho_2]:\qquad
    \rho_1 = -(\sqrt{\beta}+\sqrt{\alpha\gamma})^2,\quad
    \rho_2 = -(\sqrt{\beta}-\sqrt{\alpha\gamma})^2.
\end{equation}
Deforming the contour $\mathcal{C}_1\mapsto\mathcal{C}_2$ as shown in Fig.~\ref{fig:S_survival_2exp}, and accounting for the contribution of the pole yields 
\begin{equation}\label{eq:S_T(t|X)=res+int 2exp}
    S_T(\tau\,\vert\,X_0) = 
    \Res_{\rho=0}\left[
        \frac{1}{\rho}\left(1 - Q(\rho,1\,\vert\,X_0)\right)
    \right]
    + \frac{1}{2\pi \ii} \int_{\mathcal{C}_2} \dd \rho \; e^{\rho \tau} 
        \frac{1}{\rho}\left(1 - Q(\rho,1\,\vert\,X_0)\right),
\end{equation}
where $\mathcal{C}_2$ is the contour encircling the branch cut. 

\par The residue is straightforward to compute:
\begin{equation}\label{eq:S_T(t|X)=res result 2exp}
    \Res_{\rho=0}\left[
            \frac{1}{\rho}\left(1 - Q(\rho,1\,\vert\,X_0)\right)
    \right]
    = 1 - \frac{\alpha\gamma}{\beta} 
        e^{-\left(\frac{\beta}{\alpha}-\gamma\right)X_0}.
\end{equation}
Comparing this expression with \eqref{eq:S_infty=2exp} we see, that this is nothing but the survival probability $S_\infty(X_0)$. 
To obtain the asymptotic behavior of the branch cut integral in \eqref{eq:S_T(t|X)=res+int 2exp}, we rewrite it as an integral along the real axis
\begin{multline}\label{eq:S_T(t|X)=int 2exp real line}
    \frac{1}{2\pi \ii} \int_{\mathcal{C}_2} \dd \rho \; e^{\rho \tau} 
        \frac{1}{\rho}\left(1 - Q(\rho,1\,\vert\,X_0)\right)
    \\
    =
    e^{\rho_2 \tau}\;
    \frac{1}{2\pi\ii}
    \int_{\rho_1-\rho_2}^{0}
    \dd\rho\;
     \Delta\Big[ 
        \frac{e^{\rho\tau}}{\rho+\rho_2}
        (1 - Q(\rho+\rho_2, 1\,\vert\,X_0))
    \Big], 
\end{multline}
where $\Delta$ denotes the discontinuity across the cut 
\begin{equation}\label{eq:Delta[f]=disc 2exp}
    \Delta[f(\rho)] \equiv \lim_{\epsilon\to0}[f(\rho-\ii\epsilon) - f(\rho+\ii\epsilon)].
\end{equation} 
Since the discontinuity is purely imaginary and the integral in the right hand side is bounded, the second term in \eqref{eq:S_T(t|X)=res+int 2exp} decays exponentially fast as $e^{\rho_2\tau}$ at large times. Substituting \eqref{eq:S_T(t|X)=res result 2exp} into \eqref{eq:S_T(t|X)=res+int 2exp} and replacing the integral by its asymptotic behavior $e^{\rho_2\tau}$ with $\rho_2$ given by \eqref{eq:rho=branchcut 2exp}, we obtain the exponential decay of the survival probability
\begin{equation}\label{eq:S_T-Sinf=decay 2exp}
    S_T(\tau\,\vert\,X_0) - S_\infty(X_0) \underset{\tau\to\infty}{\asymp}
    \exp\left[ - \left(\sqrt{\beta} - \sqrt{\alpha\gamma}\right)^2 \tau \right].
\end{equation} 
This is the behavior presented in \eqref{eq:res_survival S=isinglike} with the decay rate $\xi_\tau(\alpha) = -(\sfrac{1}{\rho_2})$ stated in~\eqref{eq:res_corrLengths=2exp}. In other words, the survival probability $S_T(\tau\,\vert\,X_0)$ approaches its asymptotic value exponentially fast. The asymptotic value $S_\infty(X_0)$ is given by the residue of the integrand in \eqref{eq:S_T=inverseLT} at $\rho=0$ and the decay rate corresponds to the rightmost branching point.

\par Similarly, for $S_N(n\,\vert\,X_0)$, the integrand in the inverse transform \eqref{eq:S_N=inverseT} has poles at $s=0$ and $s=1$ and a branch cut
\begin{equation}\label{eq:s1=2exp}
    s\in[s_1,\infty), \qquad  s_1=\frac{\left(\beta+\alpha\gamma\right)^2}{4\alpha\beta\gamma}.
\end{equation}
Deforming the contour of integration $\mathcal{C}_1 \mapsto \mathcal{C}_2$ and accounting for the pole at $s=1$ we again represent the survival probability as a sum of residue and the integral over the branch cut
\begin{multline}\label{eq:S_N=res+int_2exp}
    S_N(n\,\vert\,X_0) = - \Res_{s=1}\left[  \frac{1}{s^{n+1}} \frac{1}{1-s} (1 - Q(0,s\,\vert\,X_0))\right]
    \\
    + \int_{\mathcal{C}_2} \frac{\dd s}{2\pi\ii}
        \frac{1}{s^{n+1}} \frac{1}{1-s} (1 - Q(0,s\,\vert\,X_0)).
\end{multline}
A straightforward computation shows that the residue at $s=1$ corresponds to the survival probability $S_\infty(X_0)$ as
\begin{equation}\label{eq:S_N=res 2exp}
    - \Res_{s=1}\left[  \frac{1}{s^{n+1}} \frac{1}{1-s} (1 - Q(0,s\,\vert\,X_0))\right]
    = 1 - \frac{\alpha\gamma}{\beta} e^{- \left(\frac{\beta}{\alpha}-\gamma\right)X_0} = S_\infty(X_0).
\end{equation}
The integral in \eqref{eq:S_N=res+int_2exp} can be represented as an integral of the discontinuity \eqref{eq:Delta[f]=disc 2exp} over the real line essentially in the same way as in \eqref{eq:S_T(t|X)=int 2exp real line}. Factoring out the branching point by changing the value of integration $s\mapsto s s_1$ we find that
\begin{multline}
    \int_{\mathcal{C}_2} \frac{\dd s}{2\pi\ii}
        \frac{1}{s^{n+1}} \frac{1}{1-s} (1 - Q(0,s\,\vert\,X_0))
    \\ =- s_1^{-n}
    \int_{1}^{\infty} \frac{\dd s}{2\pi\ii} 
    \Delta\left[\frac{1}{s^{n+1}}\frac{1}{1-s\, s_1}(1-Q(0,s \, s_1\,\vert\,X_0)) \right].
\end{multline}
The integral in the right hand side is bounded and hence for large $n$ the left hand side decays as $s_1^{-n}$ (recall that $s_1>1$). Substituting  \eqref{eq:S_N=res 2exp} into \eqref{eq:S_N=res+int_2exp} and replacing the integral in \eqref{eq:S_N=res+int_2exp} term by its asymptotic behavior $s_1^{-n}$ we find 
\begin{equation}
    S_N(n\,\vert\,X_0) - S_\infty(X_0) \underset{n\to\infty}{\asymp} s_1^{-n}.
\end{equation}
Using the explicit form of $s_1$ as in \eqref{eq:s1=2exp} we finally arrive at
\begin{equation}\label{eq:S_n-Sinf=decay 2exp}
    S_N(n\,\vert\,X_0) - S_\infty(X_0)
    \underset{n\to\infty}{\asymp} \exp\left[ - n \log\frac{(\beta+\alpha\gamma)^2}{4\alpha\beta\gamma} \right].
\end{equation}
This is the behavior \eqref{eq:res_survival S=isinglike} with the decay rate $\xi_n(\alpha)$ given by \eqref{eq:res_corrLengths=2exp}. Note, that the probability to survive for the infinite time $S_\infty(X_0)$ corresponds to the residue at $s=1$ and the decay rate is now governed by the branching point $s_1$.

\par It should be mentioned, that the result \eqref{eq:S_n-Sinf=decay 2exp} can be obtained if we note that for large $n$ the sum in \eqref{eq:S(n)=1/(1-s) (1-Q)} can be replaced by an integral essentially changing \eqref{eq:S(n)=1/(1-s) (1-Q)} for the Laplace transform of continuous variable. In what follows we will sometimes use this observation instead of analyzing the inversion \eqref{eq:S_N=inverseT} in the complex $s$ plane.

\par From \eqref{eq:S_infty=2exp_cases} it is evident that the probability to survive indefinitely decays exponentially with $X_0$. However, when $X_0$ is close to $0$, a significant fraction of the trajectories do cross the origin. The mean values of $\tau$ and $n$ for such trajectories are given by
\begin{align}
    \label{eq:con_<t>=def}
    & \mathbb{E}[\tau\,\vert\,X_0,\tau<\infty] = 
    - \pdv{}{\rho}
    \left. \left[ \frac{Q(\rho,s\,\vert\,X_0)}{1-S_\infty(X_0)} \right] \right|_{s=1,\rho=0},
    \\
    \label{eq:con_<n>=def}
    & \mathbb{E}[n\,\vert\,X_0,\tau<\infty] = 
    \pdv{}{s}
    \left. \left[ \frac{Q(\rho,s\,\vert\,X_0)}{1-S_\infty(X_0)} \right]
    \right|_{s=1,\rho=0}.
\end{align}
Similarly, for the second moments we have
\begin{align}
    \label{eq:con_<t2>=def}
    & \mathbb{E}[\tau^2\,\vert\,X_0,\tau<\infty] = 
    \pdv{^2}{\rho^2}
    \left. \left[ \frac{Q(\rho,s\,\vert\,X_0)}{1-S_\infty(X_0)} \right] \right|_{s=1,\rho=0},
    \\
    \label{eq:con_<n2>=def}
    & \mathbb{E}[n^2\,\vert\,X_0,\tau<\infty] = 
    \left(\pdv{^2}{s^2}+\pdv{}{s} \right)
    \left. \left[ \frac{Q(\rho,s\,\vert\,X_0)}{1-S_\infty(X_0)} \right]
    \right|_{s=1,\rho=0}.
\end{align}
Straightforward computation using \eqref{eq:LT P = Q = 2exp} gives the results presented in Sec.~\ref{sec:model}, namely, \eqref{eq:res_2exp_meanT_survival}, \eqref{eq:res_2exp_meanN_survival}, \eqref{eq:res_2exp_varT_survival}, and \eqref{eq:res_2exp_varN_survival}.

\subsubsection{Critical point}\label{sec:caseI_critical}
\begin{figure}[h]
    \includegraphics{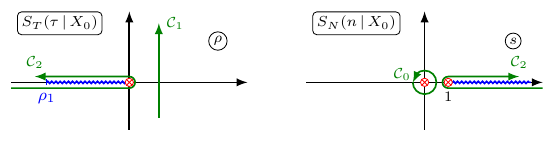}
    \caption{Analytic structure of the Laplace transform of $S_T(\tau\,\vert\,X_0)$ (left) and of the generating function of $S_N(n\,\vert\,X_0)$ (right) at the critical point for the distributions \eqref{eq:p,q=2exp}.}\label{fig:S_critical_2exp}
\end{figure}

If $\alpha=\alpha_c$ ($\beta=\alpha\gamma$), then $S_\infty(X_0)=0$ and the decay rates in \eqref{eq:S_T-Sinf=decay 2exp} and \eqref{eq:S_n-Sinf=decay 2exp} are zeros. 
At the level of the analytic structure of the integrands in the inverse transform \eqref{eq:S_T=inverseLT} and \eqref{eq:S_N=inverseT} the difference with respect to the survival regime is that the endpoints of the branch cuts coincide with the poles (see Fig.~\ref{fig:S_critical_2exp}). Thus after the transformation of the contours we cannot separate the contributions from the poles and the contributions from the cuts as was done \eqref{eq:S_T(t|X)=res+int 2exp} and \eqref{eq:S_N=res+int_2exp}. Consequently to extract the large $\tau$ (large $n$) behaviors we need to study the expansions of $Q(\rho,s\,\vert\,X_0)$ as $\rho\to0$ ($s\to1$). 

\par We start with the survival probability $S_T(\tau\,\vert\,X_0)$.
First we set $\alpha=\alpha_c$ in the exact expression for $Q(\rho,s\,\vert\,X_0)$ as given by \eqref{eq:LT P = Q = 2exp} and then use the representation \eqref{eq:S(tau)=1/rho(1-Q)} for the Laplace transform of the survival probability arriving at:
\begin{equation}\label{eq:S(tau)LT=2exp}
    \int_{0}^{\infty}\dd \tau\, e^{-\rho\tau}
    S_T(\tau\,\vert\,X_0) = 
    \frac{1}{\rho} 
    \left[
        1 - 
        \frac{\exp\left[-\frac{\gamma X_0}{2\beta}
                        \left(\rho+
                                \sqrt{\rho(4\beta+\rho)}
                        \right)
                \right]}
             {1 + \frac{\rho}{2\beta} + \frac{1}{2\beta}\sqrt{\rho(4\beta+\rho)}}
    \right].
\end{equation}
Large $\tau$ behavior of $S_T(\tau\,\vert\,X_0)$ is governed by $\rho\to0$ expansion of its Laplace transform. Expanding the right hand side of \eqref{eq:S(tau)LT=2exp} in series we obtain
\begin{equation}\label{eq:S(tau)LT=rho->0_2exp}
    \int_{0}^{\infty}\dd \tau\, e^{-\rho\tau}
        S_T(\tau\,\vert\,X_0) 
    \underset{\rho\to0}{\sim}
    \left(1 + \gamma X_0\right)
    \frac{1}{\sqrt{\beta}}
    \frac{1}{\sqrt{\rho}}.
\end{equation}
The square root divergence suggests, that $S_T(\tau\,\vert\,X_0)$ decays as $\tau^{-1/2}$. Substituting this ansatz in \eqref{eq:S(tau)LT=rho->0_2exp} yields
\begin{equation}\label{eq:S(tau)~X0_2exp}
    S_T(\tau\,\vert\,X_0) \underset{\tau\to\infty}{\sim} 
    \frac{1 + \gamma X_0}{\sqrt{\pi \beta}}
    \frac{1}{\sqrt{\tau}}.
\end{equation}
The procedure for $S_N(n\,\vert\,X_0)$ is similar. By setting $\alpha=\alpha_c$ in $Q(\rho,s\,\vert\,X_0)$ and using the representation \eqref{eq:S(n)=1/(1-s) (1-Q)} for the generating function of $S_N(n\,\vert\,X_0)$, we find that at the critical point:
\begin{equation}\label{eq:sumS(n)=2exp}
    \sum_{n=0}^{\infty} s^n\, S_N(n\,\vert\,X_0)
    = \frac{1}{1-s}\left(
    1 - 
    \left( 1 - \sqrt{1-s} \right)
    e^{ - \sqrt{1-s} \, \gamma X_0 } \right).
\end{equation}
If $n$ is large, the sum may be replaced by the integral. Then repeating the same arguments as for $S_T(\tau\,\vert\,X_0)$ with $\rho$ replaced by $-\log s$ we find the behavior of $S_N(n\,\vert\,X_0)$ at large~$n$:
\begin{equation}\label{eq:S(n)~X0_2exp}
    S_N(n\,\vert\,X_0) 
    \underset{n\to\infty}{\sim} 
    \frac{1 + \gamma X_0}{\sqrt{\pi}} \frac{1}{\sqrt{n}}.
\end{equation}
With \eqref{eq:S(tau)~X0_2exp} and \eqref{eq:S(n)~X0_2exp} we prove the power-law decay of the survival probabilities stated in~\eqref{eq:res_critical S=isinglike} and find the constants $c_\tau$ and $c_n$ given by \eqref{eq:res_constants_critical=2exp}. However, two remarks are in order.

\paragraph{Sparre-Anderson theorem} The first remark is that if $\alpha=\alpha_c$ ($\beta=\alpha\gamma$), then for $\rho=0$ the effective random walk~\eqref{eq:F(k)=2exp} is symmetric. It means, that the survival probability for $X_0=0$ can be computed by Sparre-Anderson theorem. This is a very nice universal result in the theory of the discrete-time random walks that was obtained in \cite{A-54}. The theorem states that for a symmetric random walk starting at the origin, the probability to survive up to the $n$th step is given by a simple closed expression
\begin{equation}\label{eq:SparreAnderson_Sn}
    S_N(n\,\vert\,X_0=0) = \binom{2n}{n}2^{-2n}.
\end{equation}
The universality of the result \eqref{eq:SparreAnderson_Sn} is that it holds for any distribution of jumps provided that it is symmetric and its cumulative distribution is continuous. For the effective random walk \eqref{eq:F(k)=2exp} both conditions are satisfied, and the Sparre-Anderson theorem holds.  Taking the the limit $n\to\infty$ in \eqref{eq:SparreAnderson_Sn} we find that 
\begin{equation}\label{eq:SparreAnderson_Sn_asympt}
    S_N(n\,\vert\,X_0=0)\underset{n\to\infty}{\sim}\frac{1}{\sqrt{\pi n}},
\end{equation}
which coincides with \eqref{eq:S(n)~X0_2exp}.

\paragraph{Scaling limit.} The second remark is that at the critical point the effective random walk consists of identically distributed jumps with finite moments and zero mean. Therefore it is natural to expect, that such a process converges to Brownian motion in the scaling limit. The expected scaling is $X_0\sim\sqrt{\tau}$ (or $X_0\sim \sqrt{n}$). 
In other words, if both $X_0\to\infty$ and $\tau\to\infty$ with fixed ratio $z = X_0/\sqrt{\tau}$, we should have
\begin{equation}\label{eq:S(tau)=V(z)_2exp}
    S_T(\tau\,\vert\,X_0)
    \sim 
    V_\tau\left(z_\text{c} = \frac{X_0}{\sqrt{\tau}}\right),
    \quad X_0 \to \infty, 
    \quad \tau\to\infty.
\end{equation}
The scaling function $V_\tau(z)$ can be found in a same way as was done in \cite{MMS-17}, i.e., by analyzing the Laplace transform  
\begin{equation}
    \label{eq:hatSt=def}
    \hat{S}_T(\rho\,\vert\,\lambda) \equiv \int_{0}^{\infty} \dd X_0\, e^{-\lambda X_0}
    \int_{0}^{\infty} \dd \tau\, e^{-\rho \tau} S_T(\tau\,\vert\,X_0).
\end{equation}
The behavior of $S_T(\tau\,\vert\,X_0)$ for large $X_0$ and $\tau$ can be extracted from $\lambda\to0$ and $\rho\to0$ behavior of $\hat{S}_T(\rho\,\vert\,\lambda)$. However the scaling $X_0\sim\sqrt{\tau}$ should be accounted for when constructing the expansion $\rho\to0$ and $\lambda\to0$.
Substituting the scaling form \eqref{eq:S(tau)=V(z)_2exp} into~\eqref{eq:hatSt=def} and changing the variables of integration  via $z = X_0 / \sqrt{\tau}$ and $y =\rho\tau$, we obtain
\begin{equation}\label{eq:hat(S)=1_2exp}
    \hat{S}_{T}(\rho\,\vert\,\lambda)
    \underset{\substack{\lambda\to0\\\rho\to0}}{\sim}
    \frac{1}{\rho^{3/2}}
    \int_{0}^{\infty} \dd y \, e^{-y} \sqrt{y} 
    \int_{0}^{\infty} \dd z \, 
    e^{ - \frac{\lambda}{\sqrt{\rho}} z \sqrt{y} }
    V_\tau(z).
\end{equation}
From \eqref{eq:hat(S)=1_2exp} it is clear, that the correct scaling ratio between $\lambda$  and $\rho$ is $\lambda \sim \sqrt{\rho}$.
At the same time, we can find the explicit form of $\hat{S}_T(\rho\,\vert\,\lambda)$ by performing the Laplace transform of $S_T(\rho\,\vert\,X_0)$ in \eqref{eq:S(tau)LT=2exp} with respect to $X_0$. Then by setting $\lambda=u \sqrt{\rho}$ and expanding the resulting expression in series as $\rho\to0$ we obtain
\begin{equation}\label{eq:hat(S)=2_2exp}
    \hat{S}_T(\rho\,\vert\,\lambda=u\sqrt{\rho})
    \underset{\rho\to0}{\sim}
    \frac{1}{\rho^{3/2}} 
    \frac{1}{u\left(1 + u \frac{\sqrt{\beta}}{\gamma}\right)}.
\end{equation}
Comparing \eqref{eq:hat(S)=1_2exp} and the expansion \eqref{eq:hat(S)=2_2exp} obtained from the explicit expression of $\hat{S}_{T}(\rho\,\vert\,X_0)$ yields an integral equation for the scaling function $V_\tau(z)$
\begin{equation}\label{eq:V_t int equation}
    \int_{0}^{\infty} \dd y \, e^{-y} \sqrt{y} 
    \int_{0}^{\infty} \dd z \, 
    e^{ - u z \sqrt{y} }\, 
    V_\tau(z)
    = 
    \frac{1}{u\left(1 + u \frac{\sqrt{\beta}}{\gamma}\right)}.
\end{equation}
This equation was studied in details in \cite{MMS-17} and here we will not repeat the analysis. The solution of \eqref{eq:V_t int equation} is $V_\tau(z) = \erf\left( \frac{z}{2}\frac{\gamma}{\sqrt{\beta}}\right)$.
Hence for the survival probability we have
\begin{equation}\label{eq:S(tau)~erf()_2exp}
    S_{T}(\tau\,\vert\,X_0) \sim
    \erf\left(
    \frac{1}{2}
        \frac{\gamma X_0}{\sqrt{\beta \tau}}
    \right),
    \quad \tau \to \infty, 
    \quad \frac{X_0}{\sqrt{\tau}} \text{~--- fixed}.
\end{equation}
The scaling behavior of $S_N(n\,\vert\,X_0)$ is obtained by replacing the sum by an integral in \eqref{eq:sumS(n)=2exp} and repeating the same arguments with $\rho$ replaced by $-\log s$. This procedure results in
\begin{equation}\label{eq:S(n)~erf()_2exp}
    S_N(n\,\vert\,X_0)
        \sim \erf\left(
            \frac{1}{2}
            \frac{\gamma X_0}{\sqrt{n}}
        \right),  
    \quad n\to \infty, 
    \quad \frac{X_0}{\sqrt{n}}\text{~--- fixed}.
\end{equation}

\par
\begin{figure}
\includegraphics[width=\linewidth]{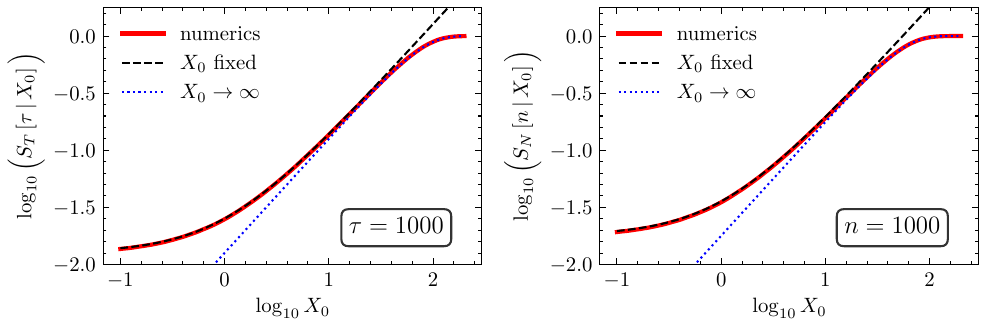}
\caption{
Survival probabilities as functions of the initial position at the \textit{critical point} with fixed $\tau = 1000$ (left) and $n=1000$ (right).  
The black dashed lines correspond to the fixed $X_0$ behavior described by \eqref{eq:S(tau)~X0_2exp} and \eqref{eq:S(n)~X0_2exp}.
The blue dotted lines correspond to the scaling limit as in \eqref{eq:S(tau)~erf()_2exp} and \eqref{eq:S(n)~erf()_2exp}. 
The model parameters are  $\alpha=2$, $\beta=2$, $\gamma=1$.
For the details of the simulation see Appendix~\ref{sec:numerics}.
}
\label{fig:S(tau)S(n)_2exp}
\end{figure}

 We conclude by numerically verifying the analytical results \eqref{eq:S(tau)~X0_2exp}, \eqref{eq:S(n)~X0_2exp} and \eqref{eq:S(tau)~erf()_2exp}, \eqref{eq:S(n)~erf()_2exp}. As is clear from Fig.~\ref{fig:S(tau)S(n)_2exp}, the numerical simulations are in perfect agreement with the analytical predictions.

\subsubsection{Absorption regime}\label{sec:caseI_absorption}
In this regime $\alpha>\alpha_c$ ($\beta<\alpha \gamma$) and $S_\infty(X_0)=0$, i.e., the drift is strong enough and the process eventually crosses the origin.
In terms of the analytic structure of the integrands in the inverse transforms \eqref{eq:S_T=inverseLT} and \eqref{eq:S_N=inverseT} of the survival probabilities, the difference is that there are no poles at $\rho=0$ and $s=1$, which can be verified by using the exact form of $Q(\rho,s\,\vert\,X_0)$ as in \eqref{eq:LT P = Q = 2exp}. Consequently, we can deform the contours of integration as in \eqref{eq:S_T(t|X)=res+int 2exp} and \eqref{eq:S_N=res+int_2exp}, but this time there will be no terms with residues. As a result, the survival probabilities at large times vanish exponentially fast with the decay rates governed by the branching points $\rho_2$ in \eqref{eq:rho=branchcut 2exp} and $s_1$ in \eqref{eq:s1=2exp}. In summary, the behavior of $S_T(\tau\,\vert\,X_0)$ and $S_N(n\,\vert\,X_0)$ is given by
\begin{align}
\label{eq:St=asympt 2exp}
    & S_{T}(\tau\,\vert\,X_0) \underset{\tau\to\infty}{\asymp}
     \exp\left[-(\sqrt{\beta} - \sqrt{\alpha \gamma})^2\tau\right],\\
\label{eq:Sn=asympt 2exp}
    & S_{N}(n\,\vert\,X_0) \underset{n\to\infty}{\asymp} \exp\left[ - n \log\frac{(\beta+\alpha\gamma)^2}{4\alpha\beta\gamma} \right].
\end{align}
This is exactly the result stated in \eqref{eq:res_absorption S=isinglike} with the ``correlation lengths'' \eqref{eq:res_corrLengths=2exp}. Note, that the decay rates \eqref{eq:S_T-Sinf=decay 2exp} and \eqref{eq:S_n-Sinf=decay 2exp}  in the \textit{survival regime} and the decay rates in the \textit{absorption regime} have the same functional form. 

\par In the rest of this section we provide more detailed description of the first-passage properties in the absorption regime.

\paragraph{Moments of $\tau$ and $n$.} First of all, since survival probabilities decay exponentially, we expect all moments of $n$ and $\tau$ to be finite. They can be found by expanding $Q(\rho,s\,\vert\,X_0)$ in series with respect to $\rho$ and $s$. Specifically, the first two moments of $\tau$ are given by
\begin{equation}\label{eq:<tau>=..<tau2>=..}
\mathbb{E}[\tau \,\vert\,X_0] = -
  \left. \partial_\rho Q(\rho,1\,\vert\,X_0) \right|_{\rho=0},
  \qquad
  \mathbb{E}[\tau^2 \,\vert\,X_0] = 
  \left. \partial^2_\rho Q(\rho,1\,\vert\,X_0) \right|_{\rho=0}.
\end{equation}
The first two moments of $n$ are obtained in a similar way as
\begin{equation}\label{eq:<n>=, <n2>=}
\mathbb{E}[n\,\vert\,X_0] = 
  \left. \partial_s Q(0,s\,\vert\,E_0) \right|_{\rho=0},
  \qquad
\mathbb{E}[n^2\,\vert\,X_0] = 
\left[ \strut \partial_s^2 + \partial_s\right]
  \left.  Q(\rho,s\,\vert\,E_0) \right|_{s=1}.
\end{equation}
In the scenario where both $p(t)$ and $q(M)$ are exponential we have the explicit form \eqref{eq:LT P = Q = 2exp} of $Q(\rho,s\,\vert\,X_0)$. After straightforward computation we find the mean and the variances of the first-passage time $\tau$ 
\begin{equation}\label{eq:<tau>=, Var[tau]=, 2exp}
    \mathbb{E}[\tau\,\vert\,X_0 ] 
    = \frac{1 + \gamma X_0}{ \alpha \gamma - \beta},
    \qquad
    \mathrm{Var}[\tau\,\vert\,X_0]  
    = \frac{\beta + \alpha\gamma + 2\beta\gamma X_0}{(\alpha\gamma - \beta)^3},
\end{equation}
as well as the mean and the variance of the number of jumps $n$ that have occurred prior to the first-passage 
\begin{equation}\label{eq:<n>=, Var[n]=_2exp}
\mathbb{E}[n\,\vert\,X_0 ]  = 
    \gamma \, \frac{\alpha + \beta X_0}{\alpha \gamma - \beta},
  \qquad 
    \mathrm{Var}[n\,\vert\,X_0]  
    = \beta \gamma \, \frac{  \alpha(\alpha\gamma+\beta) + (\alpha^2\gamma^2 + \beta^2)X_0}{(\alpha\gamma - \beta)^3}.
\end{equation}
In particular, the relations \eqref{eq:<tau>=, Var[tau]=, 2exp} and \eqref{eq:<n>=, Var[n]=_2exp} imply that for large $X_0$ the mean values of $\tau$ and $n$ are given by:
\begin{equation}\label{eq:<tau>,<n>~2expX0->inf}
    \mathbb{E}[ \tau\,\vert\,X_0]
        \underset{X_0\to\infty}{\sim}
        \frac{\gamma\, X_0}{\alpha\gamma-\beta},
    \qquad
    \mathbb{E}[ n \,\vert\,X_0]
        \underset{X_0\to\infty}{\sim}
        \frac{\gamma \beta\,  X_0}{\alpha\gamma-\beta}.
\end{equation}
This is exactly the behavior \eqref{eq:E[t], E[n],X->inf, expected} found by the heuristic argument in Sec.~\ref{sec:model}.

\paragraph{Correlation coefficient.} As argued in Sec.~\ref{sec:model}, $\tau$ and $n$ are correlated. One way to quantify this is to compute the correlation coefficient 
\begin{equation}
    \mathrm{corr}(\tau,n) = 
        \frac{\mathbb{E}[(n\tau) \,\vert\,X_0] - \mathbb{E}[\tau\,\vert\,X_0]\;  \mathbb{E}[n\,\vert\,X_0]}
             {\sqrt{\mathrm{Var}[\tau\,\vert\,X_0]\, \mathrm{Var}[n\,\vert\,X_0] }}.
\end{equation}
The correlation coefficient can be easily computed from the exact form of $Q(\rho,s\,\vert\,X_0)$ by using the identity
\begin{equation}
    \mathbb{E}[(n\tau)\,\vert\,X_0] = -\pd_s \pd_\rho Q(\rho,s\,\vert\,X_0)\Big|_{\rho=0,s=1}.
\end{equation}
After a simple calculation we obtain:
\begin{equation}
    \mathrm{corr}(\tau,n)  
    =
    \frac{\sqrt{\beta\gamma} \left(2\alpha+X_0(\beta+\alpha\gamma)\right)}
         {\sqrt{(\beta+\alpha\gamma + 2\beta\gamma\,X_0)(\alpha(\alpha\gamma+\beta) + (\alpha^2\gamma^2 + \beta^2)X_0)}}.
\end{equation}
The moments and the correlation coefficient characterize $\mathbb{P}[\tau,n\,\vert\,X_0]$ close to the typical values of $\tau$ and $n$. However, due to the explicit form of $Q(\rho,s\,\vert\,X_0)$ we can also obtain the marginal probability distributions $\mathbb{P}_{T}[\tau\,\vert\,X_0]$ and $\mathbb{P}_N[n\,\vert\,X_0]$ in the limit $X_0\to\infty$. This is done by applying the large deviation theory (see \cites{T-09,T-18,BCKT-25} for a pedagogical review).

\paragraph{Rate function for $\tau$.}
Both the mean and the variance of $\tau$ grow linearly with $X_0$ as we have shown in \eqref{eq:<tau>,<n>~2expX0->inf}. This suggests, that the marginal probability distribution $\mathbb{P}_T[\tau\,\vert\,X_0]$ follows the large deviation form 
\begin{equation}\label{eq:P[tau]=LDF_2exp}
  \mathbb{P}_T[\tau\,\vert\,X_0] 
  \underset{X_0\to\infty}{\asymp}
  e^{ - X_0 \Phi(z) },
  \qquad
  z = \frac{\alpha \tau}{X_0} - 1.
\end{equation}
The scaling variable $z$ captures the deviation of $\tau$ from its typical values.  Linear growth of the mean and variance of $\tau$ with $X_0$ suggests that scale of such fluctuations is proportional to $X_0$. 
Recall that even in the extreme case, where only a single jump with zero amplitude occurs at $t=0$, the process still requires a time of $X_0/\alpha$ to reach the origin. This implies that the scaling variable $z$ lies in the range $z\in[0;\infty)$. 
Note that in order for the large deviation form to recover the Gaussian distribution near the typical value, the scaling variable only needs to be proportional to $X_0$; for example, one could use $z' = \tau/X_0$. The particular choice of $z$ in \eqref{eq:P[tau]=LDF_2exp} is primarily motivated by convenience, as it simplifies the expressions that follow.

\par The approximation \eqref{eq:P[tau]=LDF_2exp} essentially means that the probability of observing a rescaled fluctuation $z$ decays exponentially with $X_0$, and is governed by the non-negative function $\Phi(z)$ usually referred to as the \textit{rate function}.

\par The easiest way to compute the rate function is to substitute the large deviation ansatz into the Laplace transform of the marginal probability distribution $\mathbb{P}_{T}[\tau\,\vert\,X_0]$
\begin{equation}\label{eq:Q(p)=int dz}
  \left. Q(\rho,s\,\vert\,X_0) \strut \right|_{s=1} = 
  \int_{0}^{\infty} \dd \tau\, e^{-\rho\tau}\,
    \mathbb{P}_T[\tau\,\vert\,X_0]
  \underset{X_0\to\infty}{\asymp}
  \int_{0}^{\infty} e^{-X_0 \left(\frac{1}{\alpha}\rho(z+1)+\Phi(z)\right)}\dd z .
\end{equation}
In the limit $X_0 \to \infty$, this integral is dominated by the minimum of the exponent due to the saddle-point approximation
\begin{equation}\label{eq:Q(p)=exp[min_z]}
  \left. Q(\rho,s\,\vert\,X_0) \strut \right|_{s=1} 
  \underset{X_0\to\infty}{\asymp}
  \exp\left[ -X_0 \min_z\left(\frac{1}{\alpha}\rho(z+1)+\Phi(z)\right)\right].
\end{equation}
At the same time, the explicit form of $Q(\rho,s,\vert,X_0)$, derived earlier in \eqref{eq:LT P = Q = 2exp}, has the asymptotic behavior
\begin{equation}\label{eq:Q(p)=exp[phi(p)]}
  \left. Q(\rho,s\,\vert\,X_0) \strut \right|_{s=1}
  \underset{X_0\to\infty}{\asymp}
  e^{ - X_0 \, \phi(\rho)},
\end{equation}
where
\begin{equation}\label{eq:phi(rho)=2exp}
    \phi(\rho) = \frac{1}{2\alpha}\left( \rho + \beta - \alpha \gamma + 
  \sqrt{(\rho+\beta-\alpha\gamma)^2 + 4\alpha\gamma \rho} \right).
\end{equation}
Equating the exponents in \eqref{eq:Q(p)=exp[min_z]} and \eqref{eq:Q(p)=exp[phi(p)]} implies the relation between $\Phi(z)$ and $\phi(\rho)$
\begin{equation}\label{eq:phi(rho)=max_z-2exp}
  \min_{z}\left( \frac{1}{\alpha}\, \rho(z+1) + \Phi(z) \right)
  = \phi(\rho).
\end{equation}
This identifies $\Phi(z)$ as the Legendre-Fenchel transform of $\phi(\rho)$:
\begin{equation}\label{eq:Phi(z)=max_p-2exp}
  \Phi(z) = \max_{\rho}\left(
    - \frac{1}{\alpha} \rho (z+1) + \phi(\rho)
  \right).
\end{equation}
Equation \eqref{eq:Phi(z)=max_p-2exp} together with \eqref{eq:phi(rho)=2exp} provide the rate function in the parametric form. However, the maximization can be carried out explicitly. Indeed, the maximum of \eqref{eq:Phi(z)=max_p-2exp} is reached at $\rho=\rho^*$ satisfying $\phi'(\rho^*) = (z+1)/\alpha$. Solving the maximization condition results in
\begin{equation}\label{eq:p^*=2exp}
  \rho^* = -\beta-\alpha\gamma + (1+2z)\sqrt{\frac{\alpha\beta\gamma}{z(z+1)}}.
\end{equation}
Finally, substituting \eqref{eq:p^*=2exp} into \eqref{eq:Phi(z)=max_p-2exp} we find the rate function $\Phi(z)$:
\begin{equation}\label{eq:Phi(z)=_2exp}
  \Phi(z) = \gamma \left( \sqrt{z} - \sqrt{(z+1) \frac{\beta}{\alpha\gamma} }\right)^2.
\end{equation}
This function is convex and non-negative, with a minimum at $z=\frac{\beta}{\alpha\gamma -\beta}$. The behavior of $\Phi(z)$ around the minimum characterizes the probability distribution $\mathbb{P}_{T}[\tau\,\vert\,X_0]$ near typical values of $\tau$. The behavior of the distribution for the atypical values of $\tau$ is governed by the expansions of $\Phi(z)$ for $z\to0$ and $z\to\infty$.
These expansions are given by:
\begin{equation}\label{eq:Phi(z)=asympt 2exp}
    \Phi(z) = 
    \left\{
    \begin{aligned}
        &\frac{\beta}{\gamma} - 2 \sqrt{\frac{\beta \gamma}{\alpha}} \sqrt{z}, 
        && z\to 0,\\
        &\frac{(\alpha \gamma - \beta )^3}
             {4\alpha^2\beta\gamma} (z-z^*)^2,
        &&z \to z^*=\frac{\beta}{\alpha\gamma -\beta},\\
        &
        \frac{1}{\alpha}\left(\sqrt{\alpha\gamma}-\sqrt{\beta}\right)^2 \,z ,
        &&z\to\infty.
    \end{aligned}\right.
\end{equation}

\par From \eqref{eq:Phi(z)=asympt 2exp}, we can deduce the behavior of the probability distribution. Close to the typical values, it is a Gaussian 
\begin{equation}
    \mathbb{P}_T[\tau\,\vert\, X_0] 
    \underset{X_0\to\infty}{\sim}
    \exp\left[ - \frac{1}{2} 
            \frac{\left( \tau - \mathbb{E}_\infty[\tau] \right)^2}{ \mathrm{Var}_\infty [\tau] } \right],
    \qquad  \tau \approx \mathbb{E}_\infty [\tau],
\end{equation}
where
\begin{equation}\label{eq:MeanVar_asymp=2exp}
    \mathbb{E}_\infty [\tau] = \frac{\gamma }{\alpha \gamma - \beta} \, X_0,
    \qquad 
    \mathrm{Var}_\infty[\tau]  = 
    \frac{2 \beta\gamma }{(\alpha\gamma - \beta)^3} \, X_0.
\end{equation}
The asymptotic mean and variance \eqref{eq:MeanVar_asymp=2exp} are consistent with the exact results \eqref{eq:<tau>=, Var[tau]=, 2exp} obtained for finite $X_0$.

\par For the left tail of $\mathbb{P}_T[\tau\,\vert\,X_0]$, i.e., atypically small values of $\tau$, we have
\begin{equation}\label{eq:P[T]=left tail 2exp}
    \mathbb{P}_T[\tau\,\vert\, X_0] 
    \underset{X_0\to\infty}{\sim} 
    \exp\left[ 
    - \frac{\beta}{\alpha} X_0
    + 2\sqrt{\beta \gamma X_0} \, \sqrt{\Delta \tau }
    \right],
    \qquad \tau = \Delta \tau + \frac{X_0}{\alpha}    \ll\mathbb{E}_\infty [\tau].
\end{equation}
Recall, that the first-passage time cannot be smaller than $X_0/\alpha$, and hence it is more natural to use $\Delta \tau$ rather than $\tau$ as a parameter in \eqref{eq:P[T]=left tail 2exp}.
\par For the right tail (atypically large values of $\tau$) we obtain
\begin{equation}\label{eq:P[t]=right tail 2exp}
    \mathbb{P}_T[\tau\,\vert\,X_0] 
    \underset{X_0\to\infty}{\sim}
    \exp\left[ - \left(\sqrt{\alpha\gamma}-\sqrt{\beta}\right)^2 \tau \right],
    \qquad
    \tau \gg \mathbb{E}_\infty [\tau].
\end{equation}
Note, that the right tail of the distribution has the same exponential decay as the survival probability \eqref{eq:St=asympt 2exp}. This is not a coincidence as according to \eqref{eq:S(n),S(tau)=def} the survival probability is given by:
\begin{equation}\label{eq:St=int P[T] 2exp}
    S_T(\tau\,\vert\,X_0) = \int_{\tau}^{\infty} \dd \tilde{\tau}\,\mathbb{P}_T[\tilde{\tau}\,\vert\,X_0].
\end{equation}
For sufficiently large $\tau$, the distribution may be replaced by the asymptotic behavior \eqref{eq:P[t]=right tail 2exp} and computing the integral results in \eqref{eq:St=asympt 2exp}. Note that the decay rate of $S_T(\tau\,\vert\,X_0)$ and the asymptotic behavior of $\mathbb{P}_T[\tau\,\vert\,X_0]$ were obtained by different arguments, therefore the fact that they coincide may be used as a simple self-consistency check.

\par
\begin{figure}[h]
\includegraphics[width=\linewidth]{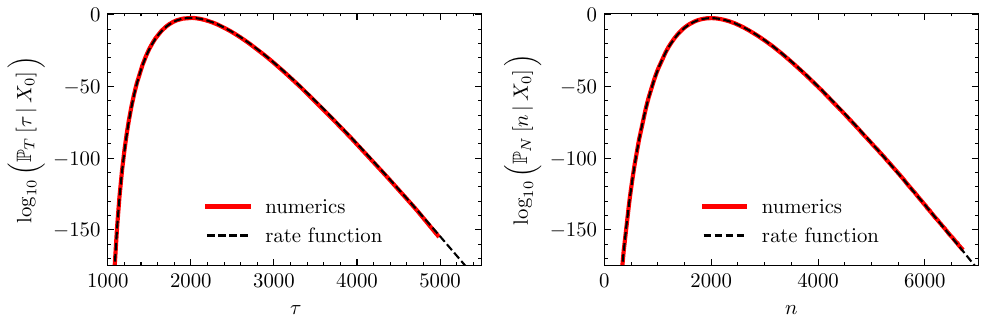}
\caption{Comparison between the numerical simulations and the analytical results for the distributions of $\tau$ (left) and $n$ (right). 
The dashed black lines denote the distributions $\mathbb{P}_T[\tau\,\vert\,X_0]$ and $\mathbb{P}_N[n\,\vert\,X_0]$ that are obtained from the large deviation forms \eqref{eq:P[tau]=LDF_2exp}, \eqref{eq:Phi(z)=_2exp} and \eqref{eq:P[n]=LDF_2exp}, \eqref{eq:Psi(c)=_2exp} respectively. Normalization constants in \eqref{eq:P[tau]=LDF_2exp} and \eqref{eq:P[n]=LDF_2exp} are restored by numerical integration. 
The solid red lines correspond to Monte-Carlo numerical simulations. The tails of the probability distributions are reached using importance sampling strategy with the exponential tilt in the observable.  The parameters of the model are $\alpha=1$, $\beta=1$, $\gamma=2$, $X_0=1000$.
For the details of the simulation see Appendix~\ref{sec:numerics}.
}
\label{fig:numerics_LDF_2exp}
\end{figure}

 The numerical verification of the large deviation approximation is shown in Fig.\ref{fig:numerics_LDF_2exp}. Note that the large deviation approximation \ref{eq:P[tau]=LDF_2exp} describes the tails of the distribution for atypical values of $\tau$, which have very small probabilities. To sample these regions, we use Importance Sampling, a strategy for generating rare events (see \cite{H-24} for an introduction), allowing us to explore probabilities as small as $10^{-150}$. Details of the simulations are given in Appendix~\ref{sec:numerics}.

\paragraph{Rate function for $n$} The mean and the variance of $n$ scale linearly with $X_0$ \eqref{eq:<n>=, Var[n]=_2exp}. Therefore it is natural to expect that the marginal probability distribution $\mathbb{P}_N[n\,\vert\,X_0]$ follows the large deviation form
\begin{equation}\label{eq:P[n]=LDF_2exp}
    \mathbb{P}_N[n\,\vert\,X_0] 
    \underset{X_0\to\infty}\asymp
    e^{ - X_0 \Psi(\nu)}, 
    \qquad
    \nu = \frac{\alpha n}{\beta X_0}.
\end{equation}
The scaling variable $\nu$ lies in the range $\nu\in[0;\infty)$. The rate function $\Psi(\nu)$ is computed in the same manner as $\Phi(z)$. Specifically, we first introduce another variable $\tilde{s}=-\log s$, substitute the large deviation ansatz into \eqref{eq:LT P = Q = 2exp} and compute the sum in the saddle-point approximation arriving at:
\begin{equation}\label{Q(s)=exp[min_c]_2exp}    
Q(\rho, e^{\tilde{s}} \vert\,X_0)\Big|_{\rho=0} = 
    \sum_{n=0}^{\infty} e^{ - \tilde{s} n } \,
  \mathbb{P}_N[n\,\vert\,X_0] 
  \underset{X_0 \to \infty}{\asymp}
  e^{ - X_0 \min\limits_{\nu}
            \left( \tilde{s}\, \frac{\beta}{\alpha} \nu 
                    + \Psi(\nu) \right)}.
\end{equation} 
Then we extract the asymptotic behavior of $Q(\rho,s\,\vert\,X_0)$ from the exact representation~\eqref{eq:LT P = Q = 2exp}
\begin{equation}\label{Q(s)=exp[psi(s)]_2exp}
    Q(\rho, e^{\tilde{s}} \vert\,X_0)\Big|_{\rho=0}
    \underset{X_0\to\infty}{\asymp}
    e^{-X_0 \psi(\tilde{s})},
\end{equation}
where 
\begin{equation}
    \psi(\tilde{s}) = \frac{1}{2\alpha}
  \left(\beta - \alpha \gamma + \sqrt{(\beta+\alpha\gamma)^2 - 4\alpha\gamma\beta \, e^{-\tilde{s}}}\right).
\end{equation}
Comparing the exponents in \eqref{Q(s)=exp[min_c]_2exp} and \eqref{Q(s)=exp[psi(s)]_2exp} we see that $\Psi(\nu)$ and $\psi(\tilde{s})$ are related by the Legendre-Fenchel transform. Inverting this transform yields the parametric representation for the rate function $\Psi(\nu)$, namely
\begin{equation}\label{eq:Psi(c)=max_s_2exp}
  \Psi(\nu) = \max_{\tilde{s}}
  \left(
    -\tilde{s} \, \nu \, \frac{\beta}{\alpha} + \psi(\tilde{s})
  \right).
\end{equation}
The maximization in \eqref{eq:Psi(c)=max_s_2exp} can be again carried out explicitly. 
This is done by first finding the solution $\tilde{s}^*$ of $\psi'(\tilde{s}) = \nu \frac{\beta}{\alpha}$ and then substituting it into \eqref{eq:Psi(c)=max_s_2exp}.
Performing this algebraic manipulations we find the rate function $\Psi(\nu)$ explicitly
\begin{equation}\label{eq:Psi(c)=_2exp}
    \Psi(\nu) = \frac{1}{2}\left(\frac{\beta}{\alpha} - \gamma \right) + \gamma \, \Omega(\nu)
    + \frac{\beta\, \nu}{\alpha} \log\left[ 2 \nu \,\Omega(\nu) \right],
\end{equation}
where
\begin{equation}
    \Omega(\nu) = \sqrt{ \left( \frac{\beta\,\nu}{\alpha\gamma} \right)^2 +\frac{1}{4} \left(\frac{\beta}{\alpha\gamma}+1\right)^2 }- \frac{\beta \, \nu}{\alpha\gamma}.
\end{equation}
This is a convex non-negative function with minimum at $\nu^*=\frac{\alpha\gamma}{\alpha\gamma-\beta}$ and the asymptotic behavior
\begin{equation}\label{eq:Psi(nu)=asympt 2exp}
    \Psi(\nu) = \left\{
    \begin{aligned}
        &\frac{\beta}{\alpha} + \frac{\beta}{\alpha}\nu\log\nu,
        && \nu \to 0, \\
        &
        \frac{\beta}{2\alpha^2\gamma}
        \frac{(\alpha\gamma-\beta)^3}{\beta^2+\alpha^2\gamma^2}
        (\nu-\nu^*)^2,
        && \nu \to \nu^*=\frac{\alpha\gamma}{\alpha\gamma-\beta}, \\
        & \nu \frac{\beta}{\alpha} \log \frac{(\beta+\alpha\gamma)^2}{4\alpha\beta\gamma},
        && \nu \to \infty.
    \end{aligned}
    \right. 
\end{equation}
Similarly to the distribution of $\tau$ we now use \eqref{eq:Psi(nu)=asympt 2exp} to extract the asymptotic behavior of the marginal distribution $\mathbb{P}_N[n\,\vert\,X_0]$. Close to the typical values of $n$ this distribution is a Gaussian
\begin{equation}
    \mathbb{P}_N[n\,\vert\,X_0] 
        \underset{X_0\to\infty}{\sim}
        \exp\left[ - \frac{1}{2} 
            \frac{\left( n - \mathbb{E}_\infty[n]\right)^2}{\mathrm{Var}_\infty[n]}\right],
    \qquad
    n \approx \mathbb{E}_\infty[ n],
\end{equation}
where
\begin{equation}
    \mathbb{E}_\infty[n]
        = \frac{\beta \gamma}{\alpha\gamma - \beta} \, X_0,
    \qquad
    \mathrm{Var}_\infty[n] =  \frac{\beta^2 + \alpha^2\gamma^2}{(\alpha\gamma-\beta)^3} \beta \gamma \, X_0.
\end{equation}
The asymptotic mean and variance are consistent with the exact results \eqref{eq:<n>=, Var[n]=_2exp}. 

\par The left tail of the distribution $\mathbb{P}_N[n\,\vert\,X_0]$ (atypically small values of $n$) is given by:
\begin{equation}
    \mathbb{P}_N[n\,\vert\,X_0] 
    \underset{X_0\to\infty}{\sim}
    \exp\left[
        -\frac{\beta}{\alpha}\,X_0 - n \log n 
    \right]
    ,\qquad n \ll \mathbb{E}_\infty[n]
\end{equation}
Note, that the behavior is different from those of $\mathbb{P}_T[\tau\,\vert\,X_0]$ as in \eqref{eq:P[T]=left tail 2exp}, but the constant term is the same. This term essentially gives the probability that there is only one jump at $t=0$ and this jump has zero amplitude. For the exponential distribution $p(t)$ this probability is nothing but
\begin{equation}
    \mathbb{P}_T[\tau=0\,\vert\,X_0]\approx
    \mathbb{P}_N[n=0\,\vert\,X_0]\approx
    \int_{\frac{X_0}{\alpha}}^{\infty} p(t) \dd t = 
    e^{- \frac{\beta}{\alpha} X_0}
    .
\end{equation}
For the right tail of the distribution (atypically large values of $n$) we have
\begin{equation}\label{eq:P[n]=right tail 2exp}
    \mathbb{P}_N[n\,\vert\,X_0] 
    \underset{X_0\to\infty}{\sim}
    \exp\left[
        - n \log \frac{(\beta+\alpha\gamma)^2}{4\alpha\beta\gamma}
    \right]
    ,\qquad n \gg \mathbb{E}_\infty[n].
\end{equation}
We complete the description of the marginal probability $\mathbb{P}_N[n\,\vert\,X_{0}]$ by noting that as expected the decay in the right tail \eqref{eq:P[n]=right tail 2exp} is the same as the decay of the survival probability $S_N(n\,\vert\,X_0)$ as given in \eqref{eq:Sn=asympt 2exp}. The results of the numerical simulations are shown in Fig.~\ref{fig:numerics_LDF_2exp}.

\section{Exactly solvable case II. Exponential time distribution, fixed jumps}\label{sec:exact2}
This section is devoted to another exactly solvable case, where the time intervals are exponentially distributed while all jumps are the same,
\begin{equation}\label{eq:p,q=fixedM}
    p(t) = \beta e^{-\beta t}, \qquad
    q(M) = \delta(M-M_0).
\end{equation}
For the choice \eqref{eq:p,q=fixedM} both the renewal equation and effective random walk approaches can be carried out explicitly. Note that since the jumps are fixed, $n$ uniquely determines $\tau$ by the first-passage condition \eqref{eq:0=X0-first-passage condition},
\begin{equation}\label{eq:X(tau)=0_fixM condition}
    X_0 - \alpha\tau - n M_0 = 0.
\end{equation}
Hence the joint probability distribution must be proportional to the $\delta$-function, 
\begin{equation}\label{eq:P[tau,n]~delta_fixM}
    \mathbb{P}[\tau,n\,\vert\,X_0] \sim 
    \delta\left( \tau -  \frac{ X_0 + nM_0}{\alpha} \right).
\end{equation}
This means that we can focus on either one of the marginal probability distributions $\mathbb{P}_{T}[\tau\,\vert\,X_0]$ and $\mathbb{P}_N[n\,\vert\,X_0]$, and the other can then be easily recovered. However, we will keep track of both $n$ and $\tau$ for pedagogical purposes.

\subsection{Renewal equation approach} 
We now apply the renewal equation approach to the case where the jump amplitudes are fixed. This follows a procedure similar to that in Sec.~\ref{sec:exact1_renew}, where both $p(t)$ and $q(M)$ were exponential distributions.

\par First, we substitute the probability distributions \eqref{eq:p,q=fixedM} into the renewal equation \eqref{eq:renewal_Q}. 
Due to the fixed jump amplitudes, the integral equation simplifies, and only one integral remains instead of the double one seen in \eqref{eq:renewal_Q_2exp_1}.
Specifically, we have:
\begin{equation}\label{eq:renewal_fix_M-2}
    Q(\rho,s\,\vert\,X_0) 
        = s \, e^{ - \frac{\rho+\beta}{\alpha}(X_0 + M_0) }
     + 
    \frac{\beta s}{\alpha} 
    \int_{0}^{X_0 + M_0} 
        e^{- \frac{\beta  + \rho}{\alpha}(X_0+M_0-y)  } 
     Q(\rho,s\,\vert\, y)\dd y.
\end{equation}
To transform this integral equation into a differential form, we take the derivative with respect to $X_0$, arriving at:
\begin{equation}\label{eq:renewal_fix_M-3}
\dv{}{X_0}\left[ 
    Q(\rho,s\,\vert\,X_0) \, 
        e^{ \frac{\rho+\beta}{\alpha}(X_0 + M_0) }
    \right] 
    =   
    \frac{\beta s}{\alpha} \,
    e^{ \frac{\beta  + \rho}{\alpha} (X_0+M_0)  } \, 
        Q(\rho,s\,\vert\, X_0+M_0)
\end{equation}
or, equivalently,
\begin{equation}\label{eq:diff_fixM}
  \left[ \dv{}{X_0} + \frac{\rho+\beta}{\alpha} \right] 
    Q(\rho,s\,\vert\,X_0) = 
\frac{\beta s}{\alpha} Q(\rho,s\,\vert\,X_0+M_0).
\end{equation}
In contrast to the previous case with exponential $p(t)$ and $q(M)$, which led to a second-order differential equation, here we obtain a first-order delayed differential equation (see e.g., \cite{Smith2010}). The general solution to \eqref{eq:diff_fixM} has the form
\begin{equation}\label{eq:Q(E)=general_fixM}
  Q(\rho,s\,\vert\,X_0) = A\, e^{-\omega X_0}.
\end{equation}
Substituting this ansatz into \eqref{eq:diff_fixM}, we find that $\omega$ is determined by solving
\begin{equation}\label{eq:omega_eq_fixM}
  -\omega + \frac{\rho+\beta}{\alpha} = \frac{\beta s}{\alpha} e^{-\omega M_0}.
\end{equation}
The constant $A$ in \eqref{eq:Q(E)=general_fixM} is fixed by noting that $Q(\rho,s\,\vert\, - M_0) = s $, which is evident from~\eqref{eq:renewal_fix_M-2}. Therefore, the explicit expression for the generating function reads:
\begin{equation}\label{eq:Q(E)=_fixM}
  Q(\rho,s\,\vert\,X_0) = s \, e^{- \omega (X_0+M_0) }.
\end{equation}
Note that, once again, we are able to easily reduce the integral equation \eqref{eq:renewal_fix_M-2} to a differential equation \eqref{eq:diff_fixM} because the inter-jump intervals are exponentially distributed. 

\par
\begin{figure}[h]
\includegraphics[width=\linewidth]{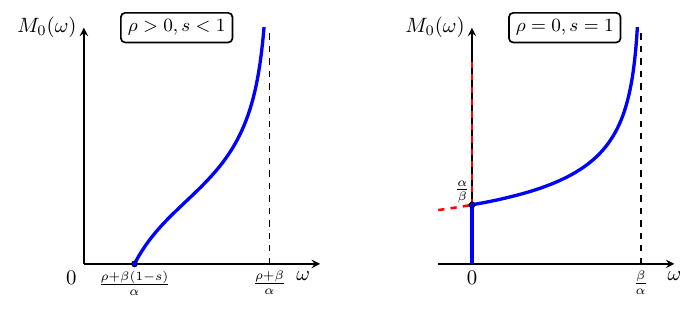}
\caption{Plots of $M_0(\omega)$ for $\rho>0$, $s<1$ (left) and for $\rho=0$, $s=1$ (right). Physical solution with $\omega\ge0$ is shown by the solid blue line. The non-physical solution for $\rho=0$, $s=1$ is shown by the dashed red line. }\label{fig:plots_w_fixM}
\end{figure}

 To analyze the solutions of the transcendental equation \eqref{eq:omega_eq_fixM}, we rewrite it as:
\begin{equation}\label{eq:M_0(omega)}
  M_0 = -\frac{1}{\omega} \log\left[
    \frac{1}{s} \left( \frac{\rho}{\beta} + 1 - \omega \frac{\alpha}{\beta} \right)
  \right].
\end{equation}
By plotting $M_0$ as a function of $\omega$ (see Fig.~\ref{fig:plots_w_fixM}), we identify two distinct scenarios:
\begin{itemize}
    \item If $s<1$ or $\rho>0$, there exists a unique positive solution of \eqref{eq:omega_eq_fixM}. This solution ensures that the generating function decays exponentially as $X_0 \to \infty$.
    \item If $\rho=0$ and $s=1$, the behavior changes.
    For $M_0 \ge \frac{\alpha}{\beta}$, there is a unique positive solution, and the generating function decays exponentially as $X_0\to\infty$. 
    For $M_0 < \frac{\alpha}{\beta}$, however, the solution for $\omega$ obtained from \eqref{eq:M_0(omega)} is negative. This would lead to an non-physical exponential growth of the generating function as $X_0\to\infty$. In this case, we select the trivial solution $\omega=0$ of \eqref{eq:omega_eq_fixM} instead.
\end{itemize}
The combined solution for $\omega$ in these cases is illustrated in Fig.~\ref{fig:plots_w_fixM}.

\par In fact, \eqref{eq:omega_eq_fixM} can be transformed into a standard transcendental equation. Rewriting it as
\begin{equation}
    \left( \omega - \frac{\rho+\beta}{\alpha} \right) M_0\; 
    e^{\left(\omega-\frac{\rho+\beta}{\alpha}\right)M_0}
    =
    -\frac{s \beta}{\alpha} M_0
    e^{-\frac{\rho+\beta}{\alpha}M_0},
\end{equation}
we obtain an equation of the form $x e^{x} = z$, where $z\in[-(\sfrac{1}{e});0]$. 
This is a well-known transcendental equation whose solutions are expressed in terms of the Lambert $W$-function \cite{CGHJK-96}. 
In particular, it has two real solutions: $x = W_0(z)$ and $x=W_{-1}(z)$. 
In our case, the physically relevant solution corresponds to the principal branch $W_0$. Specifically, we have:
\begin{equation}\label{eq:omega=LW_fixM}
  \omega = \frac{\rho+\beta}{\alpha} 
  + \frac{1}{M_0} W_0\left[
    -\frac{s\beta}{\alpha} M_0 
        e^{-\frac{\rho+\beta}{\alpha}M_0}
  \right].
\end{equation}
The other non-physical solution, represented by the dashed red line in Fig.~\ref{fig:plots_w_fixM}, that appears for $\rho=0$ and $s=1$ corresponds to the $W_{-1}$ branch.

\subsection{Effective random walk approach}
\begin{figure}[h]
\includegraphics{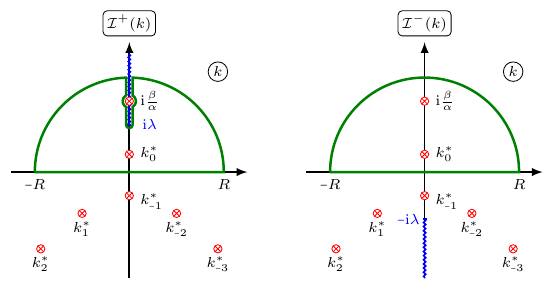}
\caption{Analytic structure of $\mathcal{I}^+(k)$ (left) and $\mathcal{I}^-(k)$ (right) in \eqref{eq:phi^pm=byparts_fixM} for the probability distributions \eqref{eq:p,q=fixedM}. The integrals over semi-circles vanish when $R\to\infty$.}
\label{fig:I^pm_fixM_anal_structure}
\end{figure}

Having determined the Laplace transform of the probability distribution using the renewal equation approach, we now derive the same result by leveraging the effective-time random walk framework introduced in Sec.~\ref{sec:framework_RW}. For the probability distributions defined in \eqref{eq:p,q=fixedM}, the effective random walk is characterized, according to \eqref{eq:c(rho)=def} and \eqref{eq:F(k)=def}, by:
\begin{equation}\label{eq:c(rho)=fixM}
    F(k;\rho) = - e^{\ii k M_0} \frac{\ii  \frac{\beta}{\alpha}}{
    k - \ii \frac{\beta}{\alpha}},
    \qquad
    c(\rho) = e^{ - \rho \frac{M_0}{\alpha} }.
\end{equation}
Our next goal is to find the functions $\phi^{\pm}(\lambda;\rho,s)$, which are essential for computing the Laplace transform of the first-passage time distribution \eqref{eq:LTQ=generalIvanov}. This procedure is similar to those presented in Sec.~\ref{sec:exact1_RW}. Specifically, we first integrate \eqref{eq:phi^pm=def} by parts
\begin{equation}\label{eq:phi^pm=byparts_fixM}
    \phi^{\pm}(\lambda;\rho, s) = \exp\left[  
    \frac{1}{2\pi \ii} \int_{-\infty}^{\infty}
    \dd k\;  \mathcal{I}^{\pm}(k)
    \right],
    \quad
    \mathcal{I}^\pm(k) = \mp \log[ k \mp \ii \lambda] 
    \frac{\partial_k F(k;\rho)}{\frac{1}{s\, c(\rho)}-  F(k;\rho)}.
\end{equation}
The integrals are then computed by analyzing their analytic structure in the complex $k$-plane. 
Specifically, the integrands $\mathcal{I}^{\pm}(k)$ have a branch cut originating from the logarithmic term $\log[ k \mp \ii \lambda]$, and a pole at $k=\ii(\beta/\alpha)$ arising from the numerator $\partial_k F(k;\rho)$. Furthermore, the denominator gives rise to infinitely many poles, which are solutions to the transcendental equation:
\begin{equation}\label{eq:k*i=LW}
    \frac{1}{s \, c(\rho)} = F(k;\rho) :
    \qquad
    k_\ell^{*}(\rho,s) =
        \ii \left( \frac{\beta}{\alpha} +  
            \frac{1}{M_0} W_\ell\left[
                    - s \frac{\beta}{\alpha} M_0 
                        e^{ - \frac{\rho+\beta}{\alpha}M_0}
            \right]
            \right),
\end{equation}
where $W_\ell$ denotes the $\ell$th branch of the Lambert $W$-function. The analytic structure of $\mathcal{I}^{\pm}(k)$ is shown in Fig.~\ref{fig:I^pm_fixM_anal_structure}. 

\par A key observation is that all poles except for $k_0^{*}$, corresponding to the principal branch of the Lambert $W$-function, lie in the lower half-plane. 
Therefore only the pole at $k=k_0^{*}(\rho,s)$ contributes when closing the contour in the upper half-plane. 

\par 
Following the same procedure as in Sec.~\ref{sec:exact1_RW}, which involves computing the residues at $k=\ii(\beta/\alpha)$ and $k=k_0^*$, and carefully accounting for the contribution from the branch cut integral in $\mathcal{I}^+(k)$, we arrive at the explicit forms of the functions $\phi^{\pm}(\lambda;\rho,s)$:
\begin{equation}\label{eq:phi^pm=fixM}
    \phi^+(\lambda;\rho,s) = 
        \frac{\lambda+\ii k_0^*(\rho,s)}{\lambda-\beta/\alpha}
        \frac{1}
             {1-s\,c(\rho)\, F(\ii\lambda; \rho)}
    ,
    \quad
    \phi^-(\lambda;s,\rho) = 
        \frac{\lambda+\beta/\alpha}{\lambda - \ii k_0^*(\rho,s)}.
\end{equation}
Substituting these expressions into the generalized Pollaczek-Spitzer formula for the effective random walk \eqref{eq:LTQ=generalIvanov}, we obtain the triple Laplace transform \eqref{eq:Q(rho,s|lambda)=def} of $\mathbb{P}[\tau,n\,\vert\,X_0]$:
\begin{equation}\label{eq:hat(Q)=fixM}
    \hat{Q}(\rho,s\,\vert\,\lambda)
    = \frac{1}{\lambda+\frac{\rho}{\alpha}} 
    + \frac{1}{\lambda+\frac{\rho}{\alpha}} \frac{\ii k_0^{*}(\rho,s)}{\beta/\alpha}
    \frac{\lambda+\frac{\rho+\beta}{\alpha}}
         {\lambda+\frac{\rho}{\alpha}-\ii k_0^*(\rho,s)}.
\end{equation}
Note, that this transform has the same functional form as the one obtained in the first exactly solvable case \eqref{eq:dEs^nP[n|E]=2exp}, a consequence of the integrands $\mathcal{I}^{\pm}(k)$ in \eqref{eq:I^pm=byparts} and \eqref{eq:phi^pm=byparts_fixM} corresponding to the Fourier transforms of the effective random walks \eqref{eq:F(k)=2exp} and \eqref{eq:c(rho)=fixM} having the same analytic structure in the upper part of the complex $k$-plane.

\par Inverting the transform $\lambda\mapsto X_0$ in \eqref{eq:hat(Q)=fixM}, we arrive at:
\begin{equation}\label{eq:Q=answer_RW_fixM}
    Q(\rho,s\,\vert\,X_0) = 
    \left( 1 + \ii k_{0}^{*}(\rho,s) \frac{\alpha}{\beta}\right)
    e^{- \left(\frac{\rho}{\alpha} - \ii k_{0}^{*}(\rho,s)\right) X_0}.
\end{equation}
To verify that we have obtained the same result as in the renewal equation approach \eqref{eq:Q(E)=_fixM}, note that the Lambert function satisfies $W(x)e^{W(x)} = x$, hence
\begin{equation}\label{eq:(1+ik)e^kM=s}
     \left( 1 + \ii k_{0}^{*}(\rho,s) \frac{\alpha}{\beta}\right)
    e^{\left(\frac{\rho}{\alpha} - \ii k_{0}^{*}(\rho,s)\right) M_0}
    = s.
\end{equation}
Applying this identity to \eqref{eq:Q=answer_RW_fixM} we obtain
\begin{equation}\label{eq:Q(rho,s|X0)=fixM_RW}
    Q(\rho,s\,\vert\,X_0) = 
    s\,
    e^{- \left(\frac{\rho}{\alpha} - \ii k_{0}^{*}(\rho,s)\right) (X_0 + M_0)}.
\end{equation}
Finally, by comparing \eqref{eq:omega=LW_fixM} and \eqref{eq:k*i=LW}, it becomes clear that $\omega = \frac{\rho}{\alpha} - \ii k_0^{*}$, and that \eqref{eq:Q(rho,s|X0)=fixM_RW} is indeed equivalent to \eqref{eq:Q(E)=_fixM}. This equivalence confirms that the renewal equation approach and the effective random walk method yield identical results.

\subsection{First-passage properties}
We have shown that both the renewal equation approach and the effective random walk approach yield identical expressions for the Laplace transform of the joint probability distribution
\begin{equation}\label{eq:LTP[tau,n|x]=fixM_0}
    \int_{0}^{\infty} \dd \tau\, e^{-\rho\tau} 
    \sum_{n=0}^{\infty} s^{n}\,
    \mathbb{P}[\tau,n\,\vert\,X_0]
    = Q(\rho,s\,\vert\,X_0).
\end{equation}
Specifically, from either \eqref{eq:Q(E)=_fixM} or \eqref{eq:Q(rho,s|X0)=fixM_RW}, we obtain
\begin{equation}\label{eq:LTP[tau,n|x]=fixM}
    Q(\rho,s\,\vert\,X_0) = 
    s \exp\left[
        - 
        \left(
        \frac{\rho+\beta}{\alpha}
        + \frac{1}{M_0} 
        W_{0}\left[
            -s \frac{\beta M_0}{\alpha} e^{-\frac{\rho+\beta}{\alpha}M_0}
        \right]
        \right) (X_0+M_0)
    \right].
\end{equation}
Below, we analyze \eqref{eq:LTP[tau,n|x]=fixM} and extract the first-passage properties, obtaining the results stated in Sec.~\ref{sec:Res_case2}.

\par Analogously to the case where both $p(t)$ and $q(M)$ are exponential distributions, the transform \eqref{eq:LTP[tau,n|x]=fixM_0} can be inverted explicitly if the process starts at the origin. The inversion relies on the expansion of the Lambert $W$-function for small arguments~\cite{CGHJK-96}:
\begin{equation}\label{eq:W(z)=expansion}
    W_{0}(z) = \sum_{n=1}^{\infty} \frac{(-n)^{n-1}}{n!} z^n, \qquad z\to0.
\end{equation}
First, we use \eqref{eq:W(z)=expansion} to expand \eqref{eq:LTP[tau,n|x]=fixM} in series with respect to $s$, which allows us to invert the transform $s\mapsto n$.  The subsequent inversion $\rho\mapsto\tau$ is then straightforward. The resulting joint distribution takes the form:
\begin{equation}
    \mathbb{P}[\tau,n\,\vert\,X_0=0]
    = \frac{1}{n!} e^{-\beta \tau}
    \left( n\,\frac{\beta M_0}{\alpha}\right)^{n-1}
    \delta\left(
        \tau - n \frac{M_0}{\alpha}
    \right).
\end{equation}
The $\delta$-function highlights the deterministic relation \eqref{eq:X(tau)=0_fixM condition} between the first-passage time $\tau$ and the number of jumps $n$ before the first-passage.

\begin{figure}[h]
    \includegraphics{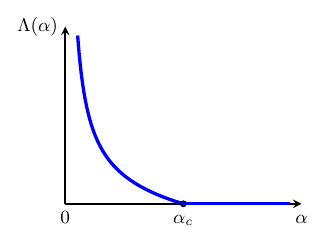}
    \caption{Plot of $\Lambda(\alpha)$ from \eqref{eq:Lambda(a)=def}, which determines the survival probability $S_\infty(X_0)$ as in \eqref{eq:Sinf(X)=cases fixM}. For $\alpha \ge \alpha_c$, we have $\Lambda(\alpha) = 0$ and thus $S_\infty(X_0) = 0$, indicating that the process eventually crosses the origin. For $\alpha < \alpha_c$, $\Lambda(\alpha) > 0$, resulting in a finite probability to survive indefinitely.}
    \label{fig:Lambda(a)_plt}
\end{figure}

\par
\begin{figure}[h]
\includegraphics[width=.5\linewidth]{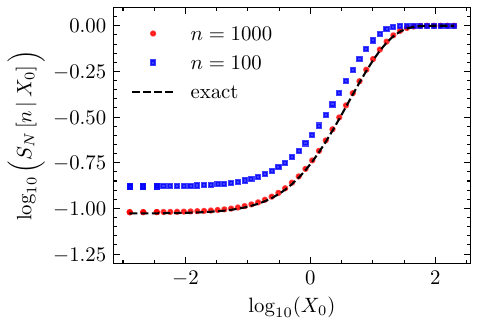}
\caption{
Survival probability $S_N(n\,\vert\,X_0)$ as a function of $X_0$ for $n=1000$ (red circles) and $n=100$ (blue squares) computed numerically and exact expansion for $S_\infty(X_0)$ as in \eqref{eq:Sinf(X)=cases fixM} (black dashed line).
 The model parameters are $\alpha=2$,  $\beta=2$, $M_0=1.05$. For the details of the simulations see Appendix~\ref{sec:numerics}.}\label{fig:Sinf_fixM}
\end{figure}

 The first quantity we compute is the probability to survive at the infinite time \eqref{eq:S_infty(X0)}. It is readily computed from \eqref{eq:LTP[tau,n|x]=fixM} by setting $\rho=0$ and $s=1$:
\begin{equation}\label{eq:Sinf(X)=cases fixM}
  S_\infty(X_0) = 
    1 - \exp\left[ - \Lambda(\alpha) \frac{X_0+M_0}{M_0} \right],
\end{equation}
where
\begin{equation}\label{eq:Lambda(a)=def}
    \Lambda(\alpha) = \frac{\alpha_c}{\alpha}
      + W_0\left[
        -\frac{\alpha_c}{\alpha}
            e^{-\frac{\alpha_c}{\alpha}}
      \right],
      \qquad
      \alpha_c = \beta M_0.
\end{equation}
The behavior of the survival probability is governed by the function $\Lambda(\alpha)$, plotted in Fig.~\ref{fig:Lambda(a)_plt}. This function reveals the existence of three distinct regimes:
\begin{enumerate}
    \item $\alpha<\alpha_c$, 
        \textit{survival regime},
    \item $\alpha=\alpha_c$,
        \textit{critical point}, 
    \item $\alpha>\alpha_c$,
        \textit{absorption regime}.
\end{enumerate}
The numerical verification of \eqref{eq:Sinf(X)=cases fixM} is shown in Fig.~\ref{fig:Sinf_fixM}. 

\par To gain a more refined description of the first-passage properties, we analyze the survival probabilities for the finite times and number of jumps. These are given by the inverse Laplace transforms \eqref{eq:S_T=inverseLT} and \eqref{eq:S_N=inverseT}:
\begin{align}
\label{eq:S_T(tau)=inverseLT_fixM}
    & S_T(\tau\,\vert\,X_0) = \frac{1}{2\pi \ii} \int_{\mathcal{C}_1} \dd \rho \; e^{\rho \tau} 
    \left. 
        \frac{1}{\rho}\left(1 - Q(\rho,s\,\vert\,X_0)\right)
    \right|_{s=1},\\
\label{eq:S_N(n)=inverseT_fixM}
    & S_N(n\,\vert\,X_0) =
    \frac{1}{2\pi\ii} 
    \oint_{\mathcal{C}_0} \dd s\, 
    \frac{1}{s^{n+1}} \frac{1}{1-s} (1 - Q(\rho,s\,\vert\,X_0))
    \Big|_{\rho=0}.
\end{align}
We analyze these survival probabilities separately for each regime. The computations closely follow those in Sec.~\ref{sec:exact1_fpp}, so we briefly outline the main steps and present the results.

\subsubsection{Survival regime}
When the drift is weak ($\alpha<\alpha_c$), the process has a finite probability of remaining above the origin indefinitely. In this regime, we analyze the survival probabilities $S_T(\tau\,\vert\,X_0)$ and $S_N(n\,\vert\,X_0)$, characterizing how they approach the asymptotic value $S_\infty(X_0)$ and obtaining the results \eqref{eq:res_survival S=isinglike} with the ``correlation lengths'' given by~\eqref{eq:res_corrLengths=fixM}. 

\par
\begin{figure}[h]
    \includegraphics{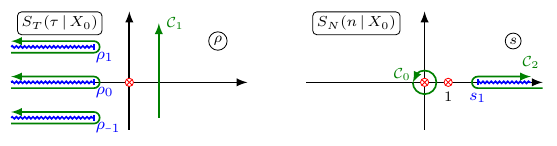}
    \caption{
    Analytic structure of the Laplace transform $S_T(\tau\,\vert\,X_0)$ (left) and of the generating function $S_N(n\,\vert\,X_0)$ for the distributions \eqref{eq:p,q=fixedM} in the survival regime. 
    }\label{fig:S_survival_fixM}
\end{figure}

 We commence with the survival probability $S_T(\tau\,\vert\,X_0)$. To analyze its behavior at large times, we need to understand the analytic structure of the Laplace transform given by \eqref{eq:S_T(tau)=inverseLT_fixM}. A key element is the Lambert $W$-function, $W_0(z)$, which appears in the expression for $Q(\rho, s\,\vert\,X_0)$ as in \eqref{eq:LTP[tau,n|x]=fixM}.  The function $W_0(z)$ has a branch cut along the negative real axis for $z<-1/e$. This branch cut induces branch cuts in $Q(\rho, 1\,\vert\,X_0)$ that have a form:
\begin{equation}
    \rho \in \left(-\infty+ 2\pi\ii \; \frac{\alpha}{M_0}; \rho_\ell\right],
    \quad
    \rho_\ell =  -\beta +\frac{\alpha}{M_0}\left(1+\log \frac{\beta M_0}{\alpha}\right) + 2\pi \ii \frac{\alpha}{M_0} \ell, 
\end{equation}  
where $\ell$ is an integer. Consequently the integrand \eqref{eq:S_T(tau)=inverseLT_fixM} has infinitely many cuts and a pole at $\rho=0$ (see Fig.~\ref{fig:S_survival_fixM}).

\par To find the asymptotic behavior of the inverse Laplace transform \eqref{eq:S_T(tau)=inverseLT_fixM}, we deform the contour $\mathcal{C}_1$ to encircle the branch cuts. This deformation separates the contribution of the pole at $\rho=0$ from the contributions of the branch cuts. Representing the integrals along the cuts as integrals over the real line, we obtain:
\begin{multline}\label{eq:S_t=res+int fixM}
    S_T(\tau\,\vert\,X_0)= \Res_{\rho=0}
    \left[
        \frac{e^{\rho\tau}}{\rho}(1-Q(\rho,1\,\vert\,X_0))
    \right]
    \\
    +\sum_{\ell=-\infty}^{\infty} \int_{-\infty}^{\rho^*}\frac{\dd\rho}{2\pi\ii} \Delta\left[
        \frac{\exp\left(\rho\tau + 2\pi\ii \ell \frac{\alpha}{M_0} \right)}
        {\rho+ 2\pi\ii \ell \frac{\alpha}{M_0} }
        \left(1 - Q\left(\rho+2\pi\ii\ell \frac{\alpha}{M_0}, 1 \,\vert\,X_0\right)\right)
    \right],
\end{multline}
where $\Delta[f(z)]$ is the discontinuity across the cut as defined in \eqref{eq:Delta[f]=disc 2exp}, and  $\rho^*$ is the real part of the branching points
\begin{equation}\label{eq:rho*=fixM}
 \rho^*=-\beta+\frac{\alpha}{M_0}\left(1+\log\frac{\beta M_0}{\alpha}\right)   
\end{equation}
Computing the residue in \eqref{eq:S_t=res+int fixM}, we see that it gives exactly the survival probability $S_\infty(X_0)$,
\begin{equation}
    \Res_{\rho=0}
    \left[
        \frac{e^{\rho\tau}}{\rho}(1-Q(\rho,1\,\vert\,X_0))
    \right]
    =
    1 - e^{ - \Lambda(\alpha) \frac{X_0+M_0}{M_0} } = S_\infty(X_0),
\end{equation}
where $\Lambda(\alpha)$ is given by \eqref{eq:Lambda(a)=def}. 

\par The remaining term in \eqref{eq:S_t=res+int fixM} involves the sum of integrals along the branch cuts. Since the discontinuity across each cut is purely imaginary and originates from the $(1-Q)$ term, we can rewrite \eqref{eq:S_t=res+int fixM} as:
\begin{multline}\label{eq:S_T=Sinf+SumInt2}
    S_T(\tau\,\vert\,X_0) - S_\infty(X_0) 
     \\
         = e^{\rho^*\tau}
        \sum_{\ell=-\infty}^{\infty} \int_{-\infty}^{0}
        \frac{\dd\rho}{2\pi\ii}
        \frac{(\rho+\rho^*) \, e^{\rho\tau}\Delta\left[
        \left(1 - Q\left(\rho+\rho^*+2\pi\ii\ell \frac{\alpha}{M_0}, 1 \,\vert\,X_0\right)\right)
        \right]}
             {(\rho+\rho^*)^2 + \left(2\pi \ell \frac{\alpha}{M_0}\right)^2}
\end{multline} 
The sum of integrals in \eqref{eq:S_T=Sinf+SumInt2} converges. This can be verified by first replacing the integrands by their absolute values and then bounding the discontinuity by $\left| \Delta[1-Q] \right|<2\pi$. Then the leading-order behavior of the difference between $S_T(\tau\,\vert\,X_0)$ and its long-time limit $S_\infty(X_0)$ is determined by the prefactor $e^{\rho^*\tau}$. Using the explicit form of $\rho^*$ as in \eqref{eq:rho*=fixM} we finally obtain
\begin{equation}\label{eq:S_T-S_inf=decay fixM}
    S_T(\tau\,\vert\,X_0) -S_\infty(X_0) 
        \underset{\tau\to\infty}{\asymp} 
        \exp\left[-\tau\left(
            \beta - \frac{\alpha}{M_0}-\frac{\alpha}{M_0}
                \log\frac{\beta M_0}{\alpha}
            \right)
        \right].
\end{equation}
This is exactly the result stated in \eqref{eq:res_survival S=isinglike} with $\xi_\tau(\alpha)=-(\sfrac{1}{\rho^*})$ given by \eqref{eq:res_corrLengths=fixM}. 

\par Now, let us turn to the survival probability $S_N(n\,\vert\,X_0)$. The analytic structure of the integrand in \eqref{eq:S_N(n)=inverseT_fixM} is qualitatively the same as that encountered in the case where both $p(t)$ and $q(M)$ are exponential distributions (see Fig.~\ref{fig:S_survival_fixM}). Namely, there are poles at $s=0$ and $s=1$ and a branch cut along the real axis
\begin{equation}\label{eq:s1=fixM}
    s\in[s_1\;\infty), \qquad s_1 = \frac{\alpha}{\beta M_0} e^{\frac{\beta M_0}{\alpha} - 1}.
\end{equation}
Repeating the very same arguments as in the case where both distributions are exponential, we find that the survival probability asymptotically tends to a constant 
\begin{equation}\label{eq:S_n-S_inf=decay fixM}
    S_N(n\,\vert\,X_0) - S_\infty(X_0) \underset{n\to\infty}{\asymp}
    \exp\left[
        -n \left( \frac{\beta M_0}{\alpha}  - 1
            - \log \frac{\beta M_0}{\alpha} \right)
    \right].
\end{equation}
This is the result stated in \eqref{eq:res_survival S=isinglike} with $\xi_n(\alpha)=-\frac{1}{\log s_1}$ as in \eqref{eq:res_corrLengths=fixM}.

\par Conditional mean and variances for $\tau$ and $n$ presented in \eqref{eq:res_fixM_meanT_survival}, \eqref{eq:res_fixM_meanN_survival}, \eqref{eq:res_fixM_varT_survival}, and \eqref{eq:res_fixM_varN_survival} are obtained by substituting \eqref{eq:Sinf(X)=cases fixM} and \eqref{eq:LTP[tau,n|x]=fixM} into \eqref{eq:con_<t>=def}, \eqref{eq:con_<n>=def}, \eqref{eq:con_<t2>=def} and \eqref{eq:con_<n2>=def}.

\subsubsection{Critical point.}
At the critical point ($\alpha=\alpha_c$), $\Lambda(\alpha)=0$, hence the survival probability at infinite time \eqref{eq:Sinf(X)=cases fixM} is zero. The decay rates of the survival probabilities we have computed in the survival regime \eqref{eq:S_T-S_inf=decay fixM}, \eqref{eq:S_n-S_inf=decay fixM} are zeros meaning that there is no exponential decay. 

\par The key difference in the analysis of the inverse transforms \eqref{eq:S_T(tau)=inverseLT_fixM} and \eqref{eq:S_N(n)=inverseT_fixM} at the critical point is that the pole at $\rho=0$ (or $s=1$) coincides with the endpoint of the branch cut. This confluence of singularities alters the asymptotic behavior, and the large $\tau$ (large $n$) behavior is now governed by the $\rho\to0$ (or $s\to1$) expansions of the transforms.

\par We start with $S_T(\tau\,\vert\,X_0)$. Using the exact form of $Q(\rho,s\,\vert\,X_0)$ as in \eqref{eq:LTP[tau,n|x]=fixM} and substituting it into \eqref{eq:S(tau)=1/rho(1-Q)} we find the Laplace transform of the survival probability:
\begin{equation}
    \int_{0}^{\infty} \dd \tau\, e^{-\rho \tau}
    S_T(\tau\,\vert\,X_0) =
    \frac{1}{\rho}\left(1 -  
    \exp\left[
    - \left(1 + \frac{X_0}{M_0}\right)
      \left( 1 + \frac{\rho}{\beta} + 
        W_0\left[-e ^{-\frac{\rho+\beta}{\beta}} \right]
      \right)
    \right]
    \right).
\end{equation}
Expanding this expression for small $\rho$ reveals a square root singularity
\begin{equation}\label{eq:int S ~ 1/sqrt(rho)}
    \int_{0}^{\infty} \dd \tau\, e^{-\rho \tau}\, 
    S_T(\tau\,\vert\,X_0) \sim 
    \left( 1 + \frac{X_0}{M_0} \right)
    \sqrt{\frac{2}{\beta}} \, \frac{1}{\sqrt{\rho}},
    \qquad \rho \to 0.
\end{equation}
This divergence suggests that the survival probability decays as $1/\sqrt{\tau}$. Substituting this ansatz into \eqref{eq:P[tau,n|0]=2exp} we obtain the large time behavior:
\begin{equation}\label{eq:S(tau)~X0_fixM}
    S_{T}(\tau\,\vert\,X_0) 
    \underset{\tau \to \infty}{\sim}
    \left( 1 + \frac{X_0}{M_0} \right)
    \sqrt{\frac{2}{\pi \beta}} \, \frac{1}{\sqrt{\tau}}.
\end{equation}
Large $n$ behavior of $S_N(n\,\vert\,X_0)$ is found in the similar way. First we substitute the exact form of $Q(\rho,s\,\vert\,X_0)$ as in \eqref{eq:LTP[tau,n|x]=fixM} into the representation \eqref{eq:S(n)=1/(1-s) (1-Q)} for the generating function of $S_N(n\,\vert\,X_0)$ and arrive at
\begin{equation}
    \sum_{n=0}^{\infty}s^{n} S_{N}(n\,\vert\,X_0)
    =\frac{1}{1-s}
    \left(1 - s\,
         \exp\left[-\left(1 + \frac{X_0}{M_0}\right) 
             \left(1 + W_0\left[-\frac{s}{e}\right]\right)
         \right]
    \right).
\end{equation}
Then we replace the sum with an integral, and repeat the same procedure as for $S_T(\tau\,\vert\,X_0)$ with $\rho$ replaced by $-\log s$, we find the large $n$ behavior of $S_N(n\,\vert\,X_0)$:
\begin{equation}\label{eq:S(n)~X0_fixM}
    S_{N}(n\,\vert\,X_0) \underset{n\to\infty}{\sim}
    \left(1 + \frac{X_0}{M_0}\right) 
    \sqrt{\frac{2}{\pi}}
    \frac{1}{\sqrt{n}}.
\end{equation}

\paragraph{Sparre-Anderson theorem.}
It is instructive to briefly revisit the scenario where both $p(t)$ and $q(M)$ are exponential. In that case, we were able to compute the survival probability $S_N(n\,\vert\,X_0=0)$ using the Sparre-Anderson theorem \eqref{eq:SparreAnderson_Sn}, which predicted a decay of $\frac{1}{\sqrt{\pi n}}$.
However, in the present case of fixed jump amplitudes, as is clear from \eqref{eq:S(n)~X0_fixM}, we find a different asymptotic behavior:
\begin{equation}\label{eq:S_n~sqrt(2/pi n)}
    S_{N}(n\,\vert\,X_0=0) \underset{n\to\infty}{\sim} 
    \sqrt{\frac{2}{\pi n}}.
\end{equation}
This decay, while still a power law, differs from the decay predicted by the Sparre-Anderson theorem \eqref{eq:SparreAnderson_Sn_asympt} by a factor of $\sqrt{2}$.

\par 
This discrepancy, however, represents neither a contradiction nor an error in our calculations. 
The Sparre-Andersen theorem \eqref{eq:SparreAnderson_Sn} applies to symmetric random walks with a continuous cumulative distribution function. However, the effective random walk described by \eqref{eq:c(rho)=fixM} at the critical point, despite having zero mean displacement, is genuinely asymmetric. This asymmetry makes the Sparre-Anderson theorem inapplicable.

\par Interestingly, the decay in \eqref{eq:S_n~sqrt(2/pi n)} also appears in the survival probability of the symmetric lattice random walk (see, e.g., \cite{MajumdarSchehr2024}). In this case, the process is symmetric but does violate the assumption of a continuous cumulative distribution function.

\paragraph{Scaling limit.}
At the critical point, the system exhibits a scaling limit analogous to that observed in the double exponential case (Sec.~\ref{sec:caseI_critical}). As the underlying calculations are essentially identical, we directly state the resulting scaling behaviors:
\begin{align}
\label{eq:S(tau)~erf()_fixM}
    & S_T(\tau\,\vert\,X_0) \sim
    \erf\left( \frac{X_0}{M_0} \frac{1}{\sqrt{2\beta\tau}} \right),
    \quad \tau\to\infty,
    \quad \frac{X_0}{\sqrt{\tau}}\text{~--- fixed},
    \\
\label{eq:S(n)~erf()_fixM}
    & S_N(n\,\vert\,X_0)\sim
     \erf\left( \frac{X_0}{M_0} \frac{1}{ \sqrt{2 n} }  \right)
     ,
    \quad n\to\infty,
    \quad \frac{X_0}{\sqrt{n}}\text{~--- fixed}.
\end{align}
Comparing these results to their counterparts obtained in the case where both $p(t)$ and $q(M)$ are exponential, we observe that in both cases, the scaling behavior can be uniformly characterized as
\begin{align}
& S_T(\tau\,\vert\,X_0) \sim
    \erf\left( 
        \frac{X_0}{\sqrt{\tau}} 
        \sqrt{\frac{1}{2}
                \frac{\langle t\rangle}{\left\langle (M-\alpha_c t)^2\right\rangle}} 
    \right)
    ,
    \quad \tau\to\infty,
    \quad \frac{X_0}{\sqrt{\tau}}\text{~--- fixed},
    \\
& S_N(n\,\vert\,X_0) \sim
    \erf\left( 
        \frac{X_0}{\sqrt{n}} 
        \sqrt{\frac{1}{2}
                \frac{1}{\left\langle (M-\alpha_c t)^2\right\rangle}} 
    \right),
    \quad n\to\infty,
    \quad \frac{X_0}{\sqrt{n}}\text{~--- fixed}.
\end{align}
Thus, the scaling limit is essentially the same for both cases. This universality arises because, in the scaling limit, the details of the jump distribution become irrelevant, and the process converges to a Brownian motion. As long as the effective random walk has zero mean and finite moments, we expect the Brownian limit to be universally valid, regardless of the specific forms of $p(t)$ and $q(M)$.

\par
\begin{figure}
\includegraphics[width=\linewidth]{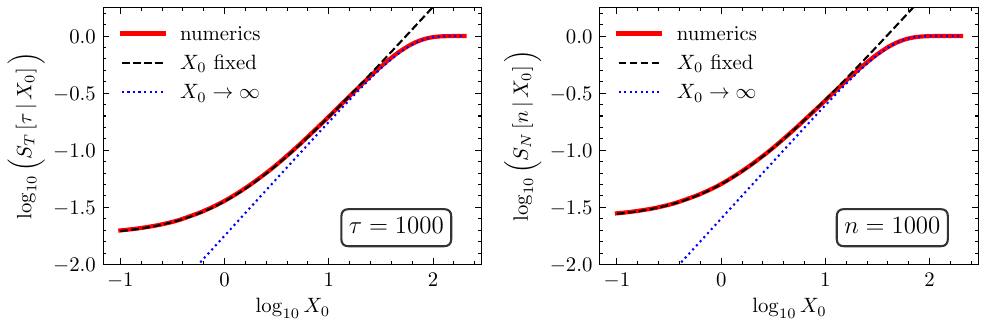}
\caption{Survival probabilities as functions of the initial position with fixed $\tau=1000$ (left) and $n=1000$ (right). The black dashed lines correspond to the fixed $X_0$ behavior described by \eqref{eq:S(tau)~X0_fixM} and \eqref{eq:S(n)~X0_fixM}. The blue dotted lines correspond to the scaling limit as in \eqref{eq:S(tau)~erf()_fixM} and \eqref{eq:S(n)~erf()_fixM}. The model parameters are $\alpha=2$, $\beta=2$, $M_0=1$. For the details of the simulations see Appendix~\ref{sec:numerics}.}\label{fig:S(tau)S(n)_fixM}
\end{figure}

 We conclude the analysis at the critical point by verifying the analytical results \eqref{eq:S(tau)~X0_fixM}, \eqref{eq:S(n)~X0_fixM}, \eqref{eq:S(tau)~erf()_fixM}, and \eqref{eq:S(n)~erf()_fixM} by performing numerical simulations (see Fig.~\ref{fig:S(tau)S(n)_fixM}).

\subsubsection{Absorption regime} 
If the drift is strong ($\alpha>\alpha_c$), then the process is strongly pulled toward the origin and the first-passage event eventually occurs. In the analytic structure of the inverse transforms \eqref{eq:S_T(tau)=inverseLT_fixM}, \eqref{eq:S_N(n)=inverseT_fixM} of the survival probabilities this is manifested in the absence of the poles at $\rho=0$ and $s=1$. The disappearance of the poles implies that in the leading order the behavior of the survival probabilities is governed  by the branching points $\rho^*$ as in \eqref{eq:rho*=fixM} and $s_1$ as in \eqref{eq:s1=fixM}. Specifically, we have 
\begin{align}
\label{eq:St=asympt fixM}
    & S_T(\tau\,\vert\,X_0) \underset{\tau\to\infty}{\asymp}
        \exp\left[-\tau\left(
            \beta - \frac{\alpha}{M_0}-\frac{\alpha}{M_0}
                \log\frac{\beta M_0}{\alpha}
            \right)
        \right],\\
\label{eq:Sn=asympt fixM}
    & S_N(n\,\vert\,X_0)
    \underset{n\to\infty}{\asymp}
        \exp\left[
        -n \left( \frac{\beta M_0}{\alpha}  - 1
            - \log \frac{\beta M_0}{\alpha} \right)
    \right].
\end{align}
This is exactly the exponential decay \eqref{eq:res_absorption S=isinglike} with the ``correlation lengths'' given by \eqref{eq:res_corrLengths=fixM}. 

\par Below, we analyze the probability distribution $\mathbb{P}[\tau,n\,\vert\,X_0]$ in more detail obtaining the remaining results stated in Sec.~\ref{sec:Res_case2}.
Specifically, we compute the mean and variance of both $\tau$ and $n$, and determine the large deviation forms for their marginal probability distributions.

\paragraph{Moments of $\tau$ and $n$.} 
First, we compute the first two moments of $\tau$ and $n$ to characterize their behavior near typical values. This is achieved by expanding the Laplace transform $Q(\rho,s\,\vert\,X_0)$ in series with respect to $\rho$ and $s$, then using \eqref{eq:<tau>=..<tau2>=..} and \eqref{eq:<n>=, <n2>=} to relate the expansion terms to the moments. From the exact representation \eqref{eq:LTP[tau,n|x]=fixM}, we obtain for $\tau$:
\begin{equation}\label{eq:<T>=, Var[T]=, fixM}
  \mathrm{E}[\tau\,\vert\,X_0 ]
    = \frac{M_0+X_0}{\alpha - \beta M_0},
  \qquad
  \mathrm{Var}\left[\tau\,\vert\,X_0\right] 
  = \frac{ X_0+M_0}{(\alpha-\beta M_0)^3}\, \beta M_0^2,
\end{equation}
and for $n$:
\begin{equation}\label{eq:<n>=, Var[n]=, fixM}
  \mathrm{E}[ n\,\vert\,X_0 ] = \frac{\alpha + \beta X_0}{\alpha - \beta M_0},
  \qquad 
  \mathrm{Var}[n\,\vert\,X_0] = \frac{X_0 + M_0}{(\alpha-\beta M_0)^3} \, \beta \alpha^2.
\end{equation}
The means and variances depend linearly on the initial position $X_0$, and diverge as $\alpha$ approaches its critical value $\alpha_c = \beta M_0$.

\paragraph{Correlation coefficient}
To quantify the correlation between the $\tau$ and $n$, we compute the correlation coefficient
\begin{equation}\label{eq:corr(tau,n)=fixM}
    \mathrm{corr}(\tau,n) = 
        \frac{\mathbb{E}[(\tau n)\,\vert\,X_0] - \mathbb{E}[\tau\,\vert\,X_0]\;  \mathbb{E}[n\,\vert\,X_0]}
             {\sqrt{\mathrm{Var}[\tau\,\vert\,X_0]\, \mathrm{Var}[n\,\vert\,X_0] }}.
\end{equation}
Using the exact form of $Q(\rho,s\,\vert\,X_0)$ we find that if the jumps are fixed, then
\begin{equation}
    \mathrm{corr}(\tau,n) = 1.
\end{equation}
In other words, we have a perfect linear correlation between $n$ and $\tau$. This strong relationship arises directly from the first-passage condition \eqref{eq:X(tau)=0_fixM condition}. In simpler terms, it means that for each additional jump the first-passage time increases linearly, with no variance around this value.

\paragraph{Rate function for $\tau$.}
The mean and the variance of $\tau$ and $n$ grow linearly with $X_0$. Therefore we expect the marginal probability distribution to follow the large deviation form
\begin{equation}\label{eq:P[tau]=LDF_fixM}
    \mathbb{P}_{T}[\tau\,\vert\,X_0] \underset{X_0\to\infty}{\asymp}
    e^{- X_0 \Phi(z)},
    \qquad
    z = \frac{\alpha\tau}{X_0} - 1,
\end{equation}
with the scaling parameter $z$ lying in the range $z\in[0;\infty)$. This large deviation form is the same as the one in the first exactly solvable case \eqref{eq:P[tau]=LDF_2exp} but the rate function $\Phi(z)$ is different.

\par To determine the rate function we first substitute the large deviation ansatz into  the Laplace transform \eqref{eq:LTP[tau,n|x]=fixM_0} and compute the integral in the saddle-point approximation. Then we compare the result with the exponential behavior extracted from the explicit form of $Q(\rho,s\,\vert\,X_0)$ as given in \eqref{eq:LTP[tau,n|x]=fixM} and equate the exponents arriving at a Legendre transform similar to \eqref{eq:phi(rho)=max_z-2exp}. Finally, inverting this transform we find the rate function as the solution of the maximization problem
\begin{equation}\label{eq:Phi(z)=max_p-fixM}
  \Phi(z) = \max_{\rho}\left(- \frac{1}{\alpha} \rho (z+1) + \phi(\rho) \right),
\end{equation}
where
\begin{equation}\label{eq:phi(p)=fixM}
  \phi(\rho) = 
  \frac{\rho+\beta}{\alpha} 
  + \frac{1}{M_0} W_0\left[
    -\frac{\beta}{\alpha} M_0 e^{-\frac{M_0}{\alpha}(\rho+\beta)}
  \right].
\end{equation}
This maximization can be carried out explicitly. Indeed, the maximum is reached at $\rho^*$ which is the solution of $\phi'(\rho^*) = \frac{z+1}{\alpha}$. 
Note that the Lambert $W$-function satisfies
\begin{equation}\label{eq:W'(x)}
  \dv{}{x} W_0(x) = \frac{W_0(x)}{x(1+W_0(x))}.
\end{equation}
Using this identity we find that $\rho^*$ satisfies
\begin{equation}\label{eq:phi'(p*)=0}
  -\frac{1}{\alpha}\, z - 
  \frac{W_0\left[
          -\frac{\beta}{\alpha} M_0 
            e^{-\frac{M_0}{\alpha}(\rho^* + \beta)}
          \right]
  }{ 
    1 + W_0\left[
          -\frac{\beta}{\alpha} M_0 
            e^{-\frac{M_0}{\alpha}(\rho^* + \beta)}
          \right]
  }
  = 0.
\end{equation}
Solving \eqref{eq:phi'(p*)=0} with respect to the Lambert function yields
\begin{equation}\label{eq:phi'(p*)=0_1}
  W_0\left[ -\frac{\beta}{\alpha}\,
                  M_0 e^{-M_0 \frac{p^*+\beta}{\alpha}}\right] 
  = - \frac{z}{z+1}.
\end{equation}
Recall that the Lambert function $W_0(z)$ is a solution of $x e^{x} = z$ and hence we can reduce \eqref{eq:phi'(p*)=0_1} to 
\begin{equation}\label{eq:phi'(p*)=0_2}
  -\frac{\beta}{\alpha}\,
                  M_0 e^{-M_0 \frac{\rho^*+\beta}{\alpha}} 
  = - \frac{z}{z+1} \exp\left[ -\frac{z}{z+1}\right],
\end{equation}
which is readily solved, yielding
\begin{equation}\label{eq:p^*=fixM}
  \rho^* = -\beta + \frac{\alpha}{M_0} 
  \left( 
    \frac{z}{z+1} 
    - \log \left[ \frac{z}{z+1} \frac{\alpha}{\beta M_0} \right]
  \right).
\end{equation}
Substituting \eqref{eq:p^*=fixM} into \eqref{eq:Phi(z)=max_p-fixM} and using \eqref{eq:phi'(p*)=0_1}, after some simplification, we obtain the exact form of the rate function:
\begin{equation}\label{eq:Phi(z)=fixM}
  \Phi(z) = \frac{\beta}{\alpha} + z \left(
      \frac{\beta}{\alpha} - \frac{1}{M_0} + \frac{1}{M_0}
      \log \left[ \frac{z}{z+1} \frac{\alpha}{\beta M_0}\right]
  \right).
\end{equation}
Note that the rate function is very different from the one obtained in the case where both $p(t)$ and $q(M)$ are exponential distributions \eqref{eq:Phi(z)=_2exp}. 
This is also a non-negative convex function with a minimum at $z^*=\frac{\beta M_0}{\alpha-\beta M_0}$ and the asymptotic behaviors
\begin{equation}\label{eq:Phi(z)=asympt fixM}
    \Phi(z) = \left\{
    \begin{aligned}
        & \frac{\beta}{\alpha} + \frac{1}{M_0}\, z\log z,
        && z\to0, \\
        & \frac{(\alpha-\beta M_0)^3}{2\alpha^2\beta M_0^2} (z-z^*)^2,
        && z\to z^*=\frac{\beta M_0}{\alpha-\beta M_0}, \\
        & z \left(\frac{\beta}{\alpha}-\frac{1}{M_0} - \frac{1}{M_0} \log \frac{\beta M_0}{\alpha}\right),
        && z\to \infty.
    \end{aligned}
    \right.
\end{equation}
From \eqref{eq:Phi(z)=asympt fixM} we extract the asymptotic behaviors of the probability distribution $\mathbb{P}_T[\tau\,\vert\,X_0]$. Specifically, close to the typical values of $\tau$, the distribution is Gaussian
\begin{equation}\label{eq:P[t]=gaussian fixM}
    \mathbb{P}_{T}[\tau\,\vert\,X_0]
    \underset{X_0\to\infty}{\sim}
    \exp\left[
        -\frac{1}{2} 
            \frac{(\tau-\mathbb{E}_\infty[\tau])^2}
                 {\mathrm{Var}_\infty[\tau]}
    \right],
    \qquad
    \tau \approx \mathbb{E}_\infty[\tau],
\end{equation}
where the asymptotic mean and variance are
\begin{equation}
    \mathbb{E}_\infty[\tau] = 
    \frac{X_0}{\alpha -\beta M_0},
    \qquad
    \mathrm{Var}_\infty[\tau] = 
    \frac{\beta M_0^2\,X_0}{(\alpha-\beta M_0)^3}.  
\end{equation}
This Gaussian approximation describes the typical fluctuations of the first-passage time. Both the asymptotic mean and the variance are in agreement with the results obtained in~\eqref{eq:<T>=, Var[T]=, fixM} for an arbitrary $X_0$.

\par The left tail of the distribution (atypically small values of $\tau$) is given by:
\begin{equation}\label{eq:P[t]=left tail fixM}
    \mathbb{P}_T[\tau\,\vert\,X_0] 
    \underset{X_0\to\infty}{\sim}
    \exp\left[
    -X_0 \frac{\beta}{\alpha}
    +\frac{\alpha}{M_0} \,\Delta \tau 
     \log \Delta \tau
    \right],\quad
    \tau = \Delta \tau + \frac{X_0}{\alpha} \ll \mathbb{E}_\infty[\tau].
\end{equation}
Comparing this with the left tail obtained in the first exactly solvable case \eqref{eq:P[T]=left tail 2exp}, we observe a notable difference in the functional form, yet the constant term, $-X_0 \frac{\beta}{\alpha}$, remains identical. This shared constant term reflects a fundamental property of the system: the probability of reaching the origin with minimal delay is governed by the initial waiting time distribution, $p(t) = \beta e^{-\beta t}$, irrespective of the specific jump size distribution.

\par
\begin{figure}[h]
\includegraphics[width=\linewidth]{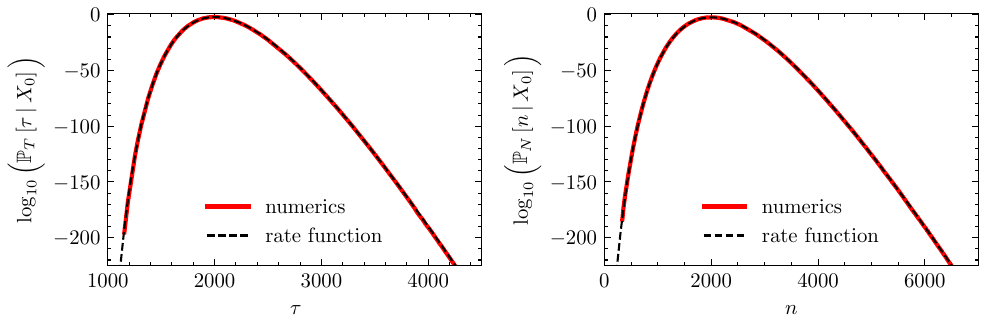}
\caption{Comparison between numerical simulation the and analytical results for the distributions of $\tau$ (left) and $n$ (right). 
The dashed black lines denote the distributions $\mathbb{P}_T[\tau\,\vert\,X_0]$ and $\mathbb{P}_N[n\,\vert\,X_0]$ that are obtained from the large deviation forms \eqref{eq:P[tau]=LDF_fixM}, \eqref{eq:Phi(z)=fixM} and \eqref{eq:P[n]=LDF_fixM}, \eqref{eq:Psi(nu)=ans_fixM} respectively. Normalization constant are restored by numerical integration. 
The solid red lines correspond to the Monte-Carlo numerical simulations. The tails of the probability distributions are reached with the help of the importance sampling with the exponential tilt in the observable. The parameters of the model are $\alpha=1$, $\beta=1$, $M_0=0{.}5$, $X_0=1000$. For the details of the simulations see Appendix~\ref{sec:numerics}.}\label{fig:numerics_LDF_fixM}
\end{figure}

 Consider the event where  $\Delta \tau =  0$. In this scenario, the process must have reached the origin very quickly, implying that the initial jump was immediately followed by the drift carrying the process to the origin. In other words, to reach the origin at $\tau = X_0/\alpha$, the process must take effectively no jumps (strictly speaking, only the jump at $ t=0$ with zero amplitude). Thus, the probability of this event depends only on the initial waiting time, yielding:
\begin{equation}
    \mathbb{P}_T[\tau=0\,\vert\,X_0] \approx \int_{\frac{X_0}{\alpha}}^{\infty} p(t) \dd t = e^{- \frac{\beta}{\alpha}X_0}.
\end{equation}
For the right tail (atypically large values of $\tau$), we obtain an exponential decay with the same rate as in the survival probability \eqref{eq:St=asympt fixM}, essentially confirming \eqref{eq:res_absorption S=isinglike}. Specifically, we have:
\begin{equation}\label{eq:P[t]=right tail fixM}
    \mathbb{P}_T[\tau\,\vert\,X_0]
    \underset{X_0\to\infty}{\sim}
    \exp\left[
        -\left(\beta-\frac{\alpha}{M_0} - \frac{\alpha}{M_0}\log \frac{\beta M_0}{\alpha} \right) \tau
    \right],
    \qquad \tau \gg \mathbb{E}_\infty[\tau].
\end{equation}
This completes the description of the marginal probability $\mathbb{P}_T[\tau\,\vert\,X_{0}]$ (see Fig.~\ref{fig:numerics_LDF_fixM} for the comparison with the numerical simulations). Trajectories with atypical values of $\tau$ are again generated using the Importance Sampling strategy as explained in more detail in Appendix~\ref{sec:numerics}.

\paragraph{Rate function for $n$.} 
Given the linear growth of the mean and variance of $n$ with $X_0$, we expect the probability distribution for the number of jumps $n$ to follow the large deviation form:
\begin{equation}\label{eq:P[n]=LDF_fixM}
  \mathbb{P}_N[n\,\vert\,X_0] 
  \underset{X_0\to\infty}{\asymp}
  e^{ -X_0 \Psi(\nu) }, 
  \qquad \nu = \frac{\alpha\, n}{\beta X_0},
\end{equation}
where the scaling parameter $\nu$ lies in the range $\nu\in[0;\infty)$. This large deviation form mirrors that of the first exactly solvable case \eqref{eq:P[n]=LDF_2exp}, albeit with a different rate function~$\Psi(\nu)$.

\par 
The rate function $\Psi(\nu)$ can be determined in two ways. One approach is to  use the fact that, for fixed jump amplitudes, $\tau$ and $n$ are deterministically related via the first-passage condition  \eqref{eq:X(tau)=0_fixM condition}. Alternatively, we could perform the derivation used to obtain $\Psi(\nu)$ in the first exactly solvable case. However, since  $\tau$ and $n$ are linearly related, this derivation would be almost identical to the one we just provided for $\Phi(z)$. Therefore, we choose the more direct approach and use the first-passage condition \eqref{eq:X(tau)=0_fixM condition} to  determine~$\Psi(\nu)$.  

\par Taking  the limit $X_0\to\infty$ in \eqref{eq:P[tau,n]~delta_fixM} leads to the relationship between the scaling variables $z$ and $\nu$, i.e., $z = \nu \, \frac{\beta M_0}{\alpha}$ which allows $\Psi(\nu)$ to be obtained from $\Phi(z)$ via  
\begin{equation}
    \Psi\left( \nu\right) = \Phi\left(\nu \frac{\beta M_0}{\alpha} \right).
\end{equation}
Using the exact form \eqref{eq:Phi(z)=fixM} of the rate function $\Phi(z)$, we  
find: 
\begin{equation}\label{eq:Psi(nu)=ans_fixM}
  \Psi(\nu) = \frac{\beta}{\alpha} 
      \left(
          1 + \nu \left(M_0 \frac{\beta}{\alpha} - 1\right)
          + \nu \log\left[ \frac{\nu}{1 + \nu \frac{\beta}{\alpha} M_0}\right]
      \right).
\end{equation}
The asymptotic behavior of $\Psi(\nu)$ and the marginal probability distribution  $\mathbb{P}_N[n\,\vert\,X_0]$ then follow directly from  \eqref{eq:Phi(z)=asympt fixM}, \eqref{eq:P[t]=gaussian fixM}, \eqref{eq:P[t]=left tail fixM}, and \eqref{eq:P[t]=right tail fixM}.  For completeness, we provide a summary of these expressions below. 

\par The rate function $\Psi(\nu)$ is convex, reaching its minimum at $\nu^*=\frac{\alpha}{\alpha-\beta M_0}$. Its asymptotic behavior is given by:
\begin{equation}\label{eq:Psi(nu)=asympt fixM}
    \Psi(\nu) = \left\{
    \begin{aligned}
        & \frac{\beta}{\alpha} + \frac{\beta}{\alpha}\, \nu\log\nu,
        && \nu\to0, \\
        & \frac{\beta(\alpha-\beta M_0)^3}{2\alpha^4} (\nu-\nu^*)^2,
        && \nu\to \nu^*=\frac{\alpha}{\alpha-\beta M_0}, \\
        & \frac{\beta}{\alpha}\left(\frac{\beta M_0}{\alpha} - 1 - \log \frac{\beta M_0}{\alpha}\right) \nu,
        && \nu\to \infty.
    \end{aligned}
    \right.
\end{equation}
Near the typical values of $n$, the distribution $\mathbb{P}_N[n\,\vert\,X_0]$ follows a Gaussian:
\begin{equation}
    \mathbb{P}_{N}[n\,\vert\,X_0]
    \underset{X_0\to\infty}{\sim}
    \exp\left[
        -\frac{1}{2} 
            \frac{(n-\mathbb{E}_\infty[n])^2}
                 {\mathrm{Var}_\infty[n]}
    \right],
    \qquad
    n \approx \mathbb{E}_\infty[n],
\end{equation}
where the mean and the variance are
\begin{equation}
    \mathbb{E}_\infty[n] = 
    \frac{\beta\, X_0}{\alpha -\beta M_0},
    \qquad
    \mathrm{Var}_\infty[n] = 
    \frac{\alpha^2 \beta\,X_0}{(\alpha-\beta M_0)^3}.  
\end{equation}
The left tail (atypically small values of $n$) is given by:
\begin{equation}
    \mathbb{P}_N[n\,\vert\,X_0] 
    \underset{X_0\to\infty}{\sim}
    \exp\left[
    - X_0 \frac{\beta}{\alpha} - n \log n
    \right],
    \quad
    n  \ll \mathbb{E}_\infty[n].
\end{equation}
For the right tail (atypically large values of $n$) we have an exponential decay with the same rate as for the survival probability  \eqref{eq:Sn=asympt fixM}, i.e., 
\begin{equation}
    \mathbb{P}_N[n\,\vert\,X_0]
    \underset{X_0\to\infty}{\sim}
    \exp\left[
        - n \left( 
            \frac{\beta M_0}{\alpha} - 1 - \log \frac{\beta M_0}{\alpha}
        \right)
    \right],
    \qquad n \gg \mathbb{E}_\infty[n].
\end{equation}
This concludes the analysis of the marginal distribution $\mathbb{P}_{N}[n\,\vert\,X_0]$ in the second exactly solvable case (see Fig.~\ref{fig:numerics_LDF_fixM} for the comparison with the numerical simulations). With this, we have now completed the analysis of the first-passage properties for both exactly solvable cases.


\section{Conclusion}\label{sec:conclusion}

\par In this paper, we have studied the first-passage properties of a jump process with a constant drift. By mapping this process onto a discrete-time random walk and applying the generalized Pollaczek-Spitzer formula, we have derived a closed-form expression for the (triple) Laplace transform of the joint probability of $n$ and $\tau$. To exemplify this approach, we have performed a detailed analysis of two exactly solvable cases, where calculations could be carried out in a well-controlled manner, and a more standard renewal equation approach is also applicable.

\par The main advantage of our approach is that, unlike the renewal equation method, it does not require solving integral equations, which is generally a challenging task unless the inter-jump distribution is exponential.
Instead, it provides a closed-form expression for the Laplace transform of the joint distribution of $n$ and $\tau$, thereby enabling a systematic asymptotic analysis of systems with arbitrary inter-jump and jump amplitude distributions. This analysis can be performed using techniques developed in \cites{CM-05,MCZ-05,ZMC-07,MSV-12,MMS-13,MMS-14,MMS-17,MMS-18,BMS-21,BMS-21b}. We emphasize that, although we map the original process onto a discrete-time random walk, this mapping retains information about both the first-passage time $\tau$ and the number of jumps $n$ before absorption. This level of detail would be lost if one were to directly apply the Pollaczek-Spitzer formula to the original process.

\par We have shown that the model exhibits two distinct regimes: the \textit{survival regime} and the \textit{absorption regime}, separated by the \textit{critical point}. In the absorption regime, the drift is strong enough to drive the process toward the origin, leading to an exponential decay of the survival probabilities. In the survival regime, where the drift is weak, there exists a finite probability that first-passage never occurs, and the process remains positive indefinitely. In this case, at large times, the survival probabilities asymptotically approach a constant, decaying exponentially. The transition between these two regimes defines the critical point, where the survival probability follows a power-law decay rather than an exponential one. This behavior is reminiscent of phase transitions in the two-dimensional Ising model, with the drift velocity playing a role analogous to that of the temperature.

\par A natural question that arises is to what extent the results obtained in this paper for two exactly solvable cases can be generalized to arbitrary distributions $p(t)$ and $q(M)$. We believe that the general structure illustrated in Fig.~\ref{fig:StSn=scheme}, as well as the asymptotic behaviors of the survival probabilities, remains valid, at least for light-tailed distributions. However, a more thorough analytical investigation is required.

\par Another interesting direction for future work is to extend this study to systems with multiple particles, with or without interactions. In particular, one could focus on the fraction of trajectories that have not crossed the origin up to a fixed observation time and analyze how this quantity is influenced by the initial conditions. We expect to observe the memory effects similar to those reported in \cites{BM-07,DG-09,KM-12,BMRS-20,BJC-22,BMR-23,SD-23,DMS-23,JRR-23,JRR-23b,BHMMRS-23,HMS-23,BMR-24,MDS-24,SM-24}.

\section*{Acknowledgements}
\noindent We acknowledge support from ANR Grant No. ANR-23-CE30-0020-01 EDIPS. We also thank M.\,J.\,Kearney for helpful discussions and for pointing out the references \cites{P-59,Z-91,S-93,DP-97,PSZ-99}.


\begin{appendix}

\section{Numerical strategy}\label{sec:numerics}

\par This appendix provides details on the numerical simulations.

\par The plots in the survival regime (Fig.\ref{fig:Sinf_2exp}, Fig.\ref{fig:Sinf_fixM}) and at the critical point (Fig.\ref{fig:S(tau)S(n)_2exp}, Fig.\ref{fig:S(tau)S(n)_fixM}) are obtained through direct trajectory generation. In this method, we fix the number of jumps, $N$, and generate $N$ pairs $(t_j, M_j)$ by independently sampling from the distributions $p(t)$ and $q(M)$. This process is repeated to generate $10^6$ trajectories. The survival probabilities are then calculated as the fraction of trajectories that remain positive up to the $n$th jump (or up to the time $\tau$).

\par When analyzing the survival probability $S_T(\tau\,\vert\,X_0)$, it is crucial to choose a sufficiently large value of $N$ to ensure that all trajectories either cross the origin or reach the total time greater than $\tau$. Indeed, if the total time at the $N$th step is smaller than $\tau$, it becomes unclear how such trajectories contribute to the survival probability $S_T(\tau\,\vert\,X_0)$.
We set $N = 1000$ for the plots shown in Fig.~\ref{fig:Sinf_2exp} and Fig.~\ref{fig:Sinf_fixM} (survival regime), and $N=10\,000$  for the plots shown in Fig.~\ref{fig:S(tau)S(n)_2exp} and Fig.~\ref{fig:S(tau)S(n)_fixM} (critical point).

\par The plots in the absorption regime (Fig.~\ref{fig:numerics_LDF_2exp}, Fig.~\ref{fig:numerics_LDF_fixM}) display the probability distributions up to very small probabilities ($\approx 10^{-150}$). To generate the trajectories with such atypical values of $\tau$ (or $n$) we employ Importance Sampling Monte Carlo (see \cite{H-24} for a pedagogical introduction). This is a widely used strategy for studying rare event statistics in various systems \cite{H-02,EMH-04,H-11,H-14,MFH-23,MHW-24,BMR-24b}. The key idea behind the Importance Sampling is to sample trajectories from a biased distribution. Here we focus on the distribution $\mathbb{P}_T[\tau\,\vert\,X_0]$ as the strategy is the same for both observables.

\par In this paper, we use Importance Sampling with an exponential tilt in the observable. Specifically, we sample $\tau$ from the biased distribution
\begin{equation}\label{eq:tildeP=is}
    \tilde{\mathbb{P}}_{T}[\tau\,\vert\,X_0]
    = \frac{e^{\theta \tau }}{Z(\theta)} 
    \mathbb{P}_{T}[\tau\,\vert\,X_0],
\end{equation}
where $\theta$ is a parameter (quasi-temperature), and $Z(\theta)$ is a normalization constant. The positive (negative) values of $\theta$ correspond to increasing the statistical weight of the trajectories with atypically large (small) values of $\tau$.

\par To sample $\tau$ from the biased distribution \eqref{eq:tildeP=is}, we apply the Metropolis-Hastings algorithm on the space of the trajectories. First, we generate a trajectory $\{(t_j, M_j)\}$ with a fixed number of jumps $N$. At each step, the algorithm proposes a new trajectory  $\{(t_j', M_j')\}$ by randomly selecting $r$ out of the $N$ pairs $(t_j, M_j)$ and resampling them from $p(t)$ and $q(M)$. The proposed trajectory is then accepted with the probability
\begin{equation}
\{(t_j, M_j)\} \mapsto \{(t_j', M_j')\} :
    \qquad
    p_\text{acc}=\min\left(1, \exp\left[ \theta (\tau' -\tau)\right]\right),
\end{equation}
where $\tau$ and $\tau'$ are the first-passage times corresponding to the trajectories $\{(t_j, M_j)\}$ and $\{(t_j', M_j')\}$. The parameter $r$ is chosen to maintain an acceptance rate above $70\%$, with typical values for $r$ in the range $[1, 10]$. The length of the trajectory $N$ is chosen large enough to ensure that all generated trajectories cross the origin. This is particularly important when studying the tail of the distribution for atypically large values of $\tau$ (or $n$). 
In our simulations we set $N = 10\,000$. 
For the quasi-temperatures $\theta$ we typically use $100$ values in the range $[-2,.5]$ and perform $10^{7}$ metropolis steps for each value of $\theta$. 

\par Once the biased trajectories are generated, their statistical weight is rescaled by a factor of $Z(\theta) e^{-\theta \tau}$. The normalization constant $Z(\theta)$ is then determined by fitting the histograms obtained for different values of $\theta$ in the region of overlap, where the biased distributions with different $\theta$ align (see \cite{BMR-24b} for details of the fitting procedure). This procedure results in the plots shown in Fig.~\ref{fig:numerics_LDF_2exp} and Fig.~\ref{fig:numerics_LDF_fixM}.

\end{appendix}

\sloppy
\printbibliography

\end{document}